\pdfoutput=1
\documentclass[prd,twocolumn,eqsecnum,nofootinbib,floatfix]{revtex4}
\usepackage{amsmath}
\usepackage{bm}
\usepackage{graphicx} 
\usepackage{float} 
\newcommand{\A}{{\cal A}} 
\newcommand{\R}{{\cal R}}
\newcommand{\N}{{\cal N}}
\newcommand{\E}{{\cal E}} 
\newcommand{\B}{{\cal B}} 
\newcommand{\stf}[1]{{\langle {#1} \rangle}}
\newcommand{\Aa}[1]{{\cal A}^{\scriptscriptstyle \sf #1}} 
\newcommand{\EE}[1]{{\cal E}^{\scriptscriptstyle \sf #1}} 
\newcommand{\EEup}[2]{{\cal E}^{{\scriptscriptstyle \sf #1}{\, #2}}}  
\newcommand{\EEd}[1]{\dot{\cal E}^{\scriptscriptstyle \sf #1}}  
\newcommand{\EEdup}[2]{\dot{\cal E}^{{\scriptscriptstyle \sf #1}{\, #2}}}  
\newcommand{\EEdd}[1]{\ddot{\cal E}^{\scriptscriptstyle \sf #1}} 
\newcommand{\BB}[1]{{\cal B}^{\scriptscriptstyle \sf #1}} 
\newcommand{\BBup}[2]{{\cal B}^{{\scriptscriptstyle \sf #1}{\, #2}}}  
\newcommand{\BBd}[1]{\dot{\cal B}^{\scriptscriptstyle \sf #1}}  
\newcommand{\BBdup}[2]{\dot{\cal B}^{{\scriptscriptstyle \sf #1}{\, #2}}}  
\newcommand{\BBdd}[1]{\ddot{\cal B}^{\scriptscriptstyle \sf #1}} 
\newcommand{\PP}[1]{{\cal P}^{\scriptscriptstyle \sf #1}} 
\newcommand{\PPd}[1]{\dot{\cal P}^{\scriptscriptstyle \sf #1}} 
\newcommand{\PPdup}[2]{\dot{\cal P}^{{\scriptscriptstyle \sf #1}{\, #2}}}  
\newcommand{\QQ}[1]{{\cal Q}^{\scriptscriptstyle \sf #1}} 
\newcommand{\QQd}[1]{\dot{\cal Q}^{\scriptscriptstyle \sf #1}} 
\newcommand{\QQdup}[2]{\dot{\cal Q}^{{\scriptscriptstyle \sf #1}{\, #2}}}  
\newcommand{\GG}[1]{{\cal G}^{\scriptscriptstyle \sf #1}} 
\newcommand{\GGd}[1]{\dot{\cal G}^{\scriptscriptstyle \sf #1}} 
  
\newcommand{\HH}[1]{{\cal H}^{\scriptscriptstyle \sf #1}} 
\newcommand{\HHd}[1]{\dot{\cal H}^{\scriptscriptstyle \sf #1}} 
\newcommand{\HHdup}[2]{\dot{\cal H}^{{\scriptscriptstyle \sf #1}{\, #2}}}  
\newcommand{\ee}[2]{e^{\scriptscriptstyle \sf #1}_{#2}} 
\newcommand{\bb}[2]{b^{\scriptscriptstyle \sf #1}_{#2}} 
\newcommand{\pp}[2]{p^{\scriptscriptstyle \sf #1}_{#2}} 
\newcommand{\qq}[2]{q^{\scriptscriptstyle \sf #1}_{#2}} 
\newcommand{\g}[2]{g^{\scriptscriptstyle \sf #1}_{#2}} 
\newcommand{\hh}[2]{h^{\scriptscriptstyle \sf #1}_{#2}} 
\newcommand{\dilog}{\,\mbox{dilog}\,}
\allowdisplaybreaks
\begin{document}
\title{Geometry and dynamics of a tidally deformed black hole} 
\author{Eric Poisson} 
\affiliation{Department of Physics, University of Guelph, Guelph,
  Ontario, N1G 2W1, Canada;} 
\affiliation{Canadian Institute for Theoretical Astrophysics,
  University of Toronto, Toronto, Ontario, M5S 3H8, Canada} 
\author{Igor Vlasov} 
\affiliation{Department of Physics, University of Guelph, Guelph,
  Ontario, N1G 2W1, Canada} 
\date{October 7, 2009} 
\begin{abstract} 
The metric of a nonrotating black hole deformed by a tidal interaction
is calculated and expressed as an expansion in the strength of the
tidal coupling. The expansion parameter is the inverse length scale
$\R^{-1}$, where $\R$ is the radius of curvature of the external
spacetime in which the black hole moves. The expansion begins at order
$\R^{-2}$, and it is carried out through order $\R^{-4}$. The metric
is parameterized by a number of tidal multipole moments, which specify 
the black hole's tidal environment. The tidal moments are
freely-specifiable functions of time that are related to the Weyl
tensor of the external spacetime. At order $\R^{-2}$ the metric
involves the tidal quadrupole moments $\E_{ab}$ and $\B_{ab}$. At
order $\R^{-3}$ it involves the time derivative of the quadrupole
moments and the tidal octupole moments $\E_{abc}$ and $\B_{abc}$. At
order $\R^{-4}$ the metric involves the second time derivative of
the quadrupole moments, the first time derivative of the octupole
moments, the tidal hexadecapole moments $\E_{abcd}$ and $\B_{abcd}$,
and bilinear combinations of the quadrupole moments. The metric is
presented in a light-cone coordinate system that possesses a clear
geometrical meaning: The advanced-time coordinate $v$ is constant on
past light cones that converge toward the black hole; the angles
$\theta$ and $\phi$ are constant on the null generators of each light
cone; and the radial coordinate $r$ is an affine parameter on each
generator, which decreases as the light cones converge toward the
black hole. The coordinates are well-behaved on the black-hole
horizon, and they are adjusted so that the coordinate description of
the horizon is the same as in the Schwarzschild geometry: $r = 2M 
+ O(\R^{-5})$. At the order of accuracy maintained in this work, the
horizon is a stationary null hypersurface foliated by apparent
horizons; it is an isolated horizon in the sense of Ashtekar and
Krishnan. As an application of our results we examine the induced 
geometry and dynamics of the horizon, and calculate the rate at which
the black-hole surface area increases as a result of the tidal
interaction.
\end{abstract} 
\maketitle

\section{Introduction and overview} 
\label{sec:intro} 

\subsection*{This work and its context} 

Our main goal in this work is to calculate the gravitational field of
a nonrotating black hole that is deformed by a tidal interaction with
external bodies. We assume that the tidal interaction is weak, and
that it changes slowly; we make no other assumptions, and 
describe the tidal environment in the most general terms compatible
with the main assumptions. This work is a continuation of a line of
inquiry that was initiated by Manasse \cite{manasse:63} in the early
nineteen sixties; our implementation is much more general, and much
more accurate, than Manasse's original work. 

We first introduce the two length scales that are relevant to this
problem. (Throughout the paper we use relativistic units and set
$G=c=1$.) The first is $M$, the mass of the black hole. The second is
$\R$, the radius of curvature of the external spacetime associated
with the external bodies, evaluated at the black hole's position. We
assume that the scales are widely separated, so that
\begin{equation} 
M \ll \R. 
\label{small_tide} 
\end{equation} 
This condition ensures that the tidal interaction is weak, and allows
us to speak meaningfully of a black hole moving in an external
spacetime; when $M$ is comparable to $\R$, no clear distinction can be
made between the ``black hole'' and the ``external spacetime.'' We
refer to the approximation scheme based on the condition $M/\R
\ll 1$ as the {\it small-tide approximation}. 

As a concrete example we may consider a situation in which the black
hole is a member of a binary system. Let $M'$ denote the mass of the
external body, $b$ the separation between the companions, 
and $V \sim \sqrt{(M+M')/b}$ the orbital velocity. The
radius of curvature is then $\R \sim \sqrt{b^3/(M+M')}$, and    
\begin{equation}
\frac{M}{\cal R} \sim \frac{M}{M+M'}  V^3. 
\end{equation} 
We demand that this be a small quantity. There are two particular
ways to achieve this. In the {\it small-hole approximation} the
black-hole mass is assumed to be much smaller than the external mass,
so that $M/(M+M') \ll 1$; then $M/\R$ is small irrespective of the
size of $V$, and the binary system can be strongly relativistic. In
the {\it weak-field approximation} it is $V$ that is assumed to be
small, while the mass ratio is left unconstrained; here the two
companions can have comparable masses, or the black hole can be much
larger than its companion, but the mutual gravity between the bodies
must be weak. The small-hole and weak-field approximations are
particular instances of the more general requirement that 
$M/\R \ll 1$; they are both incorporated within our small-tide
approximation.  

The effects of a tidal field on the structure of spacetime around a 
black hole were first investigated by Manasse \cite{manasse:63}, in
the specific context of the small-hole approximation. Using techniques
similar to those exploited in this paper, Manasse calculated the
metric of a small black hole that falls radially toward a much larger
black hole. Each black hole was taken to be nonrotating, and the small
hole was taken to move on a geodesic of the (unperturbed)
Schwarzschild spacetime of the large hole. The case of circular motion 
around a large Schwarzschild black hole was treated much later by
Poisson \cite{poisson:04a}, and Comeau and Poisson 
\cite{comeau-poisson:09} examined the case of circular 
motion around a Kerr black hole.   

The methods employed by Manasse could be applied beyond the small-hole 
approximation. Alvi \cite{alvi:00, alvi:03} realized that they could
be seamlessly extended to the more general context of the small-tide
approximation of Eq.~(\ref{small_tide}). Alvi exploited this insight to
calculate the tidal fields acting on a black hole in a post-Newtonian
binary system. In Alvi's work, the two bodies have comparable masses
and the black hole has a significant influence on the geometry of the
external spacetime. Alvi calculated the tidal fields to leading
(Newtonian) order in the post-Newtonian approximation to general
relativity, and specialized the orbital motion to circular orbits. His
work was later generalized to first post-Newtonian order, and to
generic orbits, by Taylor and Poisson \cite{taylor-poisson:08}.  

Alvi's work motivated an effort to improve our understanding of the
tidal interaction of black holes by constructing the black-hole
metric to high order in the coupling strength $M/\R$. More precisely,  
the metric is calculated in the black hole's local neighborhood, and
expressed as an expansion in powers of $r/\R \ll 1$, where $r$ is the
distance to the black hole; the metric is valid to all orders in
$M/r$. This program was initiated by Detweiler \cite{detweiler:01,
detweiler:05}, who calculated the metric through order $(r/\R)^2$, the
lowest order at which tidal effects appear. It was pursued by Poisson 
\cite{poisson:05}, who calculated the metric through order
$(r/\R)^3$. Here we calculate the metric through order $(r/\R)^4$, and 
present the most accurate version ever produced of the metric of a
tidally deformed black hole.  

Other authors have also contributed to this effort. Frolov and his
collaborators \cite{frolov-shoom:07, abdolrahimi-frolov-shoom:09}
examined the internal geometry of a tidally deformed black hole, and
Damour and Lecian \cite{damour-lecian:09} characterized the tidal 
deformation in terms of a polarizability and ``shape Love numbers;''
these works were restricted to axisymmetric tidal fields, a
restriction that is not made in this paper. In a genuine 
{\it tour de force}, Yunes and Gonzalez \cite{yunes-gonzalez:06}
calculated the tidal deformation of a rapidly rotating black hole to
order $(r/\R)^2$. The tidal deformation of neutron stars (and other 
types of compact bodies) has also been the subject of recent
investigations \cite{flanagan-hinderer:08, hinderer:08,
damour-nagar:09, binnington-poisson:09}. 

\subsection*{Tidal quadrupole moments} 

In the work of Detweiler \cite{detweiler:01, detweiler:05} and Poisson
\cite{poisson:05} reviewed previously, and in the work presented here,
the tidal environment is described in the most general terms
compatible with the Einstein field equations. The metric is
parameterized by freely-specifiable functions of time that serve the
specific purpose of specifying the tidal environment; these are
packaged in symmetric tracefree (STF) tensors   
$\E_{a_1 a_2 \cdots a_l}$ and $\B_{a_1 a_2 \cdots a_l}$
\cite{zhang:86}, which we refer to as the {\it tidal multipole
moments} of the black-hole spacetime. (The tidal moments can be
thought of as Cartesian tensors; they are symmetric and tracefree in
all pairs of indices.)  At order $(r/\R)^2$ the metric involves the
quadrupole moments $\E_{ab}$ (5 functions) and $\B_{ab}$ (5
functions). At order $(r/\R)^3$ the metric involves the time
derivative of the quadrupole moments, as well as the octupole moments
$\E_{abc}$ (7 functions) and $\B_{abc}$ (7 functions). And at order
$(r/\R)^4$ the metric involves the second time derivative of the
quadrupole moments, the first time derivative of the octupole
moments, the hexadecapole moments $\E_{abcd}$ (9 functions) and
$\B_{abcd}$ (9 functions), and bilinear combinations of the quadrupole
moments. This generality is a substantial virtue of our work, which
unifies and extends earlier works \cite{suen:86b, alvi:00, alvi:03,
fang-lovelace:05} that examined special cases.   

The metric of a tidally deformed black hole is obtained by integrating
the vacuum field equations in the local neighborhood of the black
hole. The field equations leave the tidal moments undetermined, and
the metric is presented as a functional of these arbitrary moments. In
applications of this framework the tidal environment must be
specified, and this is achieved by making appropriate choices for the
tidal moments. In practice this typically requires matching the local
metric to a global metric that includes the black hole and the
external bodies that are responsible for the tidal interaction. For
example, Taylor and Poisson \cite{taylor-poisson:08} and 
Johnson-McDaniel {\it et al.}\ \cite{johnsonmcdaniel-etal:09} carried 
out such a matching in the context of the slow-motion approximation,
in which the mutual gravity between the black hole and the external
bodies is weak. The global metric was expressed as a post-Newtonian
expansion, within which the black hole can justifiably be represented
as a point particle. These authors determined the quadrupole moments 
$\E_{ab}$ and $\B_{ab}$ to first post-Newtonian order, and their work
could easily be extended to obtain the higher multipole moments that
also appear in the black-hole metric. In general the nonlinearities of
the field equations imply that tidal fields depend on the black-hole
mass $M$. In the small-hole approximation, however, the black hole can
be treated as a test body, and this dependence disappears; in this case 
the determination of the tidal moments is simplified \cite{manasse:63,
poisson:04a, comeau-poisson:09}. 

\subsection*{Light-cone coordinates} 

The choice of coordinates is often critical in the construction of a
metric and the exploration of its properties. This is all the more
true in the case of black-hole spacetimes, which require coordinates
that are well-behaved on the horizon. We have given a lot of attention
to the selection of coordinates, and have chosen to work with a system 
$(v,r,\theta,\phi)$ that is specifically tailored to describe the
geometry of past light cones. We refer to these as  
{\it light-cone coordinates}. The coordinates have a clear geometrical
meaning: The advanced-time coordinate $v$ is constant on past light
cones that converge toward the black hole; the angles $\theta$ and
$\phi$ (which we collectively denote $\theta^A$, with the upper-case 
Latin index $A$ running from 2 to 3) are constant on the null
generators of each light cone; and the radial coordinate $r$ is an
affine parameter on each generator, which decreases as the light cones
converge toward the black hole. Through order $(r/\R)^3$ the radial 
coordinate doubles as an areal radius, in the sense that the area of
each two-surface $(v,r) = \mbox{constant}$ is equal to $4\pi r^2 [1 +
O(r^4/\R^4)]$; this property is lost at order $(r/\R)^4$. In addition,
the radial coordinate is tuned so that the coordinate description of
the black-hole horizon is the same as in the Schwarzschild spacetime:
$r = 2M[1 + O(M^5/\R^5)]$. The light-cone coordinates are well-behaved
across the horizon.  

The choice of coordinates is inspired from the work of Bondi 
{\it et al.} \cite{bondi-etal:62} and Sachs~\cite{sachs:62}, in which
light-cone coordinates were utilized to construct the metric of an  
asymptotically-flat spacetime. (Here the coordinates were based on
future light cones that expand toward future null infinity.) It is
inspired also by the work of Ellis and collaborators on observational
cosmology \cite{ellis-etal:85, ellis-etal:92a, ellis-etal:92b,
ellis-etal:92c, ellis-etal:92d, ellis-etal:94}, in which the metric of
an expanding universe is constructed from observations made by a
typical cosmological observer. Light-cone coordinates were also
introduced by Synge in his classic textbook on general relativity
\cite{synge:60}; we followed his methods closely in this work and 
its precursors \cite{poisson:04a, poisson:04b, poisson:05,
preston-poisson:06a}.     

In addition to the quasi-spherical system of light-cone coordinates
$(v,r,\theta^A)$, we find it convenient to introduce also a
quasi-Cartesian variant $(v, x^a)$. The spatial coordinates $x^a$
(with the lower-case Latin index $a$ running from 1 to 3) are
constructed in the usual way from the quasi-spherical coordinates $(r,
\theta^A)$; we have the relations $x = r\sin\theta\cos\phi$,
$y=r\sin\theta\sin\phi$, and $z = r\cos\theta$, which we collectively
denote $x^a = r\Omega^a(\theta^A)$. 

\subsection*{Black-hole metric} 

The metric of a tidally deformed black hole is obtained by
constructing a perturbation of the Schwarzschild solution, which 
describes a nonrotating black hole in complete isolation. The
Schwarzschild metric is presented in the Eddington-Finkelstein
coordinates $(v,r,\theta^A)$, and the perturbation is presented in a
light-cone gauge that preserves the geometrical meaning of the
background coordinates. To construct the perturbation we rely on the
covariant and gauge-invariant formalism of Martel and Poisson
\cite{martel-poisson:05}, and on the formulation of the light-cone
gauge by Preston and Poisson \cite{preston-poisson:06b}. At orders
$(r/\R)^2$ and $(r/\R)^3$ the perturbation satisfies the vacuum field
equations linearized with respect to the Schwarzschild solution. At
order $(r/\R)^4$ the perturbation satisfies nonlinear field
equations. 

In the quasi-Cartesian coordinates $(v, x^a)$ the building blocks of
the metric are the tidal potentials introduced in
Tables~\ref{tab:E_cart}, \ref{tab:B_cart}, \ref{tab:EE_cart},
\ref{tab:BB_cart}, \ref{tab:EBeven_cart}, and \ref{tab:EBodd_cart} of
Sec.~\ref{sec:tidal}; these are generated by the tidal multipole
moments introduced previously. The black-hole metric appears in 
Eqs.~(\ref{blackhole_metric_cart}) of Sec.~\ref{subsec:blackhole}, and
it is expressed in terms of the radial functions listed in
Tables~\ref{tab:linear_functions} and \ref{tab:nonlinear_functions}.
In the quasi-spherical coordinates $(v,r,\theta^A)$ the tidal
potentials are listed in Tables~\ref{tab:E_ang}, \ref{tab:B_ang},
\ref{tab:EE_ang}, \ref{tab:BB_ang}, \ref{tab:EBeven_ang}, and
\ref{tab:EBodd_ang} of Sec.~\ref{sec:tidal}, and expressed as
expansions in spherical-harmonic functions (see
Table~\ref{tab:Ylm}). The black-hole metric appears in 
Eqs.~(\ref{blackhole_metric_ang}) of Sec.~\ref{subsec:blackhole}, and  
it involves the same set of radial functions. 

An important property of the black-hole metric is the fact, mentioned
previously, that in our light-cone coordinates, the position of the
horizon is given by 
\begin{equation} 
r_{\rm horizon} =2M[1 + O(M^5/\R^5)], 
\label{horizon_radius} 
\end{equation} 
the same relation as in the unperturbed Schwarzschild spacetime. This
information allows us to investigate the nature and dynamics of the
horizon's geometry, which is determined by the horizon's induced
metric as well as the expansion and shear of the horizon's 
generators. An important outcome of this investigation is the
statement that  
\begin{equation} 
\Theta = O(M^5/\R^6). 
\label{horizon_expansion} 
\end{equation} 
This represents the rate at which the congruence of null generators 
expand, and we find that within the degree of accuracy maintained in
this work, the generators are stationary. This implies that the
horizon of a tidally deformed black hole is foliated by apparent
horizons. The black-hole horizon is an {\it isolated horizon}
in the sense of Ashtekar and Krishnan \cite{ashtekar-krishnan:02,
ashtekar-krishnan:03, ashtekar-krishnan:04}.   

\subsection*{Tidal heating} 

A dynamical consequence of the tidal interaction is the fact that the
black hole grows in size at a rate described by 
\begin{align} 
\frac{\kappa_0}{8\pi} \dot{\cal A}  &= 
\frac{16}{45} M^6 \Bigl( \dot{\E}_{ab} \dot{\E}^{ab} 
 + \dot{\B}_{ab} \dot{\B}^{ab} \Bigr) 
\nonumber \\ & \quad \mbox{} 
+ \frac{16}{4725} M^8 \Bigl( \dot{\E}_{abc} \dot{\E}^{abc} 
 + \frac{16}{9} \dot{\B}_{abc} \dot{\B}^{abc} \Bigr) 
\nonumber \\ & \quad \mbox{} 
+ O(M^9/\R^9).
\label{heating} 
\end{align} 
Here $\cal A$ is the surface area of a cross-section
$v=\mbox{constant}$ of the black-hole horizon, and $\kappa_0 :=
(4M)^{-1}$ is the surface gravity of the unperturbed horizon. An
overdot indicates differentiation with respect to $v$, and the
right-hand side of the equation involves the tidal moments 
$\E_{ab}$, $\E_{abc}$, $\B_{ab}$, and $\B_{abc}$. We refer to the
growth of area that results from the tidal interaction as the 
{\it tidal heating} of the black hole by the external bodies. This
choice of terminology deserves an explanation.  

The changes in black-hole parameters that result from time-dependent, 
external processes have been the subject of investigation by many
researchers, starting from the pioneering work of Teukolsky and Press
\cite{teukolsky-press:74}. It is useful to classify the physical
processes that alter the configuration of a black hole as {\it fast
processes} on the one hand, and {\it slow processes} on the
other. Fast processes occur on a time scale that is shorter than or
comparable to the black-hole mass $M$, and these produce
(electromagnetic and/or gravitational) radiation that is partially
absorbed by the black hole; in this case the changes in black-hole
parameters that result from the interaction are radiative
changes. Slow processes, on the other hand, occur on a time scale that
is long compared with the black-hole mass; while these continue to
produce changes in the black-hole parameters, the phenomenon no longer
possesses a radiative character, and the phrase ``tidal heating''
captures the physics better than the phrase ``black-hole absorption.''
It is good to point out that the mathematical formulation of the
phenomenon by Teukolsky and Press \cite{teukolsky-press:74} (see also
Ref.~\cite{poisson:04d}) is valid both for fast and slow processes,
and is insensitive to matters of interpretation.   

The tidal heating of black holes was recently investigated by Alvi
\cite{alvi:01}, Poisson \cite{poisson:04d, poisson:05}, Taylor and  
Poisson \cite{taylor-poisson:08}, and Comeau and Poisson
\cite{comeau-poisson:09}, building on earlier work by Poisson
and Sasaki \cite{poisson-sasaki:95} and Tagoshi, Mano, and Takasugi  
\cite{tagoshi-mano-takasugi:97}. The notion of tidal work, tidal
torque, and tidal heating was put on a firm relativistic footing by
Purdue \cite{purdue:99}, Favata \cite{favata:01}, and  
Booth and Creighton \cite{booth-creighton:00}. In a recent work
\cite{poisson:09}, Poisson compared the equations that describe the 
rate of change of the black-hole mass, angular momentum, and surface
area that result from a tidal interaction with external bodies, with
the equations that describe how tidal forces do work, torque, and
produce heat in a Newtonian, viscous body; the equations are
strikingly similar, and the correspondence between the Newtonian-body
and black-hole results is revealed to hold in near-quantitative
detail. The tidal heating and torquing of black holes was incorporated
in an effective theory of point particles by Goldberger and Rothstein
\cite{goldberger-rothstein:06} and Porto \cite{porto:08}. 

In favorable circumstances the tidal heating and torquing of a black
hole can be relevant to astrophysical sources of gravitational
waves \cite{price-whelan:01}. In particular, it is likely to be
significant in the generation of low-frequency waves that would be 
measured by a space-based detector such as LISA \cite{LISA}. For
example, Martel \cite{martel:04} showed that during a close encounter
between a massive black hole and a compact body of a much smaller
mass, up to approximately five percent of the lost orbital energy goes
toward the tidal heating of the black hole; the rest is carried off to
infinity by the gravitational waves. Hughes \cite{hughes:01}
calculated that when the massive black hole is rapidly rotating, the
tidal heating slows down the inspiral of the orbiting body, and
therefore increases the duration of the gravitational-wave
signal. These conclusions are supported by 
Hughes {\it et al.} \cite{hughes-etal:05}
and Drasco and Hughes \cite{drasco-hughes:06}.  

There are indications that the tidal heating and torquing of
black holes may have been seen in accurate numerical simulations of
the inspiral and merger of binary black holes \cite{boyle-etal:07,
chu-pfeiffer-scheel:09}. It is conceivable that as the precision of 
these simulations continue to improve in the future, the tidal heating
will be exploited as a diagnostic of numerical accuracy; the surface
area of each simulated black hole should be seen to grow in accordance
to Eq.~(\ref{heating}) instead of staying constant in time.  

\subsection*{Other applications}   

The work presented here can serve as a foundation for the construction
of initial data sets for the numerical simulation of black-hole
inspirals (see Ref.~\cite{pretorius:09} for a review of the current
state of the art). Simulations carried out thus far have largely relied
on initial data \cite{brandt-bruegmann:97, cook-pfeiffer:04} that were   
adopted more for their flexibility and convenience than their  
astrophysical realism; these initial data tend to contain a large
amount of spurious radiation and produce unwanted eccentric orbital
motion. A promising alternative strategy is to rely on a
post-Newtonian metric to construct initial data sets that give a
faithful (though approximate) description of two widely separated
black holes. Early implementations \cite{tichy-etal:03,
nissanke:06, kelly-etal:07} of this idea treated the black holes as
post-Newtonian point masses, and produced initial data that were
unreliable close to each body. An improvement was proposed by Alvi
\cite{alvi:00, alvi:03}, who recognized that the post-Newtonian
metric should be replaced by a different representation of the
gravitational field in the local neighborhood of each black hole; he
therefore patched the post-Newtonian metric to the metric of a tidally
deformed black hole in a buffer region $M \ll r \ll \R$, in which each
metric gives an acceptable representation of the true gravitational 
field. Alvi's construction was perfected by Yunes and his
collaborators \cite{yunes-tichy:06, yunes-etal:06, yunes:07}, and the
most mature implementation of this idea is contained in the recent
work of Johnson-McDaniel {\it et al.}\ \cite{johnsonmcdaniel-etal:09}.  
Because the black-hole metric presented in this paper is more accurate
than the Detweiler metric \cite{detweiler:01, detweiler:05} used by 
Johnson-McDaniel {\it et al.}, it could be involved in an improved
version of their construction. 

Another interesting avenue of application for our metric would be to
involve it in the dynamical-horizon formalism of Ashtekar and Krishnan
\cite{ashtekar-krishnan:02, ashtekar-krishnan:03,
ashtekar-krishnan:04}, which provides a purely local characterization
of the structure and dynamics of black-hole horizons. Our work could
contribute new insights to this effort by allowing the horizon
quantities to be expressed in terms of the tidal moments, which encode   
information about the external universe. In this way, our metric would
be play the role of messenger between the outside world and the
horizon. Initial steps along those lines were taken by Kavanagh and
Booth \cite{kavanagh-booth:06}. A particularly interesting question to
investigate is whether the dynamical-horizon notions of mass and
current multipole moments \cite{ashtekar-etal:04} are compatible with
the recent observation that the tidal Love numbers of a nonrotating
black hole must all be zero \cite{binnington-poisson:09}.  

\subsection*{Organization of the paper} 

The remainder of the paper is divided into five sections and four
appendices. We describe our results in Secs.~\ref{sec:tidal}, 
\ref{sec:blackhole}, and \ref{sec:horizon}, and provide derivations of 
these results in Secs.~\ref{sec:background_derivation},
\ref{sec:blackhole_derivation} and the Appendices. 

We begin in Sec.~\ref{sec:tidal} by providing definitions for the
tidal multipole moments $\E_{ab}$, $\E_{abc}$, $\E_{abcd}$, 
$\B_{ab}$, $\B_{abc}$, and $\B_{abcd}$. The tidal moments allow us to 
introduce the length scale $\R$, and they give rise to tidal
potentials that form the building blocks for the construction of the
black-hole metric.    

In Sec.~\ref{sec:blackhole} we display our expressions for the metric
of a tidally deformed black hole. We proceed in two steps. We first
consider a smooth timelike geodesic $\gamma$ in a vacuum region 
of an arbitrary spacetime, and we construct the metric of this
spacetime in a neighborhood of the world line. We refer to this
spacetime as the {\it background spacetime}, and its metric is
displayed in Eqs.~(\ref{background_metric_cart}) and
(\ref{background_metric_ang}); the metric is presented in the
light-cone coordinates $(v,r,\theta^A)$ and $(v,x^a)$ introduced
previously. We next insert a black hole of mass $M$ into the
background spacetime, place it on the world line $\gamma$, and
re-calculate the metric. The metric of the black-hole spacetime is
displayed in Eqs.~(\ref{blackhole_metric_ang}) and 
(\ref{blackhole_metric_cart}).  

In Sec.~\ref{sec:horizon} we describe the consequences of the tidal
deformation on the structure and dynamics of the black-hole
horizon. We establish the statements of Eqs.~(\ref{horizon_radius}),
(\ref{horizon_expansion}), and (\ref{heating}).  

In Sec.~\ref{sec:background_derivation} we provide a derivation of the 
background metric of Eqs.~(\ref{background_metric_cart}) and
(\ref{background_metric_ang}). In
Sec.~\ref{sec:blackhole_derivation} we present a derivation of the 
black-hole metric of Eqs.~(\ref{blackhole_metric_cart}) and
(\ref{blackhole_metric_ang}).  

In Appendix~\ref{app:spherical_harmonics} we describe how the tidal
potentials can be decomposed in (scalar, vector, and tensor) spherical 
harmonics. In Appendix~\ref{app:determinant} we calculate the
determinant of the induced metric on the black-hole horizon; the
result is displayed (but not derived) in Sec.~\ref{sec:horizon}. In
Appendix~\ref{app:horizon_details} we provide calculational details
relevant to the computation of the tidal heating in
Sec.~\ref{sec:horizon}. And in Appendix~\ref{app:alternative} we
sketch a second, alternative derivation of the background metric of 
Eqs.~(\ref{background_metric_cart}) and
(\ref{background_metric_ang}). 

\section{Tidal moments and potentials} 
\label{sec:tidal} 

The work of Zhang \cite{zhang:86} reveals that the metric of any
vacuum spacetime can be constructed in the neighborhood of any
geodesic world line and expressed in terms of two sets of tidal
multipole moments. We begin our exploration of the geometry of a
tidally deformed black hole in Sec.~\ref{subsec:tidal_def} with a
formal definition of these moments. We use them in
Sec.~\ref{subsec:tidal_scales} to specify the length and time scales
associated with the tidal environment, and in 
Sec.~\ref{subsec:tidal_parity} we describe their properties under
parity transformations. We conclude in
Secs.~\ref{subsec:tidal_pot_cartesian} and
\ref{subsec:tidal_pot_angular} with the introduction of potentials
that can be constructed from the tidal moments.   

\subsection{Definition of tidal moments} 
\label{subsec:tidal_def} 

We consider a vacuum region of spacetime in a neighborhood of a 
smooth timelike geodesic $\gamma$. The world line is described by 
the parametric relations $z^\alpha(\tau)$ in an arbitrary coordinate
system $x^\alpha$, and it is parameterized by proper time $\tau$. The
velocity vector $u^\alpha := dz^\alpha/d\tau$ is tangent to the world
line, and we construct a vectorial basis by adding to $u^\alpha$ an
orthonormal triad of vectors $e^\alpha_a(\tau)$, which we assume to be
orthogonal to $u^\alpha$ and parallel-transported on the world line;
the Latin index $a$ labels the three members of the triad. 

We use the basis $(u^\alpha, e^\alpha_a)$ to decompose tensors that
are evaluated on the world line. For example, 
\begin{subequations}
\label{Weyl} 
\begin{align} 
C_{a0b0} &:= C_{\alpha\mu\beta\nu} e^\alpha_a u^\mu e^\beta_b u^\nu,  
\\
C_{abc0} &:= C_{\alpha\beta\gamma\mu} e^\alpha_a e^\beta_b e^\gamma_c 
u^\mu,  \\
C_{abcd} &:= C_{\alpha\beta\gamma\delta} e^\alpha_a e^\beta_b
e^\gamma_c e^\delta_d 
\end{align}
\end{subequations}
are the frame components of the Weyl tensor evaluated on $\gamma$;
these are functions of proper time $\tau$. We shall also need the
frame components of its first covariant derivative,  
\begin{subequations}
\label{Weyl_first} 
\begin{align} 
C_{a0b0|c} &:= C_{\alpha\mu\beta\nu;\gamma} e^\alpha_a u^\mu
e^\beta_b u^\nu e^\gamma_c,  \\
C_{abc0|d} &:= C_{\alpha\beta\gamma\mu;\delta} e^\alpha_a e^\beta_b
e^\gamma_c e^\delta_d u^\mu, \\  
C_{abcd|e} &:= C_{\alpha\beta\gamma\delta;\epsilon} e^\alpha_a
e^\beta_b e^\gamma_c e^\delta_d e^\epsilon_e,  
\end{align}
\end{subequations}
as well as 
\begin{subequations}
\label{Weyl_second} 
\begin{align} 
C_{a0b0|cd} &:= C_{\alpha\mu\beta\nu;\gamma\delta} e^\alpha_a 
u^\mu e^\beta_b u^\nu e^\gamma_c e^\delta_d, \\ 
C_{abc0|de} &:= C_{\alpha\beta\gamma\mu;\delta\epsilon} e^\alpha_a  
e^\beta_b e^\gamma_c u^\mu e^\delta_d e^\epsilon_e, \\
C_{abcd|ef} &:= C_{\alpha\beta\gamma\delta;\epsilon\eta} e^\alpha_a   
e^\beta_b e^\gamma_c e^\delta_d e^\epsilon_e e^\eta_f,  
\end{align} 
\end{subequations} 
the frame components of its second covariant derivatives. We
manipulate frame indices as if they were associated with Cartesian
tensors; we lower them with the Kronecker delta $\delta_{ab}$,
and we raise them with $\delta^{ab}$. 

The symmetries of the Weyl tensor imply that it possesses 10 
algebraically-independent components, and these can be encoded in the
two symmetric-tracefree (STF) tensors   
\begin{subequations}
\label{tidalmoment_2} 
\begin{align}  
\E_{ab} &:= \bigl( C_{a0b0} \bigr)^{\rm STF}, \\ 
\B_{ab} &:= \frac{1}{2} \bigl( \epsilon_{apq} C^{pq}_{\ \ b0}
\bigr)^{\rm STF}. 
\end{align} 
\end{subequations} 
Here $\epsilon_{abc}$ is the permutation symbol, and the STF sign   
instructs us to symmetrize all free indices and remove all traces. 
We also use an angular-bracket notation to indicate the same
operation: if $A_{abcd}$ is an arbitrary tensor, then 
\begin{equation} 
A_\stf{abcd} := \bigl( A_{abcd} \bigr)^{\rm STF}. 
\end{equation} 
In the case of Eq.~(\ref{tidalmoment_2}) the STF operation is
superfluous, because $C_{a0b0}$ and 
$\epsilon_{apq} C^{pq}_{\ \ b0}$ are already symmetric and 
tracefree. We refer to $\E_{ab}$ and $\B_{ab}$ as the 
{\it tidal quadrupole moments} associated with the world line
$\gamma$. Each STF tensor contains 5 independent components, and  
these are functions of proper time $\tau$.  We use $\dot{\E}_{ab}$ and 
$\dot{\B}_{ab}$ to denote the derivative of each moment with respect
to $\tau$, and $\ddot{\E}_{ab}$ and $\ddot{\B}_{ab}$ to denote the
second derivatives. The derivatives of the Weyl tensor in the
direction of $u^\alpha$ can be expressed directly in terms of these
quantities.  

The symmetries of the Weyl tensor and the Bianchi identities imply 
that the spatial derivatives of the Weyl tensor --- those listed in
Eq.~(\ref{Weyl_first}) --- possess 24 algebraically-independent
components. These are encoded in the STF tensors 
\begin{subequations}
\label{tidalmoment_3} 
\begin{align}  
\E_{abc} &:= \bigl( C_{a0b0|c} \bigr)^{\rm STF}, \\ 
\B_{abc} &:= \frac{3}{8} \bigl( \epsilon_{apq} C^{pq}_{\ \ b0|c} 
\bigr)^{\rm STF}, 
\end{align} 
\end{subequations} 
and in $\dot{\E}_{ab}$ and $\dot{\B}_{ab}$. We refer to $\E_{abc}$ and
$\B_{abc}$ as the {\it tidal octupole moments} associated with the
world line $\gamma$. Each tensor contains 7 independent components,
and these are functions of proper time $\tau$.  We use
$\dot{\E}_{abc}$ and $\dot{\B}_{abc}$ to denote the derivative of each
moment with respect to $\tau$. The derivatives of the
spatially-differentiated Weyl tensor in the direction of $u^\alpha$
can be expressed directly in terms of these quantities. 

We shall also require the second spatial derivatives of the Weyl
tensor, as listed in Eqs.~(\ref{Weyl_second}). In this case there are
62 algebraically-independent components, and these are encoded in
$\E_{ab}$, $\B_{ab}$, $\ddot{\E}_{ab}$, $\ddot{\B}_{ab}$,
$\dot{\E}_{abc}$, $\dot{\B}_{abc}$, and in the new STF tensors 
\begin{subequations} 
\label{tidalmoment_4} 
\begin{align} 
\E_{abcd} &:= \frac{1}{2} \bigl( C_{a0b0|cd} \bigr)^{\rm STF}, \\
\B_{abcd} &:= \frac{3}{20} \bigl( \epsilon_{apq} C^{pq}_{\ \ b0|cd}  
\bigr)^{\rm STF}.
\end{align} 
\end{subequations} 
We refer to $\E_{abcd}$ and $\B_{abcd}$ as the {\it tidal hexadecapole
moments} associated with the world line $\gamma$. Each tensor contains
9 independent components, and these are functions of proper time
$\tau$.  

The decomposition of the Weyl tensor and its derivatives in terms of
tidal moments is displayed in Eqs.~(\ref{Weyl_dec}),
(\ref{Weyl_first_dec}), and (\ref{Weyl_second_dec}) of
Appendix~\ref{app:alternative}. The numerical factors inserted in
Eqs.~(\ref{tidalmoment_2}), (\ref{tidalmoment_3}), and
(\ref{tidalmoment_4}) are inherited from Zhang's choice of
normalization \cite{zhang:86}.   

The black-hole metric of Sec.~\ref{sec:blackhole} will be expressed in
terms of the tidal moments $\E_{ab}$, $\E_{abc}$, $\E_{abcd}$,
$\B_{ab}$, $\B_{abc}$, and $\B_{abcd}$, which are treated as
freely-specifiable functions of time. As we shall explain in
Sec.~\ref{sec:blackhole}, the relation between the tidal moments and
the Weyl tensor will be more subtle than what was described here.  

\subsection{Tidal scales} 
\label{subsec:tidal_scales} 

The tidal moments allow us to specify length and time scales that
characterize the tidal environment around the world line $\gamma$. We
thus introduce the length scales $\R$ and ${\cal L}$, the time scale
${\cal T}$, and the velocity scale ${\cal V}$. Our discussion here
follows Thorne and Hartle \cite{thorne-hartle:85} and
Zhang \cite{zhang:86}. 

The length $\R$ is the local radius of curvature, and represents
the strength of the Weyl tensor evaluated on the world line. We
express this as 
\begin{equation} 
\E_{ab} \sim \frac{1}{\R^2}, \qquad 
\B_{ab} \sim \frac{{\cal V}}{\R^2}.  
\label{scale1} 
\end{equation} 
The first equation indicates that a typical component of 
${\cal E}_{ab}$ will have a magnitude comparable to 
$\R^{-2}$. The second equation indicates that a typical
component of ${\cal B}_{ab}$ will differ from this by a factor 
of ${\cal V}$, and this defines the velocity scale.   

The length ${\cal L}$ is the inhomogeneity scale, and it measures the 
degree of spatial variation of the Weyl tensor. It is defined through
the relations  
\begin{equation} 
\E_{abc} \sim \frac{1}{\R^2 {\cal L}}, \qquad 
\B_{abc} \sim \frac{{\cal V}}{\R^2 {\cal L}} 
\label{scale2} 
\end{equation} 
involving the tidal octupole moments. We expect that the hexadecapole
moments will be suppressed by an additional factor of ${\cal L}$: 
\begin{equation} 
\E_{abcd} \sim \frac{1}{\R^2 {\cal L}^2}, \qquad 
\B_{abcd} \sim \frac{{\cal V}}{\R^2 {\cal L}^2}.  
\label{scale3} 
\end{equation} 
  
The time ${\cal T}$ is the scale associated with changes in the
behavior of the Weyl tensor. This is defined by 
\begin{equation} 
\dot{\E}_{ab} \sim \frac{1}{\R^2 {\cal T}}, \qquad 
\dot{\B}_{ab} \sim \frac{{\cal V}}{\R^2 {\cal T}}.  
\label{scale4} 
\end{equation} 
We also have $\ddot{\E}_{ab} = \R^{-2} {\cal T}^{-2}$ and  
$\dot{\E}_{abc} = \R^{-2} {\cal L}^{-1} {\cal T}^{-1}$, as well
as $\ddot{\B}_{ab} = {\cal V} \R^{-2} {\cal T}^{-2}$ and 
$\dot{\B}_{abc} = {\cal V} \R^{-2} {\cal L}^{-1} 
{\cal T}^{-1}$. 

To illustrate the meaning of these tidal scales, let us consider as an
example the world line of an observer moving on a circular orbit of
radius $b$ around a body of mass $M'$. In this case the velocity scale 
is ${\cal V} \sim \sqrt{M'/b}$, the radius of curvature is $\R
\sim \sqrt{b^3/M'}$, the inhomogeneity scale is ${\cal L} \sim b$, and
the time scale is ${\cal T} \sim \sqrt{b^3/M'}$. In this example we
have that ${\cal T} \sim \R$, ${\cal V} \sim {\cal L}/{\cal T}$, and 
${\cal L} \sim {\cal V} \R$. When the motion is slow we
have that ${\cal L}  \ll \R$, and there is a wide separation between
the two length scales. When the motion is relativistic, however, all
scales are comparable to each other.  

\begin{table}
\caption{Irreducible tidal potentials: type-${\cal E}$,
  even-parity. The superscripts $\sf q$, $\sf o$, and $\sf h$ stand
  for ``quadrupole,'' ``octupole,'' and ``hexadecapole,''
  respectively. The tidal moments $\E_{ab}$, $\E_{abc}$, and
  $\E_{abcd}$ are the STF tensors defined in the text. The extra
  factors of 2 in the hexadecapole potentials are inserted to respect
  Zhang's normalization convention for $\E_{abcd}$; see
  Eqs.~(\ref{tidalmoment_4}).}   
\begin{ruledtabular} 
\begin{tabular}{l} 
$ \EE{q} = \E_{cd} \Omega^c \Omega^d $ \\ 
$ \EE{q}_a = \gamma_a^{\ c} \E_{cd} \Omega^d $ \\
$ \EE{q}_{ab} = 2\gamma_a^{\ c} \gamma_b^{\ d} \E_{cd} 
  + \gamma_{a b} \EE{q} $ \\
\hline
$ \EE{o} = \E_{cde} \Omega^c \Omega^d \Omega^e$ \\ 
$ \EE{o}_a = \gamma_a^{\ c} \E_{cde} \Omega^d \Omega^e $ \\ 
$ \EE{o}_{ab} = 2\gamma_a^{\ c} \gamma_b^{\ d} \E_{cde} \Omega^e  
  + \gamma_{a b} \EE{o} $ \\
\hline
$ \EE{h} = 2\E_{cdef} \Omega^c \Omega^d \Omega^e \Omega^f $ \\ 
$ \EE{h}_a = 2\gamma_a^{\ c} \E_{cdef} \Omega^d \Omega^e \Omega^f $ \\   
$ \EE{h}_{ab} = 4\gamma_a^{\ c} \gamma_b^{\ d} \E_{cdef} \Omega^e
  \Omega^f + \gamma_{a b} \EE{h} $ 
\end{tabular}
\end{ruledtabular} 
\label{tab:E_cart} 
\end{table} 

\begin{table}
\caption{Irreducible tidal potentials: type-${\cal B}$, odd-parity. The  
  superscripts $\sf q$, $\sf o$, and $\sf h$ stand for ``quadrupole,''
  ``octupole,'' and ``hexadecapole,'' respectively. The tidal moments 
  $\B_{ab}$, $\B_{abc}$, and $\B_{abcd}$ are the STF tensors defined
  in the text. The factors of $\frac{4}{3}$ and $\frac{10}{3}$ are
  inserted to respect Zhang's normalization convention for $\B_{abc}$  
  and $\B_{abcd}$; see Eqs.~(\ref{tidalmoment_3}) and
  (\ref{tidalmoment_4}).}   
\begin{ruledtabular} 
\begin{tabular}{l} 
$ \BB{q}_a = \epsilon_{apq} \Omega^p \B^q_{\ c} \Omega^c $ \\ 
$ \BB{q}_{ab} = \epsilon_{apq} \Omega^p \B^q_{\ d} \gamma^d_{\ b} 
  + \epsilon_{bpq} \Omega^p \B^q_{\ c} \gamma^c_{\ a} $ \\ 
\hline
$ \BB{o}_a = \frac{4}{3} \epsilon_{apq} \Omega^p \B^q_{\ cd} \Omega^c
  \Omega^d $ \\ 
$ \BB{o}_{ab} = \frac{4}{3} ( \epsilon_{apq} \Omega^p \B^q_{\ de}
  \gamma^d_{\ b} + \epsilon_{bpq} \Omega^p \B^q_{\ ce}
  \gamma^c_{\ a} ) \Omega^e$ \\ 
\hline 
$ \BB{h}_a = \frac{10}{3} \epsilon_{apq} \Omega^p \B^q_{\ cde}
  \Omega^c \Omega^d \Omega^e $ \\ 
$ \BB{h}_{ab} = \frac{10}{3} ( \epsilon_{apq} \Omega^p 
  \B^q_{\ def} \gamma^d_{\ b} + \epsilon_{bpq} \Omega^p 
  \B^q_{\ cef} \gamma^c_{\ a} ) \Omega^e \Omega^f$ 
\end{tabular}
\end{ruledtabular} 
\label{tab:B_cart} 
\end{table} 

\begin{table}
\caption{Irreducible tidal potentials: type-${\cal E}{\cal E}$,
  even-parity. The superscripts $\sf m$, $\sf q$, and $\sf h$
  stand for ``monopole,'' ``quadrupole,'' and ``hexadecapole,''
  respectively.} 
\begin{ruledtabular} 
\begin{tabular}{l} 
$ \PP{m} = \E_{pq} \E^{pq} $ \\ 
\hline 
$ \PP{q} = \E_{p \langle c} \E^p_{\ d \rangle} \Omega^c \Omega^d $ \\ 
$ \PP{q}_a = \gamma_a^{\ c} \E_{p \langle c} \E^p_{\ d\rangle}
  \Omega^d $ \\
$ \PP{q}_{ab} = 2\gamma_a^{\ c} \gamma_b^{\ d} \E_{p \langle c}
  \E^p_{\ d\rangle} + \gamma_{ab} \PP{q} $ \\ 
\hline 
$ \PP{h} = \E_{\langle cd} \E_{ef \rangle} \Omega^c \Omega^d \Omega^e
  \Omega^f $ \\ 
$ \PP{h}_a = \gamma_a^{\ c} \E_{\langle cd} \E_{ef \rangle}
  \Omega^d \Omega^e \Omega^f $ \\
$ \PP{h}_{ab} = 2\gamma_a^{\ c} \gamma_b^{\ d} \E_{\langle cd}
  \E_{ef \rangle} \Omega^e \Omega^f + \gamma_{ab} \PP{h} $
\end{tabular}
\end{ruledtabular} 
\label{tab:EE_cart} 
\end{table} 

\begin{table}
\caption{Irreducible tidal potentials: type-${\cal B}{\cal B}$,
  even-parity. The superscripts $\sf m$, $\sf q$, and $\sf h$ stand
  for ``monopole,'' ``quadrupole,'' and ``hexadecapole,'' 
  respectively.} 
\begin{ruledtabular} 
\begin{tabular}{l} 
$ \QQ{m} = \B_{pq} \B^{pq} $ \\ 
\hline 
$ \QQ{q} = \B_{p \langle c} \B^p_{\ d \rangle} \Omega^c \Omega^d $ \\ 
$ \QQ{q}_a = \gamma_a^{\ c} \B_{p \langle c} \B^p_{\ d\rangle}
  \Omega^d $ \\
$ \QQ{q}_{ab} = 2\gamma_a^{\ c} \gamma_b^{\ d} \B_{p \langle c}
  \B^p_{\ d\rangle} + \gamma_{ab} \QQ{q} $ \\ 
\hline 
$ \QQ{h} = \B_{\langle cd} \B_{ef \rangle} \Omega^c \Omega^d \Omega^e
  \Omega^f $ \\ 
$ \QQ{h}_a = \gamma_a^{\ c} \B_{\langle cd} \B_{ef \rangle}
  \Omega^d \Omega^e \Omega^f $ \\
$ \QQ{h}_{ab} = 2\gamma_a^{\ c} \gamma_b^{\ d} \B_{\langle cd}
  \B_{ef \rangle} \Omega^e \Omega^f + \gamma_{ab} \QQ{h} $
\end{tabular}
\end{ruledtabular} 
\label{tab:BB_cart} 
\end{table} 

\begin{table}
\caption{Irreducible tidal potentials: type-${\cal E}{\cal B}$,
  even-parity. The superscripts $\sf d$, and $\sf o$ stand for
  ``dipole'' and ``octupole,'' respectively.} 
\begin{ruledtabular} 
\begin{tabular}{l} 
$ \GG{d} = \epsilon_{cpq} \E^p_{\ r} \B^{rq} \Omega^c $ \\ 
$ \GG{d}_a = \gamma_a^{\ c} \epsilon_{cpq} \E^p_{\ r} \B^{rq} $ \\ 
\hline 
$ \GG{o} = \epsilon_{pq\langle c} \E^p_{\ d} \B^q_{\ e \rangle} 
  \Omega^c \Omega^d \Omega^e $ \\ 
$ \GG{o}_a = \gamma_a^{\ c} \epsilon_{pq\langle c} \E^p_{\ d} 
  \B^q_{\ e \rangle} \Omega^d \Omega^e $ \\ 
$ \GG{o}_{ab} = 2\gamma_a^{\ c} \gamma_b^{\ d} 
  \epsilon_{pq\langle c} \E^p_{\ d} \B^q_{\ e \rangle} \Omega^e 
  + \gamma_{ab}\, \GG{o} $ \\ 
\end{tabular}
\end{ruledtabular} 
\label{tab:EBeven_cart} 
\end{table} 

\begin{table}
\caption{Irreducible tidal potentials: type-${\cal E}{\cal B}$,
  odd-parity. The superscripts $\sf q$, and $\sf h$ stand for
  ``quadrupole'' and ``hexadecapole,'' respectively.} 
\begin{ruledtabular} 
\begin{tabular}{l} 
$ \HH{q}_a = \epsilon_a^{\ pq} \Omega_p \E_{r \langle q} 
  \B^r_{\ c \rangle} \Omega^c $ \\
$ \HH{q}_{ab} = \epsilon_a^{\ pq} \Omega_p \E_{r \langle q} 
  \B^r_{\ d \rangle} \gamma^d_{\ b} + \epsilon_b^{\ pq} \Omega_p 
  \E_{r \langle q} \B^r_{\ c \rangle} \gamma^c_{\ a} $ \\
\hline 
$ \HH{h}_a = \epsilon_a^{\ pq} \Omega_p \E_{\langle qc} 
  \B_{de \rangle} \Omega^c \Omega^d \Omega^e $ \\ 
$ \HH{h}_{ab} = (\epsilon_a^{\ pq} \Omega_p \E_{\langle qd} 
  \B_{ef \rangle} \gamma^d_{\ b}
  + \epsilon_b^{\ pq} \Omega_p \E_{\langle qc} 
  \B_{ef \rangle} \gamma^c_{\ a}) \Omega^e \Omega^f $ \\ 
\end{tabular}
\end{ruledtabular} 
\label{tab:EBodd_cart} 
\end{table} 
 
To simplify our notation in later portions of the paper, we choose to    
eliminate the distinction between the different tidal scales. We
therefore set ${\cal V} \sim 1$, ${\cal L} \sim {\cal T} \sim 
\R$, and adopt $\R$ as the single scale associated with the tidal
environment. In this sloppy notation we write 
\begin{subequations}
\label{sloppyscale} 
\begin{align}  
& \hspace*{-3pt} \E_{ab} \sim \B_{ab} \sim \frac{1}{\R^2}, \\  
& \hspace*{-3pt} \E_{abc} \sim \B_{abc} \sim \dot{\E}_{ab} 
\sim \dot{\B}_{ab}  \sim \frac{1}{\R^3}, \\ 
& \hspace*{-3pt} \E_{abcd} \sim \B_{abcd} \sim \dot{\E}_{abc} 
\sim \dot{\B}_{abc} \sim \ddot{\E}_{ab} \sim \ddot{\B}_{ab} 
\sim \frac{1}{\R^4}.  
\end{align} 
\end{subequations} 
We emphasize that this relaxation of our notation is simply to save
writing. For example, an error term that should be written $O(r^5
\R^{-2} {\cal L}^{-2} {\cal T}^{-1})$ will be condensed to the
simpler expression $O(r^5/\R^5)$. The form of the equations will 
always allow us to determine the order of magnitude of each term in  
relation to the complete set of scaling quantities.  

\subsection{Parity rules} 
\label{subsec:tidal_parity} 

In the context of this work, a parity transformation is a change of
tetrad vectors described by $u^\alpha \to u^\alpha$ and 
$e^\alpha_a \to -e^\alpha_a$; the transformation keeps the permutation
symbol unchanged: $\epsilon_{abc} \to \epsilon_{abc}$. Under the
transformation the frame components of the Weyl tensor change
according to $C_{a0b0} \to C_{a0b0}$ and $C_{abc0} \to -C_{abc0}$. We
also have $C_{a0b0|c} \to -C_{a0b0|c}$, $C_{abc0|d} \to C_{abc0|d}$,
and $C_{a0b0|cd} \to C_{a0b0|cd}$, $C_{abc0|de} \to -C_{abc0|de}$. 
From Eqs.~(\ref{tidalmoment_2}), (\ref{tidalmoment_3}), and
(\ref{tidalmoment_4}) we deduce the transformation rules 
\begin{subequations} 
\label{parity} 
\begin{align} 
& \E_{ab} \to \E_{ab},  \quad 
\E_{abc} \to -\E_{abc}, \quad 
\E_{abcd} \to \E_{abcd}, \\ 
& \B_{ab} \to -\B_{ab}, \quad 
\B_{abc} \to \B_{abc}, \quad 
\B_{abcd} \to -\B_{abcd}
\end{align} 
\end{subequations} 
for the tidal moments. We say that $\E_{ab}$, $\E_{abc}$, and
$\E_{abcd}$ have even parity, because they transform as ordinary
Cartesian tensors under a parity transformation. And we say that 
$\B_{ab}$, $\B_{abc}$, and $\B_{abcd}$ have odd parity, because they
transform as pseudotensors. 

\subsection{Tidal potentials: Cartesian coordinates} 
\label{subsec:tidal_pot_cartesian} 

For the purposes of writing down the black-hole metric in  
Sec.~\ref{sec:blackhole}, we involve the tidal moments in the
construction of {\it tidal potentials} that will form the main
building blocks for the metric. We first achieve this with the help of
a system of Cartesian coordinates $x^a$ that we assume is at our 
disposal. In the following subsection we convert the potentials to 
spherical coordinates $(r,\theta,\phi)$. 

\begingroup
\squeezetable
\begin{table}
\caption{Spherical-harmonic functions $Y^{lm}$ and harmonic
  components $\A^{(l)}_m$ involved in the decomposition of $\A^{(l)} 
  := \A_{k_1 \cdots k_l} \Omega^{k_1} \cdots \Omega^{k_l} = \sum_m
  \A^{(l)}_m Y^{lm}$. The functions are real, and they are listed for
  the relevant modes $l=1$ (dipole), $l=2$, (quadrupole), $l=3$
  (octupole), and $l=4$ (hexadecapole). The abstract index $m$
  describes the dependence of these functions on the angle $\phi$; for
  example $Y^{l,2s}$ is proportional to $\sin2\phi$. To simplify the
  expressions we write $C := \cos\theta$ and $S := \sin\theta$. The
  harmonic components are expressed in terms of the independent
  components of the STF tensor $\A_{k_1\cdots k_l}$.} 
\begin{ruledtabular} 
\begin{tabular}{ll} 
$ Y^{1,0} = C $ & 
$ \Aa{d}_{0} = \A_3 $ \\ 
$ Y^{1,1c} = S \cos\phi $ &
$ \Aa{d}_{1c} = \A_1 $ \\ 
$ Y^{1,1s} = S \sin\phi $ &  
$ \Aa{d}_{1s} = \A_2 $ \\ 
\hline
$ Y^{2,0} = 1-3C^2 $ &
$ \Aa{q}_{0} = \frac{1}{2} (\A_{11} + \A_{22}) $ \\ 
$ Y^{2,1c} = 2SC\cos\phi $ &
$ \Aa{q}_{1c} = \A_{13} $ \\ 
$ Y^{2,1s} = 2SC\sin\phi $ &
$ \Aa{q}_{1s} = \A_{23} $ \\ 
$ Y^{2,2c} = S^2\cos 2\phi $ &
$ \Aa{q}_{2c} = \frac{1}{2} (\A_{11} - \A_{22}) $ \\ 
$ Y^{2,2s} = S^2\sin 2\phi $ & 
$ \Aa{q}_{2s} = \A_{12} $ \\ 
\hline 
$ Y^{3,0} = C(3-5C^2) $ &  
$ \Aa{o}_{0} = \frac{1}{2} (\A_{113}+\A_{223}) $ \\  
$ Y^{3,1c} = \frac{3}{2} S(1-5C^2)\cos\phi $ & 
$ \Aa{o}_{1c} = \frac{1}{2} (\A_{111}+\A_{122}) $ \\  
$ Y^{3,1s} = \frac{3}{2} S(1-5C^2)\sin\phi $ &
$ \Aa{o}_{1s} = \frac{1}{2} (\A_{112}+\A_{222}) $ \\  
$ Y^{3,2c} = 3S^2 C \cos 2\phi $ &
$ \Aa{o}_{2c} = \frac{1}{2} (\A_{113}-\A_{223}) $ \\  
$ Y^{3,2s} = 3S^2 C \sin 2\phi $ & 
$ \Aa{o}_{2s} = \A_{123} $ \\  
$ Y^{3,3c} = S^3 \cos 3\phi $ &
$ \Aa{o}_{3c} = \frac{1}{4} (\A_{111}-3\A_{122}) $ \\  
$ Y^{3,3s} = S^3 \sin 3\phi $ & 
$ \Aa{o}_{3s} = \frac{1}{4} (3\A_{112}-\A_{222}) $ \\  
\hline
$ Y^{4,0} = \frac{1}{2} (3-30C^2+35C^4) $ &  
$ \Aa{h}_{0} = \frac{1}{4} (\A_{1111}+2\A_{1122}+\A_{2222}) $ \\ 
$ Y^{4,1c} = 2SC(3-7C^2) \cos\phi $ &
$ \Aa{h}_{1c} = \frac{1}{2} (\A_{1113}+\A_{1223}) $ \\ 
$ Y^{4,1s} = 2SC(3-7C^2) \sin\phi $ &
$ \Aa{h}_{1s} = \frac{1}{2} (\A_{1123}+\A_{2223}) $ \\ 
$ Y^{4,2c} = S^2(1-7C^2) \cos 2\phi $ &
$ \Aa{h}_{2c} = \frac{1}{2} (\A_{1111}-\A_{2222}) $ \\ 
$ Y^{4,2s} = S^2(1-7C^2) \sin 2\phi $ &
$ \Aa{h}_{2s} = \A_{1112}+\A_{1222} $ \\ 
$ Y^{4,3c} = 4S^3 C \cos 3\phi $ &
$ \Aa{h}_{3c} = \frac{1}{4} (\A_{1113}-3\A_{1223}) $ \\ 
$ Y^{4,3s} = 4S^3 C \sin 3\phi $ &
$ \Aa{h}_{3s} = \frac{1}{4} (3\A_{1123}-\A_{2223}) $ \\ 
$ Y^{4,4c} = S^4 \cos 4\phi $ &
$ \Aa{h}_{4c} = \frac{1}{8} (\A_{1111}-6\A_{1122}+\A_{2222}) $ \\ 
$ Y^{4,4s} = S^4 \sin 4\phi $ & 
$ \Aa{h}_{4s} = \frac{1}{2} (\A_{1112}-\A_{1222}) $ 
\end{tabular}
\end{ruledtabular} 
\label{tab:Ylm} 
\end{table} 
\endgroup

\begin{table}
\caption{The first column lists the harmonic components of type-$\E$,
  even-parity tidal potentials, as defined in
  Table~\ref{tab:E_cart}. The second column lists their expansions in
  scalar, vector, and tensor harmonics. The components of $\EE{h}$
  come with an additional factor of 2 to accommodate Zhang's choice of
  normalization; see Table~\ref{tab:E_cart}.}  
\begin{ruledtabular} 
\begin{tabular}{ll} 
$ \EE{q}_{0} = \frac{1}{2} (\E_{11} + \E_{22}) $ & \\
$ \EE{q}_{1c} = \E_{13} $ &
$ \EE{q} = \sum_m \EE{q}_m Y^{2,m} $ \\ 
$ \EE{q}_{1s} = \E_{23} $ &
$ \EE{q}_A = \frac{1}{2} \sum_m \EE{q}_m Y_A^{2,m} $ \\ 
$ \EE{q}_{2c} = \frac{1}{2} (\E_{11} - \E_{22}) $ & 
$ \EE{q}_{AB} = \sum_m \EE{q}_m Y_{AB}^{2,m} $ \\ 
$ \EE{q}_{2s} = \E_{12} $ & \\ 
\hline 
$ \EE{o}_{0} = \frac{1}{2} (\E_{113}+\E_{223}) $ & \\ 
$ \EE{o}_{1c} = \frac{1}{2} (\E_{111}+\E_{122}) $ & \\ 
$ \EE{o}_{1s} = \frac{1}{2} (\E_{112}+\E_{222}) $ & 
$ \EE{o} = \sum_m \EE{o}_m Y^{3,m} $ \\ 
$ \EE{o}_{2c} = \frac{1}{2} (\E_{113}-\E_{223}) $ & 
$ \EE{o}_A = \frac{1}{3} \sum_m \EE{o}_m Y_A^{3,m} $ \\ 
$ \EE{o}_{2s} = \E_{123} $ &
$ \EE{o}_{AB} = \frac{1}{3} \sum_m \EE{o}_m Y_{AB}^{3,m} $ \\ 
$ \EE{o}_{3c} = \frac{1}{4} (\E_{111}-3\E_{122}) $ & \\ 
$ \EE{o}_{3s} = \frac{1}{4} (3\E_{112}-\E_{222}) $ & \\ 
\hline 
$ \EE{h}_{0} = \frac{1}{2} (\E_{1111}+2\E_{1122}+\E_{2222}) $ & \\ 
$ \EE{h}_{1c} = \E_{1113}+\E_{1223} $ & \\ 
$ \EE{h}_{1s} = \E_{1123}+\E_{2223} $ & \\ 
$ \EE{h}_{2c} = \E_{1111}-\E_{2222} $ & 
$ \EE{h} = \sum_m \EE{h}_m Y^{4,m} $ \\ 
$ \EE{h}_{2s} = 2(\E_{1112}+\E_{1222}) $ & 
$ \EE{h}_A = \frac{1}{4} \sum_m \EE{h}_m Y_A^{4,m} $ \\ 
$ \EE{h}_{3c} = \frac{1}{2} (\E_{1113}-3\E_{1223}) $ & 
$ \EE{h}_{AB} = \frac{1}{6} \sum_m \EE{h}_m Y_{AB}^{4,m} $ \\ 
 $ \EE{h}_{3s} = \frac{1}{2} (3\E_{1123}-\E_{2223}) $ & \\ 
$ \EE{h}_{4c} = \frac{1}{4} (\E_{1111}-6\E_{1122}+\E_{2222}) $ & \\ 
$ \EE{h}_{4s} = \E_{1112}-\E_{1222} $ & \\ 
\end{tabular}
\end{ruledtabular} 
\label{tab:E_ang} 
\end{table} 

\begin{table}
\caption{The first column lists the harmonic components of type-$\B$,
  odd-parity tidal potentials, as defined in
  Table~\ref{tab:B_cart}. The second column lists their expansions in
  scalar, vector, and tensor harmonics. The components of $\BB{o}$
  come with an additional factor of $\frac{4}{3}$ to accommodate
  Zhang's choice of normalization, and those of $\BB{h}$ come with an
  additional factor $\frac{10}{3}$; see Table~\ref{tab:B_cart}.}  
\begin{ruledtabular} 
\begin{tabular}{ll} 
$ \BB{q}_{0} = \frac{1}{2} (\B_{11} + \B_{22}) $ & \\
$ \BB{q}_{1c} = \B_{13} $ &
$ \BB{q} = \sum_m \BB{q}_m Y^{2,m} $ \\ 
$ \BB{q}_{1s} = \B_{23} $ & 
$ \BB{q}_A = \frac{1}{2} \sum_m \BB{q}_m X_A^{2,m} $ \\ 
$ \BB{q}_{2c} = \frac{1}{2} (\B_{11} - \B_{22}) $ & 
$ \BB{q}_{AB} = \sum_m \BB{q}_m X_{AB}^{2,m} $ \\ 
$ \BB{q}_{2s} = \B_{12} $ & \\ 
\hline 
$ \BB{o}_{0} = \frac{2}{3} (\B_{113}+\B_{223}) $ & \\ 
$ \BB{o}_{1c} = \frac{2}{3} (\B_{111}+\B_{122}) $ & \\ 
$ \BB{o}_{1s} = \frac{2}{3} (\B_{112}+\B_{222}) $ & 
$ \BB{o} = \sum_m \BB{o}_m Y^{3,m} $ \\ 
$ \BB{o}_{2c} = \frac{2}{3} (\B_{113}-\B_{223}) $ & 
$ \BB{o}_A = \frac{1}{3} \sum_m \BB{o}_m X_A^{3,m} $ \\ 
$ \BB{o}_{2s} = \frac{4}{3} \B_{123} $ & 
$ \BB{o}_{AB} = \frac{1}{3} \sum_m \BB{o}_m X_{AB}^{3,m} $ \\ 
$ \BB{o}_{3c} = \frac{1}{3} (\B_{111}-3\B_{122}) $ & \\ 
$ \BB{o}_{3s} = \frac{1}{3} (3\B_{112}-\B_{222}) $ & \\ 
\hline 
$ \BB{h}_{0} = \frac{5}{6} (\B_{1111}+2\B_{1122}+\B_{2222}) $ & \\ 
$ \BB{h}_{1c} = \frac{5}{3}(\B_{1113}+\B_{1223}) $ & \\ 
$ \BB{h}_{1s} = \frac{5}{3}(\B_{1123}+\B_{2223}) $ & \\ 
$ \BB{h}_{2c} = \frac{5}{3}(\B_{1111}-\B_{2222}) $ &
$ \BB{h} = \sum_m \BB{h}_m Y^{4,m} $ \\ 
$ \BB{h}_{2s} = \frac{10}{3}(\B_{1112}+\B_{1222}) $ & 
$ \BB{h}_A = \frac{1}{4} \sum_m \BB{h}_m X_A^{4,m} $ \\ 
$ \BB{h}_{3c} = \frac{5}{6} (\B_{1113}-3\B_{1223}) $ & 
$ \BB{h}_{AB} = \frac{1}{6} \sum_m \BB{h}_m X_{AB}^{4,m} $ \\ 
$ \BB{h}_{3s} = \frac{5}{6} (3\B_{1123}-\B_{2223}) $ & \\ 
$ \BB{h}_{4c} = \frac{5}{12} (\B_{1111}-6\B_{1122}+\B_{2222}) $ & \\ 
$ \BB{h}_{4s} = \frac{5}{3}(\B_{1112}-\B_{1222}) $ & 
\end{tabular}
\end{ruledtabular} 
\label{tab:B_ang} 
\end{table} 

\begin{table}
\caption{Harmonic components of type-$\E\E$, even-parity tidal
  potentials, as defined in Table~\ref{tab:EE_cart}. The
  spherical-harmonic decompositions are 
  $\PP{q} = \sum_m \PP{q}_m Y^{2,m}$, $\PP{q}_A =
  \frac{1}{2} \sum_m \PP{q}_m Y_A^{2,m}$, $\PP{q}_{AB} = \sum_m
  \PP{q}_m Y_{AB}^{2,m}$, $\PP{h} = \sum_m \PP{h}_m Y^{4,m}$,
  $\PP{h}_A = \frac{1}{4} \sum_m \PP{h}_m Y_A^{4,m}$, and $\PP{h}_{AB}
  = \frac{1}{6} \sum_m \PP{h}_m Y_{AB}^{4,m}$.}
\begin{ruledtabular} 
\begin{tabular}{l} 
$ \PP{m} = 6(\EE{q}_0)^2 + 2(\EE{q}_{1c})^2 + 2(\EE{q}_{1s})^2 
+ 2(\EE{q}_{2c})^2 + 2(\EE{q}_{2s})^2 $ \\
\hline
$ \PP{q}_0 = -(\EE{q}_0)^2 - \frac{1}{6} (\EE{q}_{1c})^2 
- \frac{1}{6} (\EE{q}_{1s})^2 + \frac{1}{3} (\EE{q}_{2c})^2 
+ \frac{1}{3} (\EE{q}_{2s})^2 $ \\ 
$ \PP{q}_{1c} = -\EE{q}_0 \EE{q}_{1c} + \EE{q}_{1c} \EE{q}_{2c} 
+ \EE{q}_{1s} \EE{q}_{2s} $ \\  
$ \PP{q}_{1s} = -\EE{q}_0 \EE{q}_{1s} + \EE{q}_{1c} \EE{q}_{2s} 
- \EE{q}_{1s} \EE{q}_{2c} $ \\  
$ \PP{q}_{2c} = 2 \EE{q}_0 \EE{q}_{2c} + \frac{1}{2} (\EE{q}_{1c})^2 
- \frac{1}{2} (\EE{q}_{1s})^2 $ \\ 
$ \PP{q}_{2s} = 2 \EE{q}_0 \EE{q}_{2s} + \EE{q}_{1c} \EE{q}_{1s} $ \\ 
\hline
$ \PP{h}_0 = \frac{18}{35} (\EE{q}_0)^2 - \frac{4}{35} (\EE{q}_{1c})^2 
- \frac{4}{35} (\EE{q}_{1s})^2 + \frac{1}{35} (\EE{q}_{2c})^2
+ \frac{1}{35} (\EE{q}_{2s})^2 $ \\ 
$ \PP{h}_{1c} = \frac{6}{7} \EE{q}_0 \EE{q}_{1c} 
+ \frac{1}{7} \EE{q}_{1c} \EE{q}_{2c} 
+ \frac{1}{7} \EE{q}_{1s} \EE{q}_{2s} $ \\ 
$ \PP{h}_{1s} = \frac{6}{7} \EE{q}_0 \EE{q}_{1s} 
+ \frac{1}{7} \EE{q}_{1c} \EE{q}_{2s}  
- \frac{1}{7} \EE{q}_{1s} \EE{q}_{2c} $ \\ 
$ \PP{h}_{2c} = \frac{6}{7} \EE{q}_0 \EE{q}_{2c} 
- \frac{2}{7} (\EE{q}_{1c})^2 
+ \frac{2}{7} (\EE{q}_{1s})^2 $ \\ 
$ \PP{h}_{2s} = \frac{6}{7} \EE{q}_0 \EE{q}_{2s} 
- \frac{4}{7} \EE{q}_{1c} \EE{q}_{1s} $ \\ 
$ \PP{h}_{3c} = \frac{1}{2} \EE{q}_{1c} \EE{q}_{2c} 
- \frac{1}{2} \EE{q}_{1s} \EE{q}_{2s} $ \\ 
$ \PP{h}_{3s} = \frac{1}{2} \EE{q}_{1c} \EE{q}_{2s} 
+ \frac{1}{2} \EE{q}_{1s} \EE{q}_{2c} $ \\ 
$ \PP{h}_{4c} = \frac{1}{2} (\EE{q}_{2c})^2 
- \frac{1}{2} (\EE{q}_{2s})^2 $ \\ 
$ \PP{h}_{4s} = \EE{q}_{2c} \EE{q}_{2s} $
\end{tabular}
\end{ruledtabular} 
\label{tab:EE_ang} 
\end{table} 

\begin{table}
\caption{Harmonic components of type-$\B\B$, even-parity tidal 
  potentials, as defined in Table~\ref{tab:BB_cart}. The
  spherical-harmonic decompositions are 
  $\QQ{q} = \sum_m \QQ{q}_m Y^{2,m}$, $\QQ{q}_A =
  \frac{1}{2} \sum_m \QQ{q}_m Y_A^{2,m}$, $\QQ{q}_{AB} = \sum_m
  \QQ{q}_m Y_{AB}^{2,m}$, $\QQ{h} = \sum_m \QQ{h}_m Y^{4,m}$,
  $\QQ{h}_A = \frac{1}{4} \sum_m \QQ{h}_m Y_A^{4,m}$, and $\QQ{h}_{AB}
  = \frac{1}{6} \sum_m \QQ{h}_m Y_{AB}^{4,m}$.}
\begin{ruledtabular} 
\begin{tabular}{l} 
$ \QQ{m} = 6(\BB{q}_0)^2 + 2(\BB{q}_{1c})^2 + 2(\BB{q}_{1s})^2 
+ 2(\BB{q}_{2c})^2 + 2(\BB{q}_{2s})^2 $ \\
\hline
$ \QQ{q}_0 = -(\BB{q}_0)^2 - \frac{1}{6} (\BB{q}_{1c})^2 
- \frac{1}{6} (\BB{q}_{1s})^2 + \frac{1}{3} (\BB{q}_{2c})^2 
+ \frac{1}{3} (\BB{q}_{2s})^2 $ \\ 
$ \QQ{q}_{1c} = -\BB{q}_0 \BB{q}_{1c} + \BB{q}_{1c} \BB{q}_{2c} 
+ \BB{q}_{1s} \BB{q}_{2s} $ \\  
$ \QQ{q}_{1s} = -\BB{q}_0 \BB{q}_{1s} + \BB{q}_{1c} \BB{q}_{2s} 
- \BB{q}_{1s} \BB{q}_{2c} $ \\  
$ \QQ{q}_{2c} = 2 \BB{q}_0 \BB{q}_{2c} + \frac{1}{2} (\BB{q}_{1c})^2 
- \frac{1}{2} (\BB{q}_{1s})^2 $ \\ 
$ \QQ{q}_{2s} = 2 \BB{q}_0 \BB{q}_{2s} + \BB{q}_{1c} \BB{q}_{1s} $ \\ 
\hline
$ \QQ{h}_0 = \frac{18}{35} (\BB{q}_0)^2 - \frac{4}{35} (\BB{q}_{1c})^2 
- \frac{4}{35} (\BB{q}_{1s})^2 + \frac{1}{35} (\BB{q}_{2c})^2
+ \frac{1}{35} (\BB{q}_{2s})^2 $ \\ 
$ \QQ{h}_{1c} = \frac{6}{7} \BB{q}_0 \BB{q}_{1c} 
+ \frac{1}{7} \BB{q}_{1c} \BB{q}_{2c} 
+ \frac{1}{7} \BB{q}_{1s} \BB{q}_{2s} $ \\ 
$ \QQ{h}_{1s} = \frac{6}{7} \BB{q}_0 \BB{q}_{1s} 
+ \frac{1}{7} \BB{q}_{1c} \BB{q}_{2s}  
- \frac{1}{7} \BB{q}_{1s} \BB{q}_{2c} $ \\ 
$ \QQ{h}_{2c} = \frac{6}{7} \BB{q}_0 \BB{q}_{2c} 
- \frac{2}{7} (\BB{q}_{1c})^2 
+ \frac{2}{7} (\BB{q}_{1s})^2 $ \\ 
$ \QQ{h}_{2s} = \frac{6}{7} \BB{q}_0 \BB{q}_{2s} 
- \frac{4}{7} \BB{q}_{1c} \BB{q}_{1s} $ \\ 
$ \QQ{h}_{3c} = \frac{1}{2} \BB{q}_{1c} \BB{q}_{2c} 
- \frac{1}{2} \BB{q}_{1s} \BB{q}_{2s} $ \\ 
$ \QQ{h}_{3s} = \frac{1}{2} \BB{q}_{1c} \BB{q}_{2s} 
+ \frac{1}{2} \BB{q}_{1s} \BB{q}_{2c} $ \\ 
$ \QQ{h}_{4c} = \frac{1}{2} (\BB{q}_{2c})^2 
- \frac{1}{2} (\BB{q}_{2s})^2 $ \\ 
$ \QQ{h}_{4s} = \BB{q}_{2c} \BB{q}_{2s} $
\end{tabular}
\end{ruledtabular} 
\label{tab:BB_ang} 
\end{table} 

\begin{table*}
\caption{Harmonic components of type-$\E\B$, even-parity tidal 
  potentials, as defined in Table~\ref{tab:EBeven_cart}. The
  spherical-harmonic decompositions are 
  $\GG{d} = \sum_m \GG{d}_m Y^{1,m}$, $\GG{d}_A =
  \sum_m \GG{d}_m Y_A^{1,m}$, $\GG{o} = \sum_m \GG{o}_m Y^{3,m}$,
  $\GG{o}_A = \frac{1}{3} \sum_m \GG{o}_m Y_A^{3,m}$, and $\GG{o}_{AB}
  = \frac{1}{3} \sum_m \GG{o}_m Y_{AB}^{3,m}$.}
\begin{ruledtabular} 
\begin{tabular}{l} 
$ \GG{d}_0 = \EE{q}_{1c} \BB{q}_{1s} - \EE{q}_{1s} \BB{q}_{1c}
+ 2\EE{q}_{2c} \BB{q}_{2s} - 2\EE{q}_{2s} \BB{q}_{2c} $ \\ 
$ \GG{d}_{1c} = (3\EE{q}_0 - \EE{q}_{2c}) \BB{q}_{1s} 
- \EE{q}_{1c} \BB{q}_{2s} - \EE{q}_{1s} (3\BB{q}_{0} 
- \BB{q}_{2c}) + \EE{q}_{2s} \BB{q}_{1c} $ \\ 
$ \GG{d}_{1s} = -(3\EE{q}_0 + \EE{q}_{2c}) \BB{q}_{1c} 
+ \EE{q}_{1c} (3\BB{q}_{0} + \BB{q}_{2c})
+ \EE{q}_{1s} \BB{q}_{2s} - \EE{q}_{2s} \BB{q}_{1s} $ \\ 
\hline
$ \GG{o}_0 = -\frac{2}{5} \EE{q}_{1c} \BB{q}_{1s}
+ \frac{2}{5} \EE{q}_{1s} \BB{q}_{1c} 
+ \frac{1}{5} \EE{q}_{2c} \BB{q}_{2s}
- \frac{1}{5} \EE{q}_{2s} \BB{q}_{2c} $ \\ 
$ \GG{o}_{1c} = -\frac{2}{5} \EE{q}_0 \BB{q}_{1s} 
- \frac{1}{5} \EE{q}_{1c} \BB{q}_{2s} 
+ \frac{1}{5} \EE{q}_{1s} (2\BB{q}_{0} + \BB{q}_{2c})  
- \frac{1}{5} \EE{q}_{2c} \BB{q}_{1s} 
+ \frac{1}{5} \EE{q}_{2s} \BB{q}_{1c} $ \\ 
$ \GG{o}_{1s} = \frac{2}{5} \EE{q}_0 \BB{q}_{1c} 
- \frac{1}{5} \EE{q}_{1c} (2\BB{q}_{0} - \BB{q}_{2c}) 
+ \frac{1}{5} \EE{q}_{1s} \BB{q}_{2s} 
- \frac{1}{5} \EE{q}_{2c} \BB{q}_{1c} 
- \frac{1}{5} \EE{q}_{2s} \BB{q}_{1s} $ \\ 
$ \GG{o}_{2c} = \EE{q}_0 \BB{q}_{2s} - \EE{q}_{2s} \BB{q}_0 $ \\ 
$ \GG{o}_{2s} = -\EE{q}_0 \BB{q}_{2c} + \EE{q}_{2c} \BB{q}_0 $ \\
$ \GG{o}_{3c} = -\frac{1}{2} \EE{q}_{1c} \BB{q}_{2s}  
- \frac{1}{2} \EE{q}_{1s} \BB{q}_{2c}
+ \frac{1}{2} \EE{q}_{2c} \BB{q}_{1s}
+ \frac{1}{2} \EE{q}_{2s} \BB{q}_{1c} $ \\ 
$ \GG{o}_{3s} = \frac{1}{2} \EE{q}_{1c} \BB{q}_{2c}  
- \frac{1}{2} \EE{q}_{1s} \BB{q}_{2s}
- \frac{1}{2} \EE{q}_{2c} \BB{q}_{1c}
+ \frac{1}{2} \EE{q}_{2s} \BB{q}_{1s} $ 
\end{tabular}
\end{ruledtabular} 
\label{tab:EBeven_ang} 
\end{table*} 

\begin{table*}
\caption{Harmonic components of type-$\E\B$, odd-parity tidal 
  potentials, as defined in Table~\ref{tab:EBodd_cart}. The
  spherical-harmonic decompositions are 
  $\HH{q}_A = \frac{1}{2} \sum_m \HH{q}_m
  X_A^{2,m}$, $\HH{q}_{AB} = \sum_m \HH{q}_m X_{AB}^{2,m}$, $\HH{h}_A
  = \frac{1}{4} \sum_m \HH{h}_m X_A^{4,m}$, and $\HH{h}_{AB} =
  \frac{1}{6} \sum_m \HH{h}_m X_{AB}^{4,m}$.}
\begin{ruledtabular} 
\begin{tabular}{l} 
$ \HH{q}_0 = -\EE{q}_{0} \BB{q}_{0} 
- \frac{1}{6} \EE{q}_{1c} \BB{q}_{1c} 
- \frac{1}{6} \EE{q}_{1s} \BB{q}_{1s} 
+ \frac{1}{3} \EE{q}_{2c} \BB{q}_{2c}
+ \frac{1}{3} \EE{q}_{2s} \BB{q}_{2s} $ \\
$ \HH{q}_{1c} = -\frac{1}{2} \EE{q}_{0} \BB{q}_{1c} 
- \frac{1}{2} \EE{q}_{1c} ( \BB{q}_{0} - \BB{q}_{2c} ) 
+ \frac{1}{2} \EE{q}_{1s} \BB{q}_{2s} 
+ \frac{1}{2} \EE{q}_{2c} \BB{q}_{1c} 
+ \frac{1}{2} \EE{q}_{2s} \BB{q}_{1s} $ \\ 
$ \HH{q}_{1s} = -\frac{1}{2} \EE{q}_{0} \BB{q}_{1s} 
+ \frac{1}{2} \EE{q}_{1c} \BB{q}_{2s} 
- \frac{1}{2} \EE{q}_{1s} ( \BB{q}_{0} + \BB{q}_{2c} ) 
- \frac{1}{2} \EE{q}_{2c} \BB{q}_{1s} 
+ \frac{1}{2} \EE{q}_{2s} \BB{q}_{1c} $ \\ 
$ \HH{q}_{2c} = \EE{q}_{0} \BB{q}_{2c} 
+ \frac{1}{2} \EE{q}_{1c} \BB{q}_{1c} 
- \frac{1}{2} \EE{q}_{1s} \BB{q}_{1s} 
+ \EE{q}_{2c} \BB{q}_{0} $ \\  
$ \HH{q}_{2s} = \EE{q}_{0} \BB{q}_{2s} 
+ \frac{1}{2} \EE{q}_{1c} \BB{q}_{1s} 
+ \frac{1}{2} \EE{q}_{1s} \BB{q}_{1c} 
+ \EE{q}_{2s} \BB{q}_{0} $ \\  
\hline 
$ \HH{h}_0 = \frac{18}{35} \EE{q}_0 \BB{q}_0 
- \frac{4}{35} \EE{q}_{1c} \BB{q}_{1c} 
- \frac{4}{35} \EE{q}_{1s} \BB{q}_{1s} 
+ \frac{1}{35} \EE{q}_{2c} \BB{q}_{2c} 
+ \frac{1}{35} \EE{q}_{2s} \BB{q}_{2s} $ \\ 
$ \HH{h}_{1c} = \frac{3}{7} \EE{q}_0 \BB{q}_{1c} 
+ \frac{1}{14} \EE{q}_{1c} ( 6\BB{q}_{0} + \BB{q}_{2c} ) 
+ \frac{1}{14} \EE{q}_{1s} \BB{q}_{2s} 
+ \frac{1}{14} \EE{q}_{2c} \BB{q}_{1c}
+ \frac{1}{14} \EE{q}_{2s} \BB{q}_{1s} $ \\ 
$ \HH{h}_{1s} = \frac{3}{7} \EE{q}_0 \BB{q}_{1s} 
+ \frac{1}{14} \EE{q}_{1c} \BB{q}_{2s} 
+ \frac{1}{14} \EE{q}_{1s} ( 6\BB{q}_{0} - \BB{q}_{2c} ) 
- \frac{1}{14} \EE{q}_{2c} \BB{q}_{1s}
+ \frac{1}{14} \EE{q}_{2s} \BB{q}_{1c} $ \\ 
$ \HH{h}_{2c} = \frac{3}{7} \EE{q}_0 \BB{q}_{2c} 
- \frac{2}{7} \EE{q}_{1c} \BB{q}_{1c} 
+ \frac{2}{7} \EE{q}_{1s} \BB{q}_{1s} 
+ \frac{3}{7} \EE{q}_{2c} \BB{q}_{0} $ \\ 
$ \HH{h}_{2s} = \frac{3}{7} \EE{q}_0 \BB{q}_{2s} 
- \frac{2}{7} \EE{q}_{1c} \BB{q}_{1s} 
- \frac{2}{7} \EE{q}_{1s} \BB{q}_{1c} 
+ \frac{3}{7} \EE{q}_{2s} \BB{q}_{0} $ \\ 
$ \HH{h}_{3c} = \frac{1}{4} \EE{q}_{1c} \BB{q}_{2c} 
- \frac{1}{4} \EE{q}_{1s} \BB{q}_{2s} 
+ \frac{1}{4} \EE{q}_{2c} \BB{q}_{1c} 
- \frac{1}{4} \EE{q}_{2s} \BB{q}_{1s} $ \\
$ \HH{h}_{3s} = \frac{1}{4} \EE{q}_{1c} \BB{q}_{2s} 
+ \frac{1}{4} \EE{q}_{1s} \BB{q}_{2c} 
+ \frac{1}{4} \EE{q}_{2c} \BB{q}_{1s} 
+ \frac{1}{4} \EE{q}_{2s} \BB{q}_{1c} $ \\
$ \HH{h}_{4c} = \frac{1}{2} \EE{q}_{2c} \BB{q}_{2c} 
- \frac{1}{2} \EE{q}_{2s} \BB{q}_{2s} $ \\ 
$ \HH{h}_{4s} = \frac{1}{2} \EE{q}_{2c} \BB{q}_{2s} 
+ \frac{1}{2} \EE{q}_{2s} \BB{q}_{2c} $   
\end{tabular}
\end{ruledtabular} 
\label{tab:EBodd_ang} 
\end{table*} 
 
We introduce 
\begin{equation} 
\Omega^a := x^a/r
\label{Omega}
\end{equation} 
as the radial unit vector, with $r := \sqrt{\delta_{ab} x^a x^b}$
denoting the usual Euclidean distance. We refer to the radial
direction as the {\it longitudinal} direction, and to the orthogonal   
space as the {\it transverse} directions. 

We wish to combine the tidal moments with $\Omega^a$ so as to form
scalar, vector, and (rank-two, symmetric) tensor potentials that
satisfy the following properties:   
\begin{enumerate} 
\item Each potential is an element of an irreducible representation of
  the rotation group labeled by multipole order $l$. 
\item Each scalar potential transforms as such under a parity
  transformation. 
\item Each vector potential transforms as such under a parity
  transformation, and is purely transverse, in the sense of being
  orthogonal to $\Omega^a$. 
\item Each tensor potential transforms as such under a parity 
  transformation, and is transverse-tracefree, in the sense of
  being orthogonal to $\Omega^a$ and having a vanishing trace. 
\end{enumerate} 
The first property implies that each potential will satisfy an 
appropriate eigenvalue equation that depends on the multipole order
and the tensorial rank of the potential; these are displayed in
Eqs.~(\ref{even_eigen_cart}) and (\ref{odd_eigen_cart}) of
Appendix~\ref{app:spherical_harmonics}. Under a parity transformation
the longitudinal vector transforms as $\Omega^a \to -\Omega^a$, a
scalar potential remains invariant, a vector potential changes sign,
and a tensor potential remains invariant. To aid the construction of
the potentials we introduce  
\begin{equation}  
\gamma^a_{\ b} := \delta^a_{\ b} - \Omega^a \Omega_b
\label{projector} 
\end{equation} 
as a projector to the transverse space orthogonal to
$\Omega^a$; this transforms as $\gamma_{ab} \to \gamma_{ab}$ under a 
parity transformation.  

The required potentials are displayed in a number of tables. In 
Table~\ref{tab:E_cart} we list the tidal potentials that are
constructed from the even-parity tidal moments $\E_{ab}$, $\E_{abc}$,
and $\E_{abcd}$. In Table~\ref{tab:B_cart} we have the tidal
potentials that are constructed from the odd-parity tidal moments
$\B_{ab}$, $\B_{abc}$, and $\B_{abcd}$. In Table~\ref{tab:EE_cart} we
list the potentials that arise from the bilinear combination
$\E_{ab} \E_{cd}$ of even-parity moments. In Table~\ref{tab:BB_cart}
we have the potentials that arise from the bilinear combination
$\B_{ab} \B_{cd}$ of odd-parity moments. In
Table~\ref{tab:EBeven_cart} we list potentials that arise when we
combine $\E_{ab}$ and $\B_{ab}$ into an even-parity structure. And
finally, in Table~\ref{tab:EBodd_cart} we combine them into an
odd-parity structure.  

We illustrate the construction of the potentials by examining a few
examples. The simplest cases involve the tidal quadrupole moment
$\E_{ab}$, which transforms as a tensor under a parity
transformation. The associated scalar potential is $\E_{ab} \Omega^a 
\Omega^b$; this transforms appropriately under a parity
transformation, and because $\E_{ab}$ is tracefree this satisfies the
scalar eigenvalue equation with $l = 2$. To get a vector potential
we first form $\E_{cb} \Omega^b$, which has the appropriate number of
free indices. We next multiply this by $\gamma^c_{\ a}$ to make the
vector transverse, and verify that the final object satisfies the
vectorial eigenvalue equation with $l = 2$. To get a transverse tensor
potential we first form $\gamma^c_{\ a} \gamma^d_{\ b} \E_{cd}$, and
we remove its trace by adding $\frac{1}{2} \gamma_{ab} (\E_{cd}
\Omega^c \Omega^d)$; the result satisfies the appropriate eigenvalue 
equation with $l = 2$. The method generalizes easily to octupole
and hexadecapole potentials, and the final results are displayed in 
Table~\ref{tab:E_cart}.    

We next examine the potentials associated with the odd-parity
quadrupole moment $\B_{ab}$. The combination $\B_{ab} \Omega^a
\Omega^b$ transforms as a pseudoscalar under a parity transformation
and does not, therefore, satisfy the criteria that were formulated
previously. To obtain a suitable potential we must involve the
permutation symbol, and the simplest allowed combination is 
$\epsilon_{apq} \Omega^p \B^q_{\ c} \Omega^c$. This transforms as a
vector under a parity transformation and satisfies the appropriate
eigenvalue equation with $l=2$; we therefore include it as one of our
building blocks. To form a tensor potential we remove $\Omega^c$,
replace it with the projector $\gamma^c_{\ b}$, and symmetrize the
indices; the end result satisfies all the properties required of a
tensor potential. The method generalizes easily to octupole and
hexadecapole potentials, and these are listed in
Table~\ref{tab:B_cart}.       

The bilinear potentials are constructed by forming STF products of
the quadrupole moments $\E_{ab}$ and $\B_{ab}$ and combining these
with appropriate factors of $\Omega^a$, $\gamma^a_{\ b}$, and
$\epsilon_{abc}$. For example, the STF products that can be formed
from two factors of $\E_{ab}$ are the scalar $\E_{pq} \E^{pq}$, the
rank-two tensor $\E_{p\langle a} \E^p_{\ b\rangle}$, and the rank-four
tensor $\E_{\langle ab} \E_{cd\rangle}$; the associated potentials are
listed in Table~\ref{tab:EE_cart}, and those of Table~\ref{tab:BB_cart}
follow from similar manipulations. When $\E_{ab}$ and $\B_{ab}$ are
both involved we must be mindful of the parity rules; for example,
the hexadecapole potential $\E_{\langle ab} \B_{cd\rangle} \Omega^a 
\Omega^b \Omega^c \Omega^d$ is ruled out because it transforms as a 
pseudoscalar under a parity transformation. The appropriate
combinations must involve the permutation symbol, and these are
displayed in Tables~\ref{tab:EBeven_cart} and \ref{tab:EBodd_cart}. 

\subsection{Tidal potentials: Angular coordinates} 
\label{subsec:tidal_pot_angular} 

A transformation from Cartesian coordinates $x^a$ to spherical
coordinates $(r,\theta,\phi)$ is effected by    
\begin{equation} 
x^a = r \Omega^a(\theta^A), 
\label{trans_cart_sph} 
\end{equation} 
in which the longitudinal vector $\Omega^a$ is now parameterized by
two polar angles $\theta^A = (\theta,\phi)$.  Explicitly, we have
that $\Omega^a 
= [\sin\theta\cos\phi,\sin\theta\sin\phi,\cos\theta]$. The 
transformation implies that $\partial x^a/\partial r = \Omega^a$ and
$\partial x^a/\partial \theta^A = r \Omega^a_A$, with 
\begin{equation} 
\Omega^a_A := \frac{\partial \Omega^a}{\partial \theta^A}. 
\label{trans_matrix} 
\end{equation} 
We note the useful identities 
\begin{subequations}
\label{ang_identities1} 
\begin{align}  
& \Omega_a \Omega^a_A = 0, \\
& \Omega_{AB} = \gamma_{ab} \Omega^a_A \Omega^b_B, \\ 
& \Omega^{AB} \Omega^a_A \Omega^b_B = \gamma^{ab}. 
\end{align} 
\end{subequations}  
Here $\Omega_{AB} = \mbox{diag}[1,\sin^2\theta]$ is the metric on the
unit two-sphere, and $\Omega^{AB}$ is its inverse. We introduce $D_A$
as the covariant-derivative operator compatible with $\Omega_{AB}$,
and $\epsilon_{AB}$ as the Levi-Civita tensor on the unit two-sphere
(with nonvanishing components $\epsilon_{\theta\phi} =
-\epsilon_{\phi\theta} = \sin\theta$). We adopt the convention that
upper-case Latin indices are raised and lowered with $\Omega^{AB}$ and
$\Omega_{AB}$, respectively. Finally, we note that
$D_C \Omega_{AB} = D_C \epsilon_{AB} = 0$.  

We convert the vector and tensor potentials from their initial  
Cartesian incarnations to angular-coordinate versions by making
use of the transformation matrix $\Omega^a_A$. We thus define 
\begin{equation} 
\EE{q}_A := \EE{q}_a \Omega^a_A,\qquad 
\EE{q}_{AB} := \EE{q}_{ab} \Omega^a_A \Omega^b_A, 
\label{conversion} 
\end{equation} 
and apply the same rule to all other potentials; for example,
$\GG{o}_{AB} := \GG{o}_{ab} \Omega^a_A \Omega^b_B$. After this 
conversion the tidal potentials become scalar, vector, and tensor 
fields on the unit two-sphere, and they depend on the angular
coordinates $\theta^A$ only. It is easy to show that the conversion
can be undone; for example $\EE{q}_a = \EE{q}_A \Omega^A_a$, with
$\Omega^A_a := \delta_{ab} \Omega^{AB} \Omega^b_B$. 

The tidal potentials can all be expressed in terms of (scalar, vector,
and tensor) spherical harmonics. Let $Y^{lm}$ be real-valued 
spherical-harmonic functions (as defined in Table~\ref{tab:Ylm}). The
relevant vectorial and tensorial harmonics of even parity are
\begin{subequations} 
\label{Ylm_even} 
\begin{align} 
& Y^{lm}_A := D_A Y^{lm}, \\
& Y^{lm}_{AB} := \Bigl[ D_A D_B 
+ \frac{1}{2} l(l+1) \Omega_{AB} \Bigr] Y^{lm};   
\end{align} 
\end{subequations}
notice that $\Omega^{AB} Y^{lm}_{AB} = 0$ by virtue of the eigenvalue  
equation satisfied by the spherical harmonics. The vectorial and
tensorial harmonics of odd parity are 
\begin{subequations} 
\label{Ylm_odd} 
\begin{align} 
& X^{lm}_A := -\epsilon_A^{\ B} D_B Y^{lm}, \\
& X^{lm}_{AB} := -\frac{1}{2} \Bigl( \epsilon_A^{\ C} D_B 
+ \epsilon_B^{\ C} D_A \Bigr) D_C Y^{lm} = 0; 
\end{align} 
\end{subequations} 
the tensorial harmonics $X^{lm}_{AB}$ also are tracefree: $\Omega^{AB}
X^{lm}_{AB} = 0$. The decomposition of the tidal potentials in
spherical harmonics is presented in Tables~\ref{tab:E_ang},
\ref{tab:B_ang},  \ref{tab:EE_ang}, \ref{tab:BB_ang},
\ref{tab:EBeven_ang}, and \ref{tab:EBodd_ang}. A derivation of these
results is presented in Appendix~\ref{app:spherical_harmonics}.  

\section{Geometry of a deformed black hole}   
\label{sec:blackhole}   

We construct the metric of a tidally deformed black hole in two
steps. In the first step we continue to think of a smooth timelike
geodesic $\gamma$ in a vacuum region of an arbitrary spacetime, and we 
construct the metric of this (background) spacetime in a neighborhood  
of the world line. We denote this neighborhood by $\N$, and we require
that it be small in comparison with the length scale $\R$ that
characterizes the tidal environment; we demand that 
\begin{equation} 
r \ll \R, 
\end{equation} 
where $r$ is a radial coordinate (to be introduced below) that
measures distance to the world line.  In the second step we
insert a black hole of mass $M$ into the background spacetime, and
place it on the world line $\gamma$. For the construction to be
successful it is necessary that the world tube traced by the black
hole fit comfortably within $\N$, and this can be achieved when    
\begin{equation} 
M \ll \R. 
\end{equation} 
This condition implies both that the black hole is weakly perturbed by
the tidal environment, and that the world tube is small when viewed on
the scale $\R$ of the external spacetime. In this situation it makes
(approximate) sense to say that the black hole moves on a world line
$\gamma$. 

The metric of the background spacetime is presented in
Sec.~\ref{subsec:background}, and the metric of the black-hole
spacetime is presented in Sec.~\ref{subsec:blackhole}. These
subsections contain what is truly a presentation of the metrics, and
there the reader will find no trace of a derivation of our results. The
calculations that lead to the metrics are quite lengthy, and
derivations are relegated to Secs.~\ref{sec:background_derivation}, 
\ref{sec:blackhole_derivation}, and Appendix~\ref{app:alternative}. We
begin in Sec.~\ref{subsec:coordinates} with a description 
of our coordinate systems. 

\subsection{Light-cone coordinates} 
\label{subsec:coordinates} 

\begin{figure} 
\includegraphics[width=1.0\linewidth]{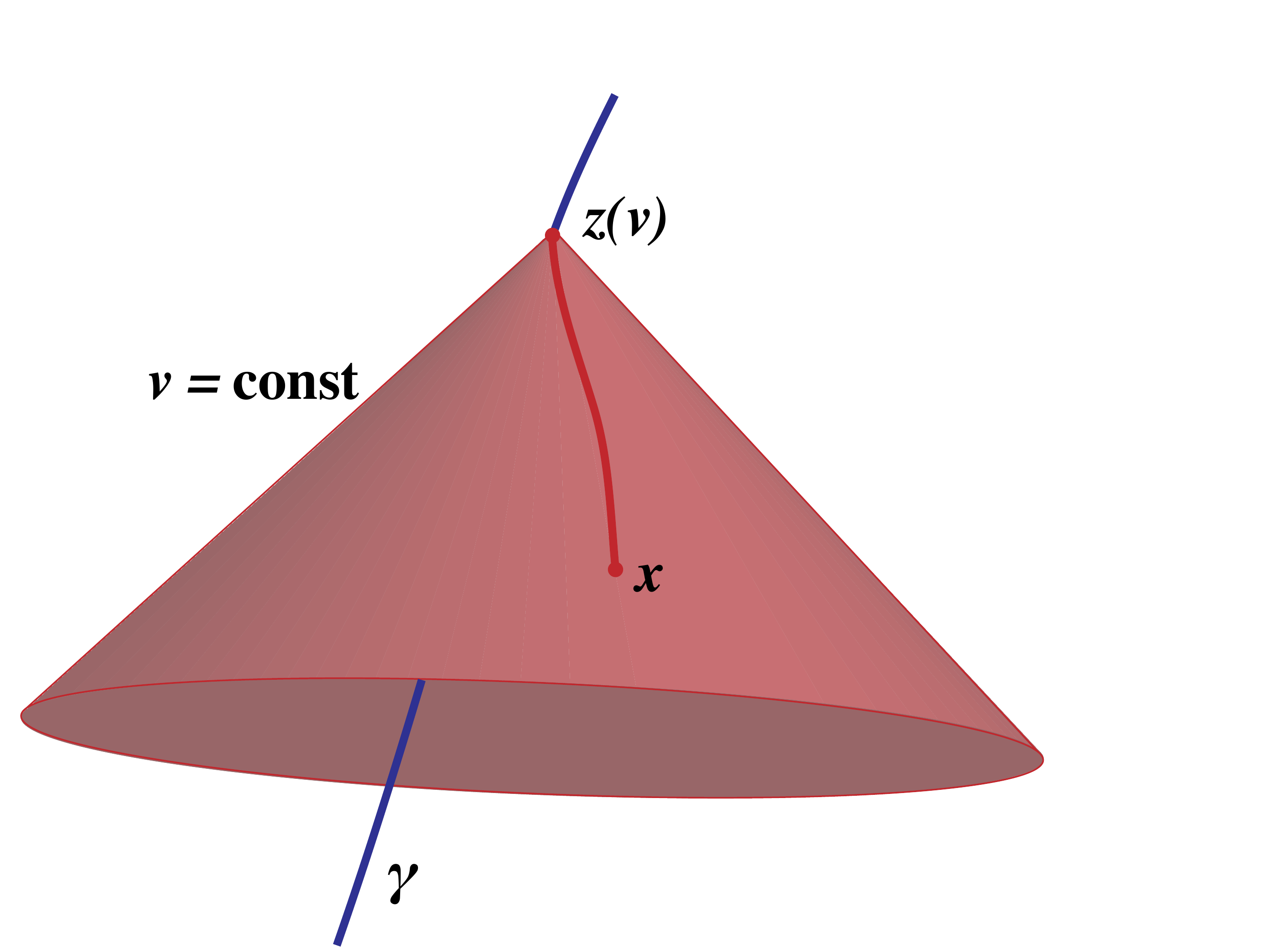}
\caption{Light-cone coordinates centered on a world line $\gamma$. The
figure shows a light cone $v=\mbox{constant}$ that intersects the 
world line at the point $z(v)$. It shows also one of the light cone's
generator, along which $\theta^A$ is constant; the affine parameter
$r$ decreases to zero as the generator approaches the world line.} 
\label{fig:background} 
\end{figure}

We work with a system $(v,r,\theta,\phi)$ that is specifically
tailored to describe the geometry of light-cone surfaces. We refer to
these as {\it light-cone coordinates}. In the case of the background
spacetime (Fig.~\ref{fig:background}), we consider past light cones
that converge toward the world line $\gamma$, so that the apex of each
cone coincides with a point on the world line. In $\N$ the light cones
provide spacetime with a foliation by null hypersurfaces, and each
light cone is generated by a congruence of null geodesics. (We assume
that caustics do not develop within $\N$, except at the apex of each
light cone.) The coordinates are intimately tied to the light cones
and their generators:   
\begin{enumerate} 
\item The advanced-time coordinate $v$ is constant on each light cone,
  and its value on a given light cone is equal to proper time $\tau$
  on the corresponding point of the world line.
\item The angular coordinates $\theta^A = (\theta,\phi)$ are constant
  on the null generators of each light cone; the angles refer to a
  set of axes that are aligned with the basis vectors $e^\alpha_a$
  introduced in Sec.~\ref{subsec:tidal_def}. 
\item The radial coordinate $r$ is an affine parameter on the null
  generators of each light cone, normalized in such a way that the
  metric takes a Minkowski form in the immediate vicinity  
  of the world line; note that $r$ decreases to zero as the
  generators converge toward the world line. 
\end{enumerate} 
To illustrate the meaning of the coordinates we examine the simple
case of an observer at rest in flat spacetime. The observer's proper
time $\tau$ is equal to coordinate time $t$, the advanced-time
coordinate is obtained by the transformation $v = t+r$, and
$(r,\theta,\phi)$ are obtained in the usual way from the Lorentzian
coordinates $x^a$. The metric is 
$ds^2 = -dv^2 + 2dvdr + r^2 d\Omega^2$, with $d\Omega^2  
:= d\theta^2 + \sin^2\theta\, d\phi^2$ denoting the metric on the
unit two-sphere. It is easy to verify that the light-cone coordinates
satisfies all the properties listed previously. 

\begin{figure} 
\includegraphics[width=1.0\linewidth]{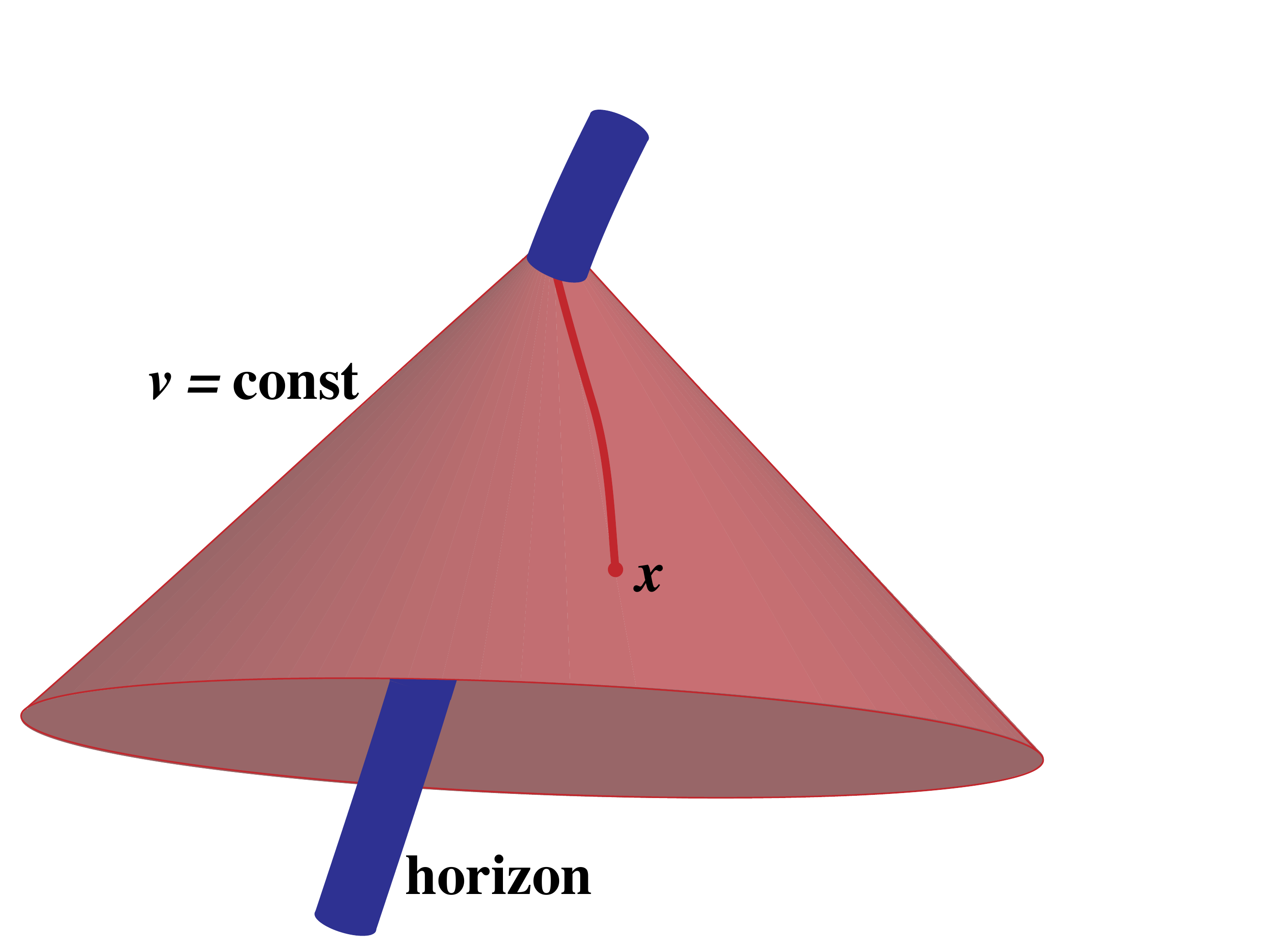}
\caption{Light-cone coordinates centered on a black hole. The figure
shows the world tube traced by the black-hole horizon. It shows also
the light cone $v=\mbox{constant}$ and one of its generators. As in
Fig.~1, the angles $\theta^A$ are constant and the affine parameter
$r$ decreases on the generator. } 
\label{fig:blackhole} 
\end{figure} 
 
In the case of the black hole spacetime (Fig.~\ref{fig:blackhole}),
the light cones no longer converge toward a world line. Instead they
converge toward the horizon, which traces a world tube in
spacetime. They still, however, provide $\N$ with a foliation by null  
hypersurfaces, and each light cone is still generated by a congruence
of null geodesics. The light-cone coordinates keep most of the
properties listed previously:  
\begin{enumerate} 
\item The advanced-time coordinate $v$ is constant on each light cone.
\item The angular coordinates $\theta^A = (\theta,\phi)$ are constant
  on the null generators of each light cone. 
\item The radial coordinate $r$ is an affine parameter on the null
  generators of each light cone.
\end{enumerate} 
Since the calibration by the world line is no longer available, the
coordinates have lost some of their rigidity. We can, however, restore
most of this rigidity by imagining two spacetime foliated by past
light cones. The first is the background spacetime, with its fully
specified set of light-cone coordinates. The second is the black-hole
spacetime, with its own set of light-cone coordinates. Far from the
black hole, where $r$ is much larger than $M$ but still much smaller 
than $\R$, the gravitational influence of the black hole is small, and
light rays behave there just as they do in the background
spacetime. We can therefore tune the black-hole coordinates so that
the asymptotic description of the null generators agree with the 
background description. This correspondence does not permit the
complete specification of the coordinates; there remains a limited
freedom to redefine the coordinates without changing their geometrical
meaning. We have thoroughly exploited this freedom to simplify the
form of the black-hole metric.  

The light-cone coordinates are best described in terms of the radial
distance $r$ and the angles $\theta^A$. It is useful, however, to
introduce also a variant of the light-cone coordinates that involves
the quasi-Cartesian system $x^a$ instead of the quasi-spherical system
$(r,\theta^A)$. The coordinates $x^a$ are defined in the usual way by  
Eq.~(\ref{trans_cart_sph}). In the flat-spacetime example examined
previously, we find that the components of the Minkowski metric are
given by $\eta_{vv} = -1$, $\eta_{va} = \Omega_a$, and $\eta_{ab} =
\gamma_{ab}$ when presented in the light-cone coordinates
$(v,x^a)$. It should be noted that the metric is mildly singular at
$x^a = 0$, because its value depends on the ambiguous direction of the 
longitudinal vector $\Omega^a$. This singularity persists in the
background spacetime, and dealing with it requires some care; but it
is not a serious obstacle to any computation.    

\subsection{Background spacetime} 
\label{subsec:background} 

The metric of a vacuum region of spacetime that surrounds a timelike
geodesic $\gamma$ can be presented in the light-cone coordinates 
$(v,x^a)$ introduced in the preceding subsection and expressed as an 
expansion in powers of $r/\R \ll 1$. It is given by 
\begin{widetext} 
\begin{subequations}
\label{background_metric_cart} 
\begin{align} 
g_{vv} &= -1 - r^2 \EE{q} + \frac{1}{3} r^3 \EEd{q}
- \frac{1}{3} r^3 \EE{o} - \frac{2}{21} r^4 \EEdd{q} 
+ \frac{1}{6} r^4 \EEd{o} - \frac{1}{12} r^4 \EE{h}
+ \frac{1}{15} r^4 \bigl( \PP{m} + \QQ{m} \bigr) 
\nonumber \\ & \quad \mbox{} 
+ \frac{2}{15} r^4 \GG{d}
+ \frac{2}{7} r^4 \QQ{q} 
+ \frac{2}{3} r^4 \GG{o} 
- \frac{1}{3} r^4 \bigl( \PP{h} + \QQ{h} \bigr)
+ O(5), \\
g_{va} &= \Omega_a 
- \frac{2}{3} r^2 \bigl( \EE{q}_a - \BB{q}_a \bigr) 
+ \frac{1}{3} r^3 \bigl( \EEd{q}_a - \BBd{q}_a \bigr) 
- \frac{1}{4} r^3 \bigl( \EE{o}_a - \BB{o}_a \bigr) 
- \frac{8}{63} r^4 \bigl( \EEdd{q}_a - \BBdd{q}_a \bigr) 
+ \frac{1}{6} r^4 \bigl( \EEd{o}_a - \BBd{o}_a \bigr) 
- \frac{1}{15} r^4 \bigl( \EE{h}_a - \BB{h}_a \bigr)
\nonumber \\ & \quad \mbox{} 
- \frac{8}{75} r^4 \GG{d}_a 
+ \frac{8}{21} r^4 \HH{q}_a
+ \frac{4}{105} r^4 \bigl( \PP{q}_a + 11 \QQ{q}_a \bigr) 
+ \frac{2}{5} r^4 \GG{o}_a 
- \frac{2}{15} r^4 \bigl( \PP{h}_a + \QQ{h}_a \bigr)
+ O(5), \\ 
g_{ab} &= \gamma_{ab} 
- \frac{1}{3} r^2 \bigl( \EE{q}_{ab} - \BB{q}_{ab} \bigr) 
+ \frac{5}{18} r^3 \bigl( \EEd{q}_{ab} - \BBd{q}_{ab} \bigr) 
- \frac{1}{6} r^3 \bigl( \EE{o}_{ab} - \BB{o}_{ab} \bigr) 
- \frac{1}{7} r^4 \bigl( \EEdd{q}_{ab} - \BBdd{q}_{ab} \bigr) 
+ \frac{3}{20} r^4 \bigl( \EEd{o}_{ab} - \BBd{o}_{ab} \bigr) 
\nonumber \\ & \quad \mbox{} 
- \frac{1}{20} r^4 \bigl( \EE{h}_{ab} - \BB{h}_{ab} \bigr)
+ \frac{8}{225} r^4 \gamma_{ab} \bigl( \PP{m} + \QQ{m} \bigr)
+ \frac{32}{225} r^4 \gamma_{ab}\, \GG{d} 
- \frac{16}{105} r^4 \gamma_{ab} \bigl( \PP{q} + \QQ{q} \bigr) 
- \frac{3}{14} r^4 \bigl( \PP{q}_{ab} - \QQ{q}_{ab} \bigr) 
\nonumber \\ & \quad \mbox{} 
+ \frac{3}{7} r^4 \HH{q}_{ab}
- \frac{8}{45} r^4 \gamma_{ab}\, \GG{o} 
+ \frac{2}{45} r^4 \gamma_{ab} \bigl( \PP{h} + \QQ{h} \bigr)
+ O(5). 
\end{align}
\end{subequations} 
\end{widetext} 
The metric features the tidal potentials encountered in
Sec.~\ref{subsec:tidal_pot_cartesian}, and the notation $O(5)$
indicates that the error terms are of order $(r/\R)^5$. The world line
is situated at $x^a = 0$, and $v$ is proper time on $\gamma$. 

The terms of order $(r/\R)^2$ involve the quadrupole tidal moments
$\E_{ab}$ and $\B_{ab}$, which are now expressed as functions of
advanced time $v$ instead of proper time $\tau$. At order $(r/\R)^3$
we find terms involving $\dot{\E}_{ab}$ and $\dot{\B}_{ab}$, the time
derivatives of the quadrupole moments, as well as the octupole tidal
moments $\E_{abc}$ and $\B_{abc}$. And at order $(r/\R)^4$ we see 
occurrences of $\ddot{\E}_{ab}$, $\ddot{\B}_{ab}$, $\dot{\E}_{abc}$,
$\dot{\B}_{abc}$, the hexadecapole tidal moments $\E_{abcd}$,
$\B_{abcd}$, and bilinear combinations of the quadrupole moments. We
see that the metric involves the majority of the tidal potentials
listed in Tables~\ref{tab:E_cart}, \ref{tab:B_cart},
\ref{tab:EE_cart}, \ref{tab:BB_cart}, \ref{tab:EBeven_cart}, and
\ref{tab:EBodd_cart}. Exceptions are $\PP{q}$, $\HH{h}_a$,
$\GG{o}_{ab}$, $\PP{h}_{ab}$, $\QQ{h}_{ab}$, and 
$\HH{h}_{ab}$; these could have occurred in the metric, but they
happen to be ruled out by the Einstein field equations.    

A useful way to view the metric of Eqs.~(\ref{background_metric_cart}) 
is to recognize that it possesses the correct number of 
freely-specifiable functions of time to describe correctly, and in
sufficient generality, the geometry of a vacuum region of spacetime in
a neighborhood $\N$ of a timelike geodesic. The functions of time
are contained in the tidal moments; we have 10 of them in $\E_{ab}$
and $\B_{ab}$, 14 in $\E_{abc}$ and $\B_{abc}$, and 18 in $\E_{abcd}$
and $\B_{abcd}$. The total is 42, the correct number for a metric
constructed through order $(r/\R)^4$. These functions encode  
meaningful information about the spacetime geometry; as we saw in 
Sec.~\ref{subsec:tidal_def} they represent components of the Weyl
tensor and its derivatives evaluated on the world line $\gamma$. 
 
In quasi-spherical coordinates $(v,r,\theta^A)$ the nonvanishing
components of the metric are 
\begin{widetext} 
\begin{subequations} 
\label{background_metric_ang} 
\begin{align} 
g_{vv} &= -1 - r^2 \EE{q} + \frac{1}{3} r^3 \EEd{q}
- \frac{1}{3} r^3 \EE{o} - \frac{2}{21} r^4 \EEdd{q} 
+ \frac{1}{6} r^4 \EEd{o} - \frac{1}{12} r^4 \EE{h}
+ \frac{1}{15} r^4 \bigl( \PP{m} + \QQ{m} \bigr) 
\nonumber \\ & \quad \mbox{} 
+ \frac{2}{15} r^4 \GG{d}
+ \frac{2}{7} r^4 \QQ{q} 
+ \frac{2}{3} r^4 \GG{o} 
- \frac{1}{3} r^4 \bigl( \PP{h} + \QQ{h} \bigr)
+ O(5), \\ 
g_{vr} &= 1, \\ 
g_{vA} &= 
-\frac{2}{3} r^3 \bigl( \EE{q}_A - \BB{q}_A \bigr) 
+ \frac{1}{3} r^4 \bigl( \EEd{q}_A - \BBd{q}_A \bigr) 
- \frac{1}{4} r^4 \bigl( \EE{o}_A - \BB{o}_A \bigr) 
- \frac{8}{63} r^5 \bigl( \EEdd{q}_A - \BBdd{q}_A \bigr) 
+ \frac{1}{6} r^5 \bigl( \EEd{o}_A - \BBd{o}_A \bigr) 
- \frac{1}{15} r^5 \bigl( \EE{h}_A - \BB{h}_A \bigr)
\nonumber \\ & \quad \mbox{} 
- \frac{8}{75} r^5 \GG{d}_A 
+ \frac{8}{21} r^5 \HH{q}_A
+ \frac{4}{105} r^5 \bigl( \PP{q}_A + 11 \QQ{q}_A \bigr) 
+ \frac{2}{5} r^5 \GG{o}_A 
- \frac{2}{15} r^5 \bigl( \PP{h}_A + \QQ{h}_A \bigr)
+ rO(5), \\ 
g_{AB} &= r^2 \Omega_{AB} 
- \frac{1}{3} r^4 \bigl( \EE{q}_{AB} - \BB{q}_{AB} \bigr) 
+ \frac{5}{18} r^5 \bigl( \EEd{q}_{AB} - \BBd{q}_{AB} \bigr) 
- \frac{1}{6} r^5 \bigl( \EE{o}_{AB} - \BB{o}_{AB} \bigr) 
- \frac{1}{7} r^6 \bigl( \EEdd{q}_{AB} - \BBdd{q}_{AB} \bigr) 
+ \frac{3}{20} r^6 \bigl( \EEd{o}_{AB} - \BBd{o}_{AB} \bigr) 
\nonumber \\ & \quad \mbox{} 
- \frac{1}{20} r^6 \bigl( \EE{h}_{AB} - \BB{h}_{AB} \bigr)
+ \frac{8}{225} r^6 \Omega_{AB} \bigl( \PP{m} + \QQ{m} \bigr)
+ \frac{32}{225} r^6 \Omega_{AB}\, \GG{d} 
- \frac{16}{105} r^6 \Omega_{AB} \bigl( \PP{q} + \QQ{q} \bigr) 
- \frac{3}{14} r^6 \bigl( \PP{q}_{AB} - \QQ{q}_{AB} \bigr) 
\nonumber \\ & \quad \mbox{} 
+ \frac{3}{7} r^6 \HH{q}_{AB}
- \frac{8}{45} r^6 \Omega_{AB}\, \GG{o} 
+ \frac{2}{45} r^6 \Omega_{AB} \bigl( \PP{h} + \QQ{h} \bigr)
+ r^2O(5). 
\end{align}
\end{subequations} 
\end{widetext} 
This can be obtained from Eq.~(\ref{background_metric_cart}) by
applying the transformation rules described in
Sec.~\ref{subsec:tidal_pot_angular}. The statements that $g_{vr} = 1$
and $g_{rA} = 0$ are exact, and they follow from the light-cone nature
of the coordinate system. Here the metric features the tidal potentials
listed in Tables~\ref{tab:E_ang}, \ref{tab:B_ang},
\ref{tab:EE_ang}, \ref{tab:BB_ang}, \ref{tab:EBeven_ang}, and
\ref{tab:EBodd_ang}. When dealing with the angular coordinates it is
convenient to rely on the decomposition of the potentials in scalar,
vector, and tensor harmonics. 

\subsection{Black-hole spacetime} 
\label{subsec:blackhole} 

The metric of a tidally deformed black hole resembles closely the
background metric presented in the preceding subsection. In the
quasi-Cartesian coordinates $(v,x^a)$ it is given by ($f := 1-2M/r$) 
\begin{widetext} 
\begin{subequations}
\label{blackhole_metric_cart} 
\begin{align}  
g_{vv} &= -f - r^2 \ee{q}{1} \EE{q} 
+ \frac{1}{3} r^3 \ee{q}{2} \EEd{q}
- \frac{1}{3} r^3 \ee{o}{1} \EE{o} 
- \frac{2}{21} r^4 \ee{q}{3} \EEdd{q} 
+ \frac{1}{6} r^4 \ee{o}{2} \EEd{o} 
- \frac{1}{12} r^4 \ee{h}{1} \EE{h}
+ \frac{1}{15} r^4 \bigl( \pp{m}{1} \PP{m} + \qq{m}{1} \QQ{m} \bigr)  
\nonumber \\ & \quad \mbox{} 
+ \frac{2}{15} r^4 \g{d}{1} \GG{d}
+ \frac{2}{7} r^4 \bigl( \pp{q}{1} \PP{q} + \qq{q}{1} \QQ{q} \bigr)  
+ \frac{2}{3} r^4 \g{o}{1} \GG{o} 
- \frac{1}{3} r^4 \bigl( \pp{h}{1} \PP{h} + \qq{h}{1} \QQ{h} \bigr)
+ O(5),  \\ 
g_{va} &= \Omega_a 
- \frac{2}{3} r^2 \bigl( \ee{q}{4} \EE{q}_a 
  - \bb{q}{4} \BB{q}_a \bigr)  
+ \frac{1}{3} r^3 \bigl( \ee{q}{5} \EEd{q}_a 
  - \bb{q}{5} \BBd{q}_a \bigr)  
- \frac{1}{4} r^3 \bigl( \ee{o}{4} \EE{o}_a 
  - \bb{o}{4} \BB{o}_a \bigr) 
- \frac{8}{63} r^4 \bigl( \ee{q}{6} \EEdd{q}_a 
  - \bb{q}{6} \BBdd{q}_a \bigr) 
\nonumber \\ & \quad \mbox{} 
+ \frac{1}{6} r^4 \bigl( \ee{o}{5} \EEd{o}_a 
  - \bb{o}{5} \BBd{o}_a \bigr) 
- \frac{1}{15} r^4 \bigl( \ee{h}{4} \EE{h}_a 
  - \bb{h}{4} \BB{h}_a \bigr)
- \frac{8}{75} r^4 \g{d}{2} \GG{d}_a 
+ \frac{8}{21} r^4 \hh{q}{2} \HH{q}_a
\nonumber \\ & \quad \mbox{} 
+ \frac{4}{105} r^4 \bigl( \pp{q}{2} \PP{q}_a 
   + 11 \qq{q}{2} \QQ{q}_a \bigr)  
+ \frac{2}{5} r^4 \g{o}{2} \GG{o}_a 
+ r^4 \hh{h}{2} \HH{h}_a 
- \frac{2}{15} r^4 \bigl( \pp{h}{2} \PP{h}_a 
   + \qq{h}{2} \QQ{h}_a \bigr) 
+ O(5), \\ 
g_{ab} &= \gamma_{ab}  
- \frac{1}{3} r^2 \bigl( \ee{q}{7} \EE{q}_{ab} 
  - \bb{q}{7} \BB{q}_{ab} \bigr) 
+ \frac{5}{18} r^3 \bigl( \ee{q}{8} \EEd{q}_{ab} 
  - \bb{q}{8} \BBd{q}_{ab} \bigr) 
- \frac{1}{6} r^3 \bigl( \ee{o}{7} \EE{o}_{ab} 
  - \bb{o}{7} \BB{o}_{ab} \bigr) 
- \frac{1}{7} r^4 \bigl( \ee{q}{9} \EEdd{q}_{ab} 
  - \bb{q}{9} \BBdd{q}_{ab} \bigr) 
\nonumber \\ & \quad \mbox{} 
+ \frac{3}{20} r^4 \bigl( \ee{o}{8} \EEd{o}_{ab} 
  - \bb{o}{8} \BBd{o}_{ab} \bigr) 
- \frac{1}{20} r^4 \bigl( \ee{h}{7} \EE{h}_{ab} 
  - \bb{h}{7} \BB{h}_{ab} \bigr)
+ \frac{8}{225} r^4 \gamma_{ab} 
   \bigl( \pp{m}{3} \PP{m} + \qq{m}{3} \QQ{m} \bigr)
+ \frac{32}{225} r^4 \gamma_{ab}\, \g{d}{3} \GG{d} 
\nonumber \\ & \quad \mbox{} 
- \frac{16}{105} r^4 \gamma_{ab} 
   \bigl( \pp{q}{3} \PP{q} + \qq{q}{3} \QQ{q} \bigr) 
- \frac{3}{14} r^4 
   \bigl( \pp{q}{4} \PP{q}_{ab} - \qq{q}{4} \QQ{q}_{ab} \bigr) 
+ \frac{3}{7} r^4 \hh{q}{3} \HH{q}_{ab}
- \frac{8}{45} r^4 \gamma_{ab}\, \g{o}{3} \GG{o} 
+ r^4 \g{o}{4} \GG{o}_{ab} 
\nonumber \\ & \quad \mbox{} 
+ \frac{2}{45} r^4 \gamma_{ab} 
   \bigl( \pp{h}{3} \PP{h} + \qq{h}{3} \QQ{h} \bigr)
+ r^4 \bigl( \pp{h}{4} \PP{h}_{ab} + \qq{h}{4} \QQ{h}_{ab} \bigr) 
+ r^4 \hh{h}{3} \HH{h}_{ab} 
+ O(5).  
\end{align} 
\end{subequations} 
\end{widetext} 
A comparison with Eqs.~(\ref{background_metric_cart}) reveals that 
there are two main differences between the metrics. The first is
that the black-hole metric involves the radial functions
$\ee{q}{n}$, $\ee{o}{n}$, $\ee{h}{n}$, $\bb{q}{n}$, $\bb{o}{n}$, and
$\bb{h}{n}$ that are listed in Table~\ref{tab:linear_functions}, as
well as the radial functions $\pp{m}{n}$, $\pp{q}{n}$, $\pp{h}{n}$,
$\qq{m}{n}$, $\qq{q}{n}$, $\qq{h}{n}$, $\g{d}{n}$, $\g{o}{n}$,
$\hh{q}{n}$, and $\hh{o}{n}$ that are listed in
Table~\ref{tab:nonlinear_functions}.  The second is that the tidal
potentials $\PP{q}$, $\HH{h}_a$, $\GG{o}_{ab}$, $\PP{h}_{ab}$,
$\QQ{h}_{ab}$, and $\HH{h}_{ab}$ now make an appearance in the
black-hole metric. (Recall that they were absent in the background
metric.) The black-hole metric contains the same number of
freely-specifiable functions of time as the background metric;
these are contained in the tidal moments $\E_{ab}$, $\B_{ab}$,
$\E_{abc}$, $\B_{abc}$, $\E_{abcd}$, and $\B_{abcd}$, which all depend
on advanced time $v$.    
  
\begin{table*} 
\caption{Radial functions: linear potentials. The radial functions
  are expressed in terms of $x := r/(2M)$ and $f := 1-1/x$. The
  dilogarithm function is defined as $\dilog x =  \int_1^x dt\,
  \log(t)/(1-t)$, with $\log x$ denoting the natural logarithm. All
  radial functions vanish at $r=2M$, except for $\ee{q}{7} =
  \frac{1}{2}$, $\ee{o}{7} = \frac{1}{10}$, $\ee{h}{7} =
  \frac{1}{42}$, $\bb{q}{7} = -\frac{1}{2}$, $\bb{o}{7} =
  -\frac{1}{10}$, $\bb{h}{7} = -\frac{1}{42}$.}   
\begin{ruledtabular} 
\begin{tabular}{l} 
$ \ee{q}{1} = f^2 $ \\ 
$ \ee{q}{2} = f \bigl[ 1 + \frac{1}{4x}( 5 + 12\log x )
  - \frac{1}{4x^2}( 27 + 12\log x ) + \frac{7}{4x^3} 
  + \frac{3}{4x^4} \bigr] $ \\
$ \ee{q}{3} = 1 + \frac{1}{24x}( 89 + 84\log x )
  + \frac{1}{160x^2} (431 + 996\log x - 1680 \dilog x)
  - \frac{1}{10x^3} (315 + 282\log x - 210 \dilog x) $ \\
$ \qquad \mbox{}
  + \frac{1}{120x^4} (4183 + 2322\log x - 1260 \dilog x) 
  - \frac{363}{40 x^5} - \frac{809}{480 x^6} $ \\
$ \ee{q}{4} = f $ \\ 
$ \ee{q}{5} = f \bigl[ 1 + \frac{1}{6x} (13 + 12 \log x) 
  - \frac{5}{2x^2} - \frac{3}{2x^3} - \frac{1}{2x^4} \bigr] $ \\ 
$ \ee{q}{6} = 1 + \frac{1}{32x} (117 + 84\log x) 
  + \frac{1}{320 x^2} ( 929 + 1836 \log x - 1680 \dilog x ) 
  - \frac{1}{40 x^3} (597 + 387 \log x - 210 \dilog x ) $ \\ 
$ \qquad \mbox{}
  + \frac{387}{80 x^4} + \frac{809}{480 x^5} 
  + \frac{809}{960 x^6} $ \\ 
$ \ee{q}{7} = 1 - \frac{1}{2x^2} $ \\ 
$ \ee{q}{8} = 1 + \frac{2}{5x} (4 + 3 \log x) - \frac{9}{5x^2} 
  - \frac{1}{5 x^3} (7 + 3 \log x) + \frac{3}{5x^4} $ \\ 
$ \ee{q}{9} = 1 + \frac{1}{216x} ( 739 + 420 \log x )
  + \frac{1}{90x^2} ( 262 + 387 \log x - 210 \dilog x )
  - \frac{1}{60 x^3} ( 317 + 70 \log x ) $ \\
$ \qquad \mbox{}
  - \frac{1}{540 x^4} ( 1511 + 1161 \log x - 630 \dilog x ) 
  + \frac{809}{1080 x^5} $ \\
  \\
$ \ee{o}{1} = f^2 \bigl( 1 - \frac{1}{2x} \bigr) $ \\
$ \ee{o}{2} = f \bigl[ 1 + \frac{1}{30x} (73 + 60\log x) 
  - \frac{1}{60x^2} ( 479 + 180 \log x ) 
  + \frac{1}{20x^3} ( 87 + 20 \log x ) 
  - \frac{3}{20x^4} + \frac{1}{60 x^5} \bigr] $ \\ 
$ \ee{o}{4} = f \bigl( 1 - \frac{2}{3x} \bigr) $ \\ 
$ \ee{o}{5} = f \bigl[ 1 + \frac{1}{40x} ( 103 + 60 \log x ) 
  - \frac{1}{60x^2} ( 233 + 60 \log x ) 
  + \frac{1}{5x^3} - \frac{1}{20x^4} - \frac{1}{120x^5} \bigr] $ \\ 
$ \ee{o}{7} = f + \frac{1}{10x^3} $ \\ 
$ \ee{o}{8} = 1 + \frac{1}{54x} ( 85 + 60 \log x ) 
  - \frac{10}{27 x^2} ( 10 + 3 \log x ) + \frac{2}{3x^3}  
  + \frac{1}{9x^4} ( 4 + \log x ) + \frac{1}{54x^5} $ \\
  \\
$ \ee{h}{1} = f^2 \bigl( f + \frac{3}{14x^2} \bigr) $ \\
$ \ee{h}{4} = f \bigl( 1 - \frac{5}{4x} + \frac{5}{14x^2} \bigr) $ \\ 
$ \ee{h}{7} = 1 - \frac{5}{3x} + \frac{5}{7x^2} - \frac{1}{42x^4} $ \\ 
  \\
$ \bb{q}{4} = f $ \\ 
$ \bb{q}{5} = f \bigl[ 1 + \frac{1}{6x} (7 + 12 \log x) 
  - \frac{3}{2x^2} - \frac{1}{2x^3} - \frac{1}{6x^4} \bigr] $ \\ 
$ \bb{q}{6} = 1 + \frac{1}{32x} (75 + 84\log x) 
  + \frac{1}{320 x^2} ( 649 + 996 \log x - 1680 \dilog x ) 
  - \frac{1}{80 x^3} (879 + 564 \log x - 420 \dilog x ) $ \\ 
$ \qquad \mbox{}
  + \frac{141}{40 x^4} + \frac{223}{160 x^5} 
  + \frac{223}{320 x^6} $ \\ 
$ \bb{q}{7} = 1 - \frac{3}{2x^2} $ \\ 
$ \bb{q}{8} = 1 + \frac{1}{5x} (5 + 6 \log x) - \frac{9}{5x^2} 
  - \frac{1}{5 x^3} (2 + 3 \log x) + \frac{1}{5x^4} $ \\ 
$ \bb{q}{9} = 1 + \frac{1}{216x} ( 529 + 420 \log x )
  + \frac{1}{360x^2} ( 593 + 1128 \log x - 840 \dilog x )
  - \frac{1}{30 x^3} ( 106 + 35 \log x ) $ \\
$ \qquad \mbox{}
  - \frac{1}{1080 x^4} ( 2357 + 1692 \log x - 1260 \dilog x ) 
  + \frac{223}{360 x^5} $ \\
  \\
$ \bb{o}{4} = f \bigl( 1 - \frac{2}{3x} \bigr) $ \\ 
$ \bb{o}{5} = f \bigl[ 1 + \frac{1}{40x} ( 97 + 60 \log x ) 
  - \frac{1}{60x^2} ( 227 + 60 \log x ) 
  + \frac{3}{10x^3} + \frac{1}{20x^4} + \frac{1}{120x^5} \bigr] $ \\ 
$ \bb{o}{7} = f - \frac{1}{10x^3} $ \\ 
$ \bb{o}{8} = 1 + \frac{1}{54x} ( 79 + 60 \log x ) 
  - \frac{1}{27 x^2} ( 97 + 30 \log x ) + \frac{2}{3x^3}  
  + \frac{1}{27x^4} ( 13 + 3\log x ) - \frac{1}{54x^5} $ \\
  \\
$ \bb{h}{4} = f \bigl( 1 - \frac{5}{4x} + \frac{5}{14x^2} \bigr) $ \\ 
$ \bb{h}{7} = 1 - \frac{5}{3x} + \frac{5}{7x^2} - \frac{1}{14x^4} $
\end{tabular}
\end{ruledtabular} 
\label{tab:linear_functions} 
\end{table*} 

\begin{table*} 
\caption{Radial functions: bilinear potentials. The radial
  functions are expressed in terms of $x := r/(2M)$ and $f :=
  1-1/x$. All radial functions vanish at $r=2M$, except for 
  $\pp{m}{3} = \qq{m}{3} = \pp{q}{3} = \qq{q}{3} = \pp{h}{3} =
  \qq{h}{3} = \frac{5}{16}$, $\g{d}{3} = \g{o}{3} = -\frac{5}{16}$, 
  $\pp{h}{4} = \frac{1}{84}$, $\qq{h}{4} = -\frac{5}{126}$, and
  $\hh{h}{3} = \frac{13}{252}$. All functions approach unity as 
  $x \to \infty$, except for $\pp{q}{1}$, $\pp{h}{4}$, $\qq{h}{4}$, 
  $\g{o}{4}$, $\hh{h}{2}$, and $\hh{h}{3}$, which all approach zero.}     
\begin{ruledtabular} 
\begin{tabular}{ll} 
$ \pp{m}{1} = f\bigl( 1 - \frac{19}{15 x} + \frac{1}{15 x^2} 
  + \frac{1}{15 x^3} + \frac{1}{5 x^4} \bigr) $ &
$ \qq{m}{1} = f\bigl( 1 - \frac{19}{15 x} + \frac{1}{15 x^2} 
  + \frac{1}{15 x^3} + \frac{1}{5 x^4} \bigr) $ \\ 
$ \pp{m}{3} = 1 - \frac{5}{4 x^2} + \frac{9}{16 x^4} $ &
$ \qq{m}{3} = 1 - \frac{15}{4 x^2} + \frac{49}{16 x^4} $ \\ 
  & \\
$ \pp{q}{1} = f^2\bigl( -\frac{26}{15 x} + \frac{19}{20 x^2} 
  + \frac{2}{15 x^3} + \frac{1}{15 x^4} \bigr) $ &
$ \qq{q}{1} = f^2\bigl( 1 + \frac{4}{15 x} - \frac{41}{20 x^2} 
  + \frac{2}{15 x^3} + \frac{1}{15 x^4} \bigr) $ \\ 
$ \pp{q}{2} = f\bigl( 1 - \frac{9}{x} + \frac{19}{4 x^2} 
  + \frac{1}{x^3} + \frac{1}{x^4} \bigr) $ &
$ \qq{q}{2} = f\bigl( 1 + \frac{1}{11 x} - \frac{101}{44 x^2} 
  + \frac{1}{11 x^3} + \frac{1}{11 x^4} \bigr) $ \\ 
$ \pp{q}{3} = 1 - \frac{5}{4 x^2} + \frac{5}{16 x^4} 
  + \frac{1}{4 x^5} $ &
$ \qq{q}{3} = 1 - \frac{15}{4 x^2} + \frac{45}{16 x^4} 
  + \frac{1}{4 x^5} $ \\ 
$ \pp{q}{4} = f\bigl( 1 + \frac{1}{x} \bigr) $ &
$ \qq{q}{4} = f\bigl( 1 + \frac{1}{x} - \frac{2}{x^2} 
  - \frac{2}{x^3} \bigr) $ \\ 
  & \\ 
$ \pp{h}{1} = f^2\bigl( 1 - \frac{121}{60 x} + \frac{39}{280 x^2}  
  + \frac{1}{30 x^3} + \frac{1}{60 x^4} \bigr) $ &
$ \qq{h}{1} = f^2\bigl( 1 + \frac{29}{60 x} - \frac{11}{280 x^2}  
  + \frac{1}{30 x^3} + \frac{1}{60 x^4} \bigr) $ \\ 
$ \pp{h}{2} = f\bigl( 1 - \frac{61}{24 x} + \frac{13}{28 x^2} 
  + \frac{1}{6 x^3} + \frac{1}{6 x^4} \bigr) $ &
$ \qq{h}{2} = f\bigl( 1 + \frac{29}{24 x} - \frac{221}{84 x^2} 
  + \frac{1}{6 x^3} + \frac{1}{6 x^4} \bigr) $ \\ 
$ \pp{h}{3} = 1 - \frac{5}{4 x^2} + \frac{5}{16 x^4} 
  + \frac{1}{4 x^5} $ &
$ \qq{h}{3} = 1 - \frac{15}{4 x^2} + \frac{45}{16 x^4} 
  + \frac{1}{4 x^5} $ \\ 
$ \pp{h}{4} = \frac{5}{36 x} - \frac{29}{252 x^2} 
  - \frac{1}{84 x^4} $ &
$ \qq{h}{4} = -\frac{5}{36 x} + \frac{19}{84 x^2} 
  - \frac{8}{63 x^4} $ \\ 
  & \\ 
$ \g{d}{1} = f^2\bigl( 1 + \frac{2}{15 x} - \frac{7}{5 x^2} 
  + \frac{1}{15 x^3} + \frac{1}{30 x^4} \bigr) $ & \\ 
$ \g{d}{2} = f\bigl( 1 - \frac{2}{3 x} + \frac{1}{6 x^2} 
  - \frac{1}{24 x^3} - \frac{1}{24 x^4} \bigr) $ &
$ \hh{q}{2} = f^2\bigl( 1 + \frac{2}{x} \bigr) $ \\ 
$ \g{d}{3} = 1 - \frac{5}{2 x^2} + \frac{5}{4 x^4} 
  - \frac{1}{16 x^5} $ & 
$ \hh{q}{3} = f^2\bigl( 1 + \frac{1}{x} \bigr)^2 $ \\  
  & \\ 
$ \g{o}{1} = f^2\bigl( 1 + \frac{79}{30 x} - \frac{53}{20 x^2} 
  + \frac{1}{15 x^3} + \frac{1}{30 x^4} \bigr) $ & \\ 
$ \g{o}{2} = f\bigl( 1 + \frac{89}{24 x} - \frac{29}{6 x^2} 
  + \frac{1}{6 x^3} + \frac{1}{6 x^4} \bigr) $ & 
$ \hh{h}{2} = -\frac{1}{2 x} + \frac{29}{42 x^2} 
  - \frac{4}{21 x^3} $ \\ 
$ \g{o}{3} = 1 - \frac{5}{2 x^2} + \frac{15}{16 x^4} 
  + \frac{1}{4 x^5} $ &
$ \hh{h}{3} = -\frac{5}{18 x} + \frac{43}{126 x^2} 
  - \frac{1}{84 x^4} $ \\ 
$ \g{o}{4} = f\bigl( \frac{7}{6 x} - \frac{1}{3 x^2} 
  - \frac{1}{3 x^3} \bigr) $ &
\end{tabular}
\end{ruledtabular} 
\label{tab:nonlinear_functions} 
\end{table*} 

The metric of Eqs.~(\ref{blackhole_metric_cart}) contains two
fundamental length scales, the black-hole mass $M$ and the tidal 
radius $\R$. It is assumed that $M \ll \R$, and the metric is valid
when $r \ll \R$. When $M \to 0$ the metric becomes the background 
metric of Eqs.~(\ref{background_metric_cart}). This can be seen from 
the fact that most of the radial functions approach unity when 
$2M/r \to 0$. Exceptions are $\pp{q}{1}$, $\g{o}{4}$, $\pp{h}{4}$,
$\qq{h}{4}$, $\hh{h}{2}$, and $\hh{h}{3}$, which all approach zero;
these functions come with the tidal potentials that were absent in the
background metric. In this limit the tidal moments acquire their
precise relation with the Weyl tensor (and its derivatives) evaluated
on the world line $x^a = 0$.    

When $M \neq 0$ the world line disappears and is replaced by the world
tube traced by the black-hole horizon. But the black-hole metric
continues to approach the background metric when $r$ is very large 
compared with $M$ (but still much smaller than $\R$). In this case the
interpretation of the tidal moments is more subtle. They still provide
a complete characterization of the tidal environment, but they are no
longer associated with the Weyl tensor evaluated on a world line. 

To elucidate the meaning of the tidal moments it is helpful to return
to the two spacetimes of Sec.~\ref{subsec:coordinates}. The first is
the background spacetime with its central world line, and the second
is the black-hole spacetime in which no such world line exists. We
assume that the spacetimes share the same set of tidal moments: the 
functions of time contained in $\E_{ab}$, $\B_{ab}$, $\E_{abc}$,
$\B_{abc}$, $\E_{abcd}$, and $\B_{abcd}$ are the same in each 
spacetime. An observer in the first spacetime can measure the tidal 
moments anywhere and relate them to the Weyl tensor (and its
derivatives) evaluated on the central world line. An observer in the
second spacetime can still measure the tidal moments, but is prevented 
by the black hole from relating the results of his measurements to a 
world-line Weyl tensor. But the second observer is free to imagine
that the black hole could be removed from the spacetime without
altering the conditions in the asymptotic region $r \gg M$, so that
his measurements could, after all, be related to a world-line Weyl
tensor.\footnote{This is a thought experiment that could not be
  realized physically. It is meant to reflect a mathematical procedure
  in which $M$ is taken to zero while keeping the tidal moments
  fixed. The procedure disregards the fact that in most physical
  applications, the tidal moments carry a dependence on $M$
  that is revealed by matching the local black-hole metric to a global
  metric that contains the black hole and the external bodies.}
This thought experiment provides him with a loose
interpretation of the tidal moments: They can still be related to a
world-line Weyl tensor, but the Weyl tensor and the world line are
fictitious extrapolations to $r = 0$ obtained from information
available in the $r \gg M$ portion of the black-hole spacetime. 

When the tidal perturbation is turned off (by putting all the tidal
moments to zero, thereby sending $\R$ off to infinity), the black-hole  
metric becomes   
\begin{equation} 
g_{vv} = -f, \qquad 
g_{va} = \Omega_a, \qquad 
g_{ab} = \gamma_{ab}, 
\label{schwarzschild_cart} 
\end{equation} 
where, we recall, $f = 1-2M/r$. This is the well-known Schwarzschild
solution expressed in the light-cone coordinates $(v,x^a$). 

In quasi-spherical coordinates $(v,r,\theta^A)$ the nonvanishing
components of the black-hole metric are 
\begin{widetext} 
\begin{subequations} 
\label{blackhole_metric_ang} 
\begin{align} 
g_{vv} &= -f - r^2 \ee{q}{1} \EE{q} 
+ \frac{1}{3} r^3 \ee{q}{2} \EEd{q}
- \frac{1}{3} r^3 \ee{o}{1} \EE{o} 
- \frac{2}{21} r^4 \ee{q}{3} \EEdd{q} 
+ \frac{1}{6} r^4 \ee{o}{2} \EEd{o} 
- \frac{1}{12} r^4 \ee{h}{1} \EE{h}
+ \frac{1}{15} r^4 \bigl( \pp{m}{1} \PP{m} + \qq{m}{1} \QQ{m} \bigr)  
\nonumber \\ & \quad \mbox{} 
+ \frac{2}{15} r^4 \g{d}{1} \GG{d}
+ \frac{2}{7} r^4 \bigl( \pp{q}{1} \PP{q} + \qq{q}{1} \QQ{q} \bigr)  
+ \frac{2}{3} r^4 \g{o}{1} \GG{o} 
- \frac{1}{3} r^4 \bigl( \pp{h}{1} \PP{h} + \qq{h}{1} \QQ{h} \bigr)
+ O(5),  \\ 
g_{vr} &= 1, \\ 
g_{vA} &= 
-\frac{2}{3} r^3 \bigl( \ee{q}{4} \EE{q}_A 
  - \bb{q}{4} \BB{q}_A \bigr)  
+ \frac{1}{3} r^4 \bigl( \ee{q}{5} \EEd{q}_A 
  - \bb{q}{5} \BBd{q}_A \bigr)  
- \frac{1}{4} r^4 \bigl( \ee{o}{4} \EE{o}_A 
  - \bb{o}{4} \BB{o}_A \bigr) 
- \frac{8}{63} r^5 \bigl( \ee{q}{6} \EEdd{q}_A 
  - \bb{q}{6} \BBdd{q}_A \bigr) 
\nonumber \\ & \quad \mbox{} 
+ \frac{1}{6} r^5 \bigl( \ee{o}{5} \EEd{o}_A 
  - \bb{o}{5} \BBd{o}_A \bigr) 
- \frac{1}{15} r^5 \bigl( \ee{h}{4} \EE{h}_A 
  - \bb{h}{4} \BB{h}_A \bigr)
- \frac{8}{75} r^5 \g{d}{2} \GG{d}_A 
+ \frac{8}{21} r^5 \hh{q}{2} \HH{q}_A
\nonumber \\ & \quad \mbox{} 
+ \frac{4}{105} r^5 \bigl( \pp{q}{2} \PP{q}_A 
   + 11 \qq{q}{2} \QQ{q}_A \bigr)  
+ \frac{2}{5} r^5 \g{o}{2} \GG{o}_A 
+ r^5 \hh{h}{2} \HH{h}_A 
- \frac{2}{15} r^5 \bigl( \pp{h}{2} \PP{h}_A 
   + \qq{h}{2} \QQ{h}_A \bigr) 
+ rO(5), \\ 
g_{AB} &= r^2 \Omega_{AB} 
- \frac{1}{3} r^4 \bigl( \ee{q}{7} \EE{q}_{AB} 
  - \bb{q}{7} \BB{q}_{AB} \bigr) 
+ \frac{5}{18} r^5 \bigl( \ee{q}{8} \EEd{q}_{AB} 
  - \bb{q}{8} \BBd{q}_{AB} \bigr) 
- \frac{1}{6} r^5 \bigl( \ee{o}{7} \EE{o}_{AB} 
  - \bb{o}{7} \BB{o}_{AB} \bigr) 
- \frac{1}{7} r^6 \bigl( \ee{q}{9} \EEdd{q}_{AB} 
  - \bb{q}{9} \BBdd{q}_{AB} \bigr) 
\nonumber \\ & \quad \mbox{} 
+ \frac{3}{20} r^6 \bigl( \ee{o}{8} \EEd{o}_{AB} 
  - \bb{o}{8} \BBd{o}_{AB} \bigr) 
- \frac{1}{20} r^6 \bigl( \ee{h}{7} \EE{h}_{AB} 
  - \bb{h}{7} \BB{h}_{AB} \bigr)
+ \frac{8}{225} r^6 \Omega_{AB} 
   \bigl( \pp{m}{3} \PP{m} + \qq{m}{3} \QQ{m} \bigr)
+ \frac{32}{225} r^6 \Omega_{AB}\, \g{d}{3} \GG{d} 
\nonumber \\ & \quad \mbox{} 
- \frac{16}{105} r^6 \Omega_{AB} 
   \bigl( \pp{q}{3} \PP{q} + \qq{q}{3} \QQ{q} \bigr) 
- \frac{3}{14} r^6 
   \bigl( \pp{q}{4} \PP{q}_{AB} - \qq{q}{4} \QQ{q}_{AB} \bigr) 
+ \frac{3}{7} r^6 \hh{q}{3} \HH{q}_{AB}
- \frac{8}{45} r^6 \Omega_{AB}\, \g{o}{3} \GG{o} 
+ r^6 \g{o}{4} \GG{o}_{AB} 
\nonumber \\ & \quad \mbox{} 
+ \frac{2}{45} r^6 \Omega_{AB} 
   \bigl( \pp{h}{3} \PP{h} + \qq{h}{3} \QQ{h} \bigr)
+ r^6 \bigl( \pp{h}{4} \PP{h}_{AB} + \qq{h}{4} \QQ{h}_{AB} \bigr) 
+ r^6 \hh{h}{3} \HH{h}_{AB} 
+ r^2O(5).  
\end{align}
\end{subequations} 
\end{widetext} 
In these coordinates the $\R \to \infty$ limit of the metric is $ds^2
= f\, dv^2 + 2\, dvdr + r^2\, d\Omega^2$, the Eddington-Finkelstein
form of the Schwarzschild solution. 

\section{Geometry and dynamics of the deformed horizon} 
\label{sec:horizon} 

In this section we extract the consequences of the black-hole metric
of Eqs.~(\ref{blackhole_metric_ang}) on the structure and dynamics of
the tidally deformed horizon. We begin in
Sec.~\ref{subsec:horizon_position} with a proof that in the coordinate
system adopted here, the deformed horizon continues to be described by
$r=2M$. In Sec.~\ref{subsec:horizon_metric} we display the components
of the induced metric on the horizon. In
Sec.~\ref{subsec:horizon_expansion} we examine the congruence of null
geodesics that generates the horizon, and derive expressions for its
expansion scalar and shear tensor. And finally, in
Sec.~\ref{subsec:heating} we integrate Raychaudhuri's equation and
calculate the rate at which the horizon grows as a result of
the tidal interaction. 

\subsection{Position of the horizon}  
\label{subsec:horizon_position} 

The black-hole metric of Eqs.~(\ref{blackhole_metric_ang}) is
presented in light-cone coordinates $(v,r,\theta^A)$ whose geometrical
meaning was described in Sec.~\ref{subsec:coordinates}. As we saw, the
coordinates are tied to the behavior of incoming light rays that mesh
together to form light cones that converge toward the black hole. And
as we saw, the light-cone coordinates are not fully specified; the
limited coordinate freedom that remains was exploited to simplify the 
form of the metric.  

Concretely, the coordinate freedom was utilized to impose a set of  
{\it horizon-locking conditions} that are designed to keep the
black-hole horizon in its usual place. As a result, {\it the horizon
of a tidally deformed  black hole is situated at} 
\begin{equation}
r = 2M \bigl[ 1 + O(5) \bigr] 
\label{2M} 
\end{equation}   
{\it in the light-cone coordinates that give rise to the metric of 
Eqs.~(\ref{blackhole_metric_ang}).} In Eq.~(\ref{2M}), and in all
equations below, the symbol $O(5)$ means that the correction terms are
of order $(M/\R)^5$. The  horizon-locking conditions are 
\begin{equation} 
g_{vv} = 0 = g_{vA} \qquad \mbox{at $r=2M$}, 
\label{horizon-locking} 
\end{equation} 
and it is easy to verify that the metric of
Eqs.~(\ref{blackhole_metric_ang}) satisfies this property.  

The horizon-locking conditions imply that the surface $r=2M$ is a null
hypersurface in the black-hole spacetime. This is seen from the fact
that $g^{rr} = 0$ at $r=2M$, which follows from the statement
that thanks to Eq.~(\ref{horizon-locking}), $g_{vr}=1$ and $g_{AB}$
are the only nonvanishing components of the metric at $r=2M$. The null 
hypersurface is generated by a congruence of null geodesics, and in
Sec.~\ref{subsec:horizon_expansion}  we show that the expansion 
of this congruence vanishes. [The expansion scalar is denoted
$\Theta$, and we show more precisely that $\Theta = O(M^5/\R^6)$. This
implies that the expansion vanishes through order $\R^{-4}$, which is
the order of accuracy maintained in the metric and the description of
the surface.] The surface $r=2M$ is therefore a 
{\it stationary null surface}, and it is in this sense that it
describes a black-hole horizon. It is an {\it isolated horizon}  in
the sense of Ashtekar and Krishnan \cite{ashtekar-krishnan:02,
  ashtekar-krishnan:03, ashtekar-krishnan:04}.   

In general we cannot state that $r=2M$ is an event horizon, because
the exact position of the event horizon depends on the entire future
history of the spacetime, and is located by tracing light rays
backwards in time from future null infinity. Because our metric may
not be accurate for all times, and because it applies only to a
neighborhood of the black hole, it does not supply us with the tools
to locate the event horizon. Under restrictive assumptions, however,
we can go beyond these limitations and prove that $r=2M$ describes the 
position of the event horizon. 

We consider a situation in which the tidal perturbation is switched
off after a time $v = v_1$, so that the metric returns to its usual
Schwarzschild form when $v > v_1$. This final Schwarzschild metric
possesses the same mass parameter $M$ as the original black-hole
metric, because (as we shall see in Sec.~\ref{subsec:heating}) the
total change in mass that can result from tidal heating over a time
$v_1 \sim {\cal R}$ is of order $M^6/{\cal R}^5$; this effect is too
small to be described by our metric. It follows that the event horizon
is described by $r=2M$ when $v >  v_1$. At earlier times the horizon is
described by a null hypersurface that joins smoothly with $r=2M$ at
$v=v_1$. Since $r=2M$ is null at all times, we conclude that the
event horizon is always situated at $r=2M$.  

\subsection{Horizon metric and surface gravity} 
\label{subsec:horizon_metric} 

The horizon metric is the specialization of the black-hole metric to
the null hypersurface $r=2M[1+O(5)]$. Because its component along the    
$v$-direction vanishes, the horizon metric is degenerate and explicitly
two-dimensional. We denote its nonvanishing components by 
$\gamma_{AB}$, and these are obtained by inserting $r=2M$ within  
Eqs.~(\ref{blackhole_metric_ang}). We get 
\begin{widetext} 
\begin{align}  
\gamma_{AB} &= 4 M^2 \Omega_{AB} 
- \frac{8}{3} M^4 \bigl( \EE{q}_{AB} + \BB{q}_{AB} \bigr) 
- \frac{8}{15} M^5 \bigl( \EE{o}_{AB} + \BB{o}_{AB} \bigr) 
- \frac{8}{105} M^6 \bigl( \EE{h}_{AB} + \BB{h}_{AB} \bigr)
\nonumber \\ & \quad \mbox{}
+ \frac{32}{45} M^6 \Omega_{AB} 
   \bigl( \PP{m} + \QQ{m} \bigr)
- \frac{128}{45} M^6 \Omega_{AB}\, \GG{d} 
- \frac{64}{21} M^6 \Omega_{AB} 
   \bigl( \PP{q} + \QQ{q} \bigr) 
+ \frac{32}{9} M^6 \Omega_{AB}\, \GG{o} 
\nonumber \\ & \quad \mbox{}
+ \frac{8}{9} M^6 \Omega_{AB} 
   \bigl( \PP{h} + \QQ{h} \bigr)
+ \frac{16}{21} M^6 \PP{h}_{AB} 
- \frac{160}{63} M^6 \QQ{h}_{AB}  
+ \frac{208}{63} M^6 \HH{h}_{AB}
+ M^2 O(5).  
\label{horizon_metric}  
\end{align}  
\end{widetext} 
In Appendix~\ref{app:determinant} we calculate the determinant
$\gamma$ of the horizon metric. We find the simple result 
\begin{equation} 
\sqrt{\gamma} = 4 M^2\sin\theta \bigl[ 1 + O(5) \bigr]; 
\label{determinant} 
\end{equation} 
the tidal potentials do not give rise to corrections to the horizon's
surface element, $\sqrt{\gamma} d\theta d\phi$.       

The surface gravity $\kappa$ of the perturbed horizon is defined by 
the relation $k^\beta \nabla_\beta k^\alpha = \kappa k^\alpha$, where
the vector $k^\alpha := [1,0,0,0]$ is tangent to the horizon's null
generators. A calculation based on the metric of
Eqs.~(\ref{blackhole_metric_ang}) reveals that  
\begin{align} 
\kappa &= \frac{1}{4M} \biggl[ 1 
+ \frac{16}{3} M^3 \dot{\E}_{ab} \Omega^a \Omega^b 
+ \frac{32}{9} M^4 \ddot{\E}_{ab} \Omega^a \Omega^b 
\nonumber \\ & \quad \mbox{} 
+ \frac{8}{9} M^4 \dot{\E}_{abc} \Omega^a \Omega^b \Omega^c 
- \frac{16}{225} M^4 \bigl( \E_{ab} \E^{ab} + \B_{ab} \B^{ab} \bigr) 
\nonumber \\ & \quad \mbox{} 
+ O(5) \biggr].
\label{surface_gravity}
\end{align}  
We see that the surface gravity is no longer uniform on the horizon; 
the tidal perturbation introduces a variation of order $(M/{\R})^3$ 
over its surface. Notice that this variation is associated with 
{\it changes} in the tidal environment. When the perturbation is
stationary the correction to the surface gravity comes from the last
term, which is of order $(M/\R)^4$ and uniform over the horizon. These
observations are compatible with the zeroth law of black-hole
mechanics.   

\subsection{Expansion scalar and shear tensor}  
\label{subsec:horizon_expansion} 

The null generators of the horizon form a congruence whose behavior
is described by an expansion scalar $\Theta$ and a shear tensor
$\sigma_{AB}$; these are defined by  (see, for example, Sec.~III of
Ref.~\cite{poisson:04d}) 
\begin{equation} 
\partial_v \gamma_{AB} = \Theta \gamma_{AB} + 2\sigma_{AB},  
\label{gamma_dot} 
\end{equation} 
together with the requirement that the shear tensor be tracefree:
$\gamma^{AB} \sigma_{AB} = 0$. The expansion scalar is then equal to
the trace of $\partial_v\gamma_{AB}$, so that $\Theta = \frac{1}{2}
\gamma^{-1} \partial_v\gamma$, where $\gamma$ is the metric
determinant. Equation (\ref{determinant}) implies that 
\begin{equation} 
\Theta = O(M^5/\R^6).  
\label{expansion} 
\end{equation} 
This means that up to this level of accuracy, the surface $r=2M$ is 
foliated by apparent horizons.   

With this we find that Eq.~(\ref{gamma_dot}) reduces to $\sigma_{AB}   
= \frac{1}{2} \partial_v\gamma_{AB} + O(M^7/\R^6)$. This is 
\begin{widetext} 
\begin{align}
 \sigma_{AB} &=  
- \frac{4}{3} M^4 \bigl( \EEd{q}_{AB} + \BBd{q}_{AB} \bigr) 
- \frac{4}{15} M^5 \bigl( \EEd{o}_{AB} + \BBd{o}_{AB} \bigr) 
- \frac{4}{105} M^6 \bigl( \EEd{h}_{AB} + \BBd{h}_{AB} \bigr)
\nonumber \\ & \quad \mbox{}
+ \frac{16}{45} M^6 \Omega_{AB} 
   \bigl( \PPd{m} + \QQd{m} \bigr)
- \frac{64}{45} M^6 \Omega_{AB}\, \GGd{d} 
- \frac{32}{21} M^6 \Omega_{AB} 
   \bigl( \PPd{q} + \QQd{q} \bigr) 
+ \frac{16}{9} M^6 \Omega_{AB}\, \GGd{o} 
\nonumber \\ & \quad \mbox{}
+ \frac{4}{9} M^6 \Omega_{AB} 
   \bigl( \PPd{h} + \QQd{h} \bigr)
+ \frac{8}{21} M^6 \PPd{h}_{AB} 
- \frac{80}{63} M^6 \QQd{h}_{AB}  
+ \frac{104}{63} M^6 \HHd{h}_{AB}
+ MO(6). 
\label{shear}  
\end{align} 
\end{widetext} 

\subsection{Tidal heating} 
\label{subsec:heating} 

More information about the expansion scalar can be obtained by
integrating Raychaudhuri's equation (see, for example, Sec.~III of 
Ref.~\cite{poisson:04d}) 
\begin{equation} 
\partial_v \Theta = \kappa \Theta - \frac{1}{2} \Theta^2 
- \sigma_{AB} \sigma^{AB}, 
\label{Ray}
\end{equation} 
in which we can neglect $\Theta^2 = O(M^{10}/\R^{12})$. The
squared shear contains terms that begin at order $M^4/\R^6$, and 
the neglected terms are of order $M^7/\R^9$ and higher. The solution 
to Eq.~(\ref{Ray}) will be an expression for $\Theta$ that begins at
order $M^5/\R^6$, and neglects terms of order $M^8/\R^9$
and higher. Given this degree of available accuracy, it is appropriate
to set 
\begin{equation} 
\kappa \simeq \kappa_0 := \frac{1}{4M} 
\end{equation} 
in Eq.~(\ref{Ray}); the fractional corrections of order 
$(M/\R)^3$ and $(M/\R)^4$ displayed in Eq.~(\ref{surface_gravity}) are 
not required in the determination of $\Theta$. 

Instead of proceeding directly with Eq.~(\ref{Ray}), we integrate it
over a cross-section of the horizon and work out a differential
equation for the surface area 
\begin{equation} 
{\cal A}(v) := \int \sqrt{\gamma}\, d\theta d\phi. 
\label{area}
\end{equation} 
Its rate of change is $\dot{\cal A} 
= \int \Theta \sqrt{\gamma}\, d\theta d\phi$, and the second rate of 
change is $\ddot{\cal A} = \int \partial_v \Theta \sqrt{\gamma}
d\theta d\phi + \int \Theta^2 \sqrt{\gamma}\, d\theta d\phi$; here we
may once more ignore the second integral involving $\Theta^2$.    

The differential equation that governs the behavior of the area
function is
\begin{equation} 
\kappa_0 \dot{\cal A} - \ddot{\cal A} = \int \sigma_{AB} \sigma^{AB}
\sqrt{\gamma}\, d\theta d\phi + O(9).
\label{Ray_integrated}
\end{equation} 
The steps required in the evaluation of the right-hand side are
described in Appendix~\ref{app:horizon_details}, and the end result is  
\begin{equation} 
\kappa_0 \dot{\cal A} - \ddot{\cal A}
= 8\pi {\cal F}, 
\label{Adeq}
\end{equation} 
where 
\begin{align} 
{\cal F}(v) &:= 
\frac{16}{45} M^6 \Bigl( \dot{\E}_{ab} \dot{\E}^{ab} 
 + \dot{\B}_{ab} \dot{\B}^{ab} \Bigr) 
\nonumber \\ & \quad \mbox{} 
+ \frac{16}{4725} M^8 \Bigl( \dot{\E}_{abc} \dot{\E}^{abc} 
 + \frac{16}{9} \dot{\B}_{abc} \dot{\B}^{abc} \Bigr) 
+ O(9) 
\label{flux}
\end{align} 
might be called the {\it flux function}. We notice that ${\cal F}$ is  
positive-definite, and that it cannot be expressed as a total
derivative with respect to $v$. 

As we show in Appendix~\ref{app:horizon_details}, the solution to
Eq.~(\ref{Adeq}) is 
\begin{equation} 
\frac{\kappa_0}{8\pi} \dot{\cal A}(v) = {\cal F}(v) 
+ \frac{\dot{\cal F}(v) }{\kappa_0} 
+ \frac{\ddot{\cal F}(v)}{\kappa_0^2}  
+ O(9). 
\label{Adot_sol} 
\end{equation} 
We observe that the second and third terms on the right-hand side  
are total time derivatives, and that it is natural to bring them over
to the left-hand side of the equation. We write our final result as 
\begin{equation} 
\frac{\kappa_0}{8\pi} \dot{\cal A}^* = {\cal F} + O(9), 
\label{tidal_heating}
\end{equation} 
in terms of a ``renormalized area function'' defined by
${\cal A}^* = {\cal A} - 8\pi {\cal F}/\kappa_0^2 
- 8\pi \dot{\cal F}/\kappa_0^3 + O(M^{10}/\R^8)$. The shift in 
area is of order $M^8/\R^6$ and corresponds to a correction
to the relation $r=2M$ of fractional order $(M/\R)^6$; this is
well beyond the level of accuracy of Eq.~(\ref{2M}). 

Equation (\ref{tidal_heating}) describes the rate at which the
renormalized black-hole area ${\cal A}^*$ increases as a result of the
tidal interaction. We refer to this phenomenon as {\it tidal heating},
recalling the deep analogy that exists between the relativistic tidal
dynamics of black holes and the Newtonian tidal dynamics of viscous
bodies \cite{hartle:73, hartle:74, poisson:09}. The phrase also
recalls the fact that in black-hole thermodynamics, the area plays the
role of entropy $S$, while the surface gravity plays the role of
(Hawking) temperature $T$; in this context we write 
$(\kappa_0/8\pi) \dot{\cal A}^* = T \dot{S}$ and
interpret Eq.~(\ref{tidal_heating}) as describing a flow of heat
across the horizon. By the first law of black-hole mechanics, $dM =
(\kappa/8\pi) dA$, the equation also describes the rate at which the
black-hole mass increases during the tidal interaction. 

\section{Background metric: Derivation} 
\label{sec:background_derivation} 

Our main results were presented in the preceding sections, and we now  
turn to a detailed derivation of these results. In this section we
provide a derivation of the background metric of
Eqs.~(\ref{background_metric_cart}) and
(\ref{background_metric_ang}); an alternative derivation
is sketched in Appendix~\ref{app:alternative}. 
In Sec.~\ref{sec:blackhole_derivation}
we present a derivation of the black-hole metric of
Eqs.~(\ref{blackhole_metric_cart}) and (\ref{blackhole_metric_ang}). 

\subsection{Kinematical properties of the metric} 
\label{subsec:kinematic} 

We construct the metric of a vacuum region of spacetime that surrounds
a timelike geodesic $\gamma$, and we adopt the light-cone coordinates  
$(v,r,\theta^A)$. These are centered on the world line, and the
geometrical meaning of the coordinates was specified in
Sec.~\ref{subsec:coordinates}. The three defining properties listed
there give rise to four important conditions on the metric.  

To spell them out we introduce the dual vector $\ell_\alpha :=
-\partial_\alpha v = [-1,0,0,0]$, which is normal to hypersurfaces of
constant advanced time $v$. According to the first property, this dual 
vector is null --- $g^{\alpha\beta} \ell_\alpha \ell_\beta = 0$ ---
and this condition immediately implies that $g^{vv} = 0$. The vector
$\ell^\alpha$ is tangent to the null generators of the light cones 
$v = \mbox{constant}$, and the second and third properties imply that
its components in the light-cone coordinates must be given by
$\ell^\alpha = [0,-1,0,0]$; that $\ell^v = \ell^A = 0$ means that $v$
and $\theta^A$ are constant on the generators, and $\ell^r = -1$ means
that $r$ is an affine parameter that decreases as the generators
converge toward the world line. The relation 
$\ell^\alpha = g^{\alpha\beta} \ell_\beta$ then implies that the
inverse metric must satisfy the conditions  
\begin{equation} 
g^{vv} = 0 = g^{vA}, \qquad g^{vr} = 1. 
\label{conditions_inverse_ang} 
\end{equation} 
Calculating the inverse, we find that the metric must satisfy 
\begin{equation} 
g_{rr} = 0 = g_{rA}, \qquad g_{vr} = 1. 
\label{consitions_ang} 
\end{equation} 
These statements are exact, and follow from the light-cone nature of
the coordinates. The other nonvanishing components of the metric are
$g_{vv}$, $g_{vA}$, and $g_{AB}$. We have not yet made a choice of
normalization for the radial coordinate $r$, nor a choice of axes for
the angular coordinates $\theta^A$. 

From the quasi-spherical coordinates $(r,\theta^A)$ we construct a
system of quasi-Cartesian coordinates $x^a = r \Omega^a(\theta^A)$ and 
transform the spatial components of the inverse metric according to
the rules described in Sec.~\ref{subsec:tidal_pot_angular}. We find
that $g^{va} = \Omega^a$, but the form of $g^{ab}$ is not
constrained by the conditions of
Eq.~(\ref{conditions_inverse_ang}). 

At this stage we impose the additional condition that the inverse
metric should be locally flat near the world line, so that
$g^{ab} = \delta^{ab} + h^{ab}$, with $h^{ab}$ going to zero (as
$r^2$) when $r \to 0$. This condition implicitly specifies a
normalization for $r$, which reduces to the usual Euclidean distance
in the immediate vicinity of the world line. And it specifies a set of
reference axes for the angular coordinates, which are aligned with the
Cartesian directions associated with the coordinates $x^a$; the fact
that $h^{ab}$ scales as $r^2$ (instead of $r$) when $r \to 0$ implies
that the world line is unaccelerated and that the axes are locally
nonrotating.    

Our expression for the inverse metric in quasi-Cartesian coordinates
is therefore 
\begin{subequations} 
\label{inverse_metric1_cart} 
\begin{align} 
g^{vv} &= 0, \\
g^{va} &= \Omega^a, \\ 
g^{ab} &= \delta^{ab} + h^{ab}, 
\end{align} 
\end{subequations} 
with 
\begin{equation} 
h^{ab} = r^2 h^{ab}_2 + r^3 h^{ab}_3 + r^4 h^{ab}_4 + O(r^5). 
\label{h_expansion} 
\end{equation} 
The last equation indicates that as $r \to 0$, $r^{-2} h^{ab}$
approaches the (direction-dependent) limit $h^{ab}_2$, $r^{-1}( r^{-2} 
h^{ab} - h^{ab}_2 )$ approaches $h^{ab}_3$, and so on.  

To proceed it is useful to introduce a decomposition of $h^{ab}$ into
longitudinal and transverse pieces. We write 
\begin{equation} 
h^{ab} = \Omega^a \Omega^b A + \Omega^a A^b + A^a \Omega^b + A^{ab} 
\label{h_decomposition} 
\end{equation} 
with 
\begin{equation} 
\Omega_a A^a = 0 = \Omega_a A^{ab}. 
\label{transverse_conditions} 
\end{equation} 
Here and below, indices on $\Omega^a$ are lowered with the Euclidean
metric $\delta_{ab}$. The first term in Eq.~(\ref{h_decomposition}) is
the longitudinal piece of $h^{ab}$, and $A := h^{ab} \Omega_a
\Omega_b$ is its component in the direction of the unit radial vector
$\Omega^a$. The last term is the transverse piece of $h^{ab}$, and 
$A^{ab} := \gamma^a_{\ c} \gamma^b_{\ d} h^{cd}$ are the components in
the directions orthogonal to $\Omega^a$; $\gamma^a_{\ c}$
is the projector of Eq.~(\ref{projector}). The second and third terms
are the longitudinal-transverse piece of $h^{ab}$, and $A^a :=
\gamma^a_{\ c} h^{cb} \Omega_b$ contains the relevant components. The
six independent components of $h^{ab}$ are contained in $A$ (one
component), $A^a$ (two independent components), and $A^{ab}$ (three
independent components). It is sometimes useful to further decompose
$A^{ab}$ into trace and tracefree pieces, but we choose not to do so
at this stage; we will find in due course that the Einstein field
equations automatically enforce $\delta_{ab} A^{ab} = 0$. The
decomposition of Eq.~(\ref{h_decomposition}) can be applied
individually to each $h^{ab}_n$ that appears in
Eq.~(\ref{h_expansion}); this defines $A_n$, $A^a_n$, and $A^{ab}_n$.  

We write the inverse metric of Eq.~(\ref{inverse_metric1_cart}) as
$g^{\alpha\beta} = \eta^{\alpha\beta} + h^{\alpha\beta}$, with
$\eta^{\alpha\beta}$ denoting the inverse of the Minkowski metric in
light-cone coordinates (with components $\eta^{vv} = 0$, 
$\eta^{va} = \Omega^a$, and $\eta^{ab} = \delta^{ab}$) and 
$h^{\alpha\beta} = O(r^2)$ denoting a perturbation (with components 
$h^{vv} = 0$, $h^{va} = 0$, and $h^{ab}$). The metric is then 
$g_{\alpha\beta} = \eta_{\alpha\beta} - h_{\alpha\beta} 
+ h_{\alpha\mu} h^\mu_{\ \beta} + O(r^5)$, where all indices are
lowered with the Minkowski metric $\eta_{\alpha\beta}$ (with
components $\eta_{vv} = -1$, $\eta_{va} = \Omega_a$, and 
$\eta_{ab} = \gamma_{ab}$). A straightforward calculation using
Eq.~(\ref{h_decomposition})  reveals that 
\begin{subequations} 
\label{metric1_cart} 
\begin{align} 
g_{vv} &= -1 - A + A_a A^a + O(r^5), \\ 
g_{va} &= \Omega_a - A_a + A_{ab} A^b + O(r^5), \\ 
g_{ab} &= \gamma_{ab} - A_{ab} + A_{ac} A^c_{\ b} + O(r^5). 
\end{align} 
\end{subequations} 
In these expressions it is understood that indices on $A^a$ and
$A^{ab}$ are lowered with $\delta_{ab}$; by virtue of
Eqs.~(\ref{transverse_conditions}) this operation is equivalent to
lowering indices with $\eta_{ab} = \gamma_{ab}$. It is also understood 
that $A$, $A^a$, and $A^{ab}$ can be expanded in powers of $r$ as in
Eq.~(\ref{h_expansion}); we have 
\begin{subequations} 
\begin{align} 
A &= r^2 A_2 + r^3 A_3 + r^4 A_4 + O(r^5), \\
A_a &= r^2 A_{2a} + r^3 A_{3a} + r^4 A_{4a} + O(r^5), \\  
A_{ab} &= r^2 A_{2ab} + r^3 A_{3ab} + r^4 A_{4ab} + O(r^5),  
\end{align} 
\end{subequations}
and a term like $A_a A^a$ reduces to $r^4 A_{2a} A^a_2 + O(r^5)$.   

We now return to the quasi-spherical coordinates $(r,\theta^A)$. We
define the angular version of the vector potential $A_a$ by 
\begin{equation} 
A_A := A_a \Omega^a_A, 
\end{equation} 
where $\Omega^a_A := \partial_A \Omega^a$, and we define the angular
version of the tensor potential $A_{ab}$ by 
\begin{equation} 
A_{AB} := A_{ab} \Omega^a_A \Omega^b_B. 
\end{equation} 
In spite of the suggestive notation, these are {\it not} the
components of the Cartesian tensors $A_a$ and $A_{ab}$ in
spherical coordinates; for these we have $A_a \partial x^a/\partial r
= A_a \Omega^a = 0$ and $A_a \partial x^a/\partial \theta^A = r A_a 
\Omega^a_A = r A_A$, with similar equations holding for $A_{ab}$. 

Using the identities of Eqs.~(\ref{ang_identities1}) we find that the
spherical-coordinate form of the metric is 
\begin{subequations} 
\label{metric1_ang} 
\begin{align} 
g_{vv} &= -1 - A + A_A A^A + O(r^5), \\ 
g_{vr} &= 1, \\ 
g_{vA} &= -r A_A + r A_{AB} A^B + O(r^6), \\ 
g_{AB} &= r^2 \Omega_{AB} - r^2 A_{AB} + r^2 A_{AC} A^C_{\ B} 
+ O(r^7). 
\end{align} 
\end{subequations} 
It is understood that in these expressions, indices on $A_A$ and
$A_{AB}$ are raised with $\Omega^{AB}$. As we saw previously, each
potential can be expanded in powers of $r$, so that 
\begin{subequations} 
\begin{align} 
A &= r^2 A_2 + r^3 A_3 + r^4 A_4 + O(r^5), \\
A_A &= r^2 A_{2A} + r^3 A_{3A} + r^4 A_{4A} + O(r^5), \\  
A_{AB} &= r^2 A_{2AB} + r^3 A_{3AB} + r^4 A_{4AB} + O(r^5). 
\end{align} 
\end{subequations}
The coefficients $A_{n}$, $A_{nA}$, and $A_{nAB}$ are assumed to
depend on $\theta^A$ only, so that the dependence of the metric on $r$
is contained explicitly in the power expansion. 

\subsection{Field equations I} 
\label{subsec:field_I} 

The metric forms of Eqs.~(\ref{metric1_cart}) and (\ref{metric1_ang})
follow directly from the light-cone nature of the coordinate systems, 
and they embody the purely kinematical requirements imposed by the
choice of coordinates. To obtain more information we must impose the
Einstein field equations. 

In this first stage we involve the metric of Eqs.~(\ref{metric1_ang}) 
in a computation of $R^{vv}$, the time-time component of the Ricci
tensor. Setting this to zero order-by-order in $r$ reveals that 
\begin{subequations}
\begin{align} 
& \Omega^{AB} A_{2AB} = 0, \\ 
& \Omega^{AB} A_{3AB} = 0, \\ 
& \Omega^{AB} A_{4AB} = \frac{3}{5} A_{2AB} A_2^{AB}. 
\end{align} 
\end{subequations} 
In quasi-Cartesian coordinates these equations read 
\begin{subequations}
\begin{align} 
& \gamma^{ab} A_{2ab} = 0, \\ 
& \gamma^{ab} A_{3ab} = 0, \\ 
& \gamma^{ab} A_{4ab} = \frac{3}{5}A_{2ab} A_2^{ab}. 
\end{align} 
\end{subequations} 
These equations mean that $A_{ab}$ is tracefree through order $r^3$, 
and that its trace is given by $\gamma^{ab} A_{ab} = \frac{3}{5} r^4 
A_{2ab} A_2^{ab} + O(r^5)$. 

We use this observation to simplify the form of the metric. Continuing
to work in the quasi-Cartesian coordinates $x^a$, we re-express
$A_{4ab}$ as a sum of trace and tracefree pieces:  
\begin{equation} 
A_{4ab} \to A_{4ab} + \frac{3}{10} \gamma_{ab}A_{2cd} A^{cd}_2, 
\end{equation} 
where the new $A_{4ab}$, like $A_{2ab}$ and $A_{3ab}$, is now known to
be tracefree. We also exploit the identity 
\begin{equation} 
A_{2ac} A^c_{2b} = \frac{1}{2} \gamma_{ab} A_{2cd} A^{cd}_2, 
\end{equation} 
which holds for any symmetric-tracefree tensor $A_{2ab}$. 

Making these substitutions in Eq.~(\ref{metric1_ang}), we arrive at 
\begin{subequations} 
\label{metric2_cart} 
\begin{align} 
g_{vv} &= -1 - r^2 A_2 - r^3 A_3 - r^4 A_4 + r^4 A_{2a} A^a_2 
\nonumber \\ & \quad \mbox{} 
+ O(r^5), \\ 
g_{va} &= \Omega_a - r^2 A_{2a} - r^3 A_{3a} - r^4 A_{4a} 
+ r^4 A_{2ab} A^b_2 
\nonumber \\ & \quad \mbox{} 
+ O(r^5), \\ 
g_{ab} &= \gamma_{ab} - r^2 A_{2ab} - r^3 A_{3ab} - r^4 A_{4ab} 
\nonumber \\ & \quad \mbox{} 
+ \frac{1}{5} r^4 \gamma_{ab} A_{2cd} A^{cd}_2 + O(r^5). 
\end{align} 
\end{subequations}  
It is understood that the vector potentials $A_{2a}$, $A_{3a}$,
$A_{4a}$, and $A_{2ab} A^b_2$ are transverse, in the sense that
they are all orthogonal to $\Omega^a$. And it is now understood that
the tensor potentials $A_{2ab}$, $A_{3ab}$, and $A_{4ab}$ are 
{\it both} transverse {\it and} tracefree; the potential $\gamma_{ab}
A_{2cd} A^{cd}_2$ is transverse and pure trace. 

In quasi-spherical coordinates the metric is 
\begin{subequations} 
\label{metric2_ang} 
\begin{align} 
g_{vv} &= -1 - r^2 A_2 - r^3 A_3 - r^4 A_4 + r^4 A_{2A} A^A_2 
\nonumber \\ & \quad \mbox{} 
+ O(r^5), \\ 
g_{vr} &=1, \\ 
g_{vA} &= - r^3 A_{2A} - r^4 A_{3A} - r^5 A_{4A} 
+ r^5 A_{2AB} A^B_2 
\nonumber \\ & \quad \mbox{} 
+ O(r^6), \\ 
g_{AB} &= r^2 \Omega_{AB} - r^4 A_{2AB} - r^5 A_{3AB} - r^6 A_{4AB}  
\nonumber \\ & \quad \mbox{} 
+ \frac{1}{5} r^6 \Omega_{AB} A_{2CD} A^{CD}_2 + O(r^7).  
\end{align} 
\end{subequations}  
Here also the tensor potentials are tracefree, except for the term
proportional to $\Omega_{AB}$, which is pure trace.  

\subsection{Field equations II} 
\label{subsec:field_II} 

The metric of Eqs.~(\ref{metric2_cart}) and (\ref{metric2_ang}) is a
partial solution to the Einstein field equations. In its
quasi-Cartesian form the metric involves the transverse potentials
$A_{2a}$,  $A_{3a}$,  $A_{4a}$,  $A_{2ab}$, $A_{3ab}$, and $A_{4ab}$,
and imposing the vacuum equation $R^{vv}=0$ has revealed the important 
fact that the tensor potentials are all tracefree. In the quasi-spherical
form of the metric, the potentials have components in the angular
directions only, and these depend on $\theta^A$ only. To obtain a
complete solution to the field equations we must now determine the
potentials.  

We rely on Zhang's work \cite{zhang:86}, which allows us to state that
{\it the metric of a vacuum region of spacetime around a timelike
geodesic is a functional of two (and only two) sets of tidal moments
$\E_{a_1 a_2 \cdots a_l}$ and $\B_{a_1 a_2 \cdots a_l}$; the tidal
moments are STF tensors that depend on proper time on the world line,
and they are related to components of the Weyl tensor (and its
derivatives) evaluated on the world line.} For a complete description
of the metric one requires an infinite number tidal moments; for an
approximate description one requires a finite number. In our case the
construction of the metric shall involve the quadrupole moments
$\E_{ab}$ and $\B_{ab}$, the octupole moments $\E_{abc}$ and
$\B_{abc}$, and the hexadecapole moments $\E_{abcd}$ and
$\B_{abcd}$. These were introduced in Sec.~\ref{subsec:tidal_def}, and
their scaling properties are described in
Sec.~\ref{subsec:tidal_scales}.  

Collecting the observations summarized in the preceding two
paragraphs, we obtain three important guiding rules for the
construction of the metric: 
\begin{enumerate} 
\item The metric is constructed from scalar potentials, vector
  potentials that are transverse, and tensor potentials that are
  transverse and tracefree. 
\item The potentials depend on the angles $\theta^A$ only; they are
  independent of $r$, which appears as a multiplicative factor in
  front of the potentials. 
\item The potentials depend on two sets of tidal moments. 
\end{enumerate}  
The rules imply that the metric must be constructed from the tidal
potentials introduced in Secs.~\ref{subsec:tidal_pot_cartesian} and
\ref{subsec:tidal_pot_angular}. There are no other possible building
blocks for the metric. 

We begin the construction of the metric with the determination of
$A_2$, $A_{2A}$, and $A_{2AB}$. (In practical matters it is convenient
to deal with the quasi-spherical representation of the metric, because
the $r$-dependence is then explicitly known.) These terms occur at
order $r^2$ in $g_{vv}$, $r^3$ in $g_{vA}$, and $r^4$ in $g_{AB}$, and
proper dimensionality requires that the potentials scale as
$\R^{-2}$. Equation (\ref{sloppyscale}) then implies that the
potentials must be constructed from $\E_{ab}$ and
$\B_{ab}$. The possible building blocks are listed in
Tables~\ref{tab:E_ang} and \ref{tab:B_ang}, and we write $A_2 = a
\EE{q}$, $A_{2A} = b \EE{q}_A + p \BB{q}_A$, $A_{2AB} = c \EE{q}_{AB}
+ q \BB{q}_{AB}$, where $a$, $b$, $c$, $p$, and $q$ are undetermined
numerical coefficients. From this we form the metric $g_{vv} = -1 -
r^2 A_2 + O(r^3)$, $g_{vr} = 1$, $g_{vA} = -r^3 A_{2A} + O(r^4)$, and
$g_{AB} = r^2 \Omega_{AB} - r^4 A_{2AB} + O(r^5)$, which we substitute
into the vacuum field equations. (At this stage of the computations
the $v$-dependence of the tidal potentials can be ignored, because
their $v$-derivatives are suppressed by a factor of order $\R^{-1}$
relative to the spatial derivatives. Another simplifying move is 
momentarily to switch off the $\phi$-dependence of the metric by
adopting $\EE{q}_0$ and $\BB{q}_0$ as the only nonvanishing components
of the tidal moments.) The exercise returns the relations
$b=\frac{2}{3} a$, $c = \frac{1}{3} a$, and $q = \frac{1}{2} p$, which
leaves $a$ and $p$ as undetermined constants. To obtain $a$ and $p$ we
compute the frame components $C_{a0b0}$ and $C_{abc0}$ of the Weyl
tensor in the limit $r\to 0$, and demand that the results agree with
Eqs.~(\ref{tidalmoment_2}). From this we find that $a = 1$ and $p =
-\frac{2}{3}$, and the metric is known through order $\R^{-2}$.   

For the computation of the frame components of the Weyl tensor we need  
\begin{subequations} 
\label{tetrad_components} 
\begin{align} 
& u^v = 1, \quad u^r = 0, \quad u^A = 0, \\ 
& e^v_a = \Omega_a, \quad e^r_a = \Omega_a, \quad 
e^A_a = \frac{1}{r} \Omega^A_a, 
\end{align}
\end{subequations} 
the components of the tetrad vectors in the light-cone
coordinates. Here $\Omega^A_a := \delta_{ab} \Omega^{AB} \Omega^b_B$,  
and the factor of $r^{-1}$ compensates for factors of $r$ in the
angular components of the metric. It is easy to show that the vectors
are orthonormal and parallel-transported along $u^\alpha$ in the limit
$r \to 0$. We observe that while some components of the Weyl tensor
are ambiguous or go to zero in the limit $r \to 0$, the frame
components are well-behaved and have a well-defined limit. 

We continue with the determination of $A_3$, $A_{3A}$, and
$A_{3AB}$. Here the potentials must scale as $\R^{-3}$, and
Eq.~(\ref{sloppyscale}) implies that they must be constructed from
$\dot{\E}_{ab}$, $\dot{\B}_{ab}$, $\E_{abc}$, and $\B_{abc}$. The
possible building blocks are listed in Tables~\ref{tab:E_ang} and
\ref{tab:B_ang}, and we express $A_3$ as a linear combination of 
$\EEd{q}$ and $\EE{o}$, $A_{3A}$ as a linear combination of
$\EEd{q}_A$, $\BBd{q}_A$, $\EE{o}_A$, and $\BB{o}_A$, and $A_{3AB}$ as 
a linear combination of $\EEd{q}_{AB}$, $\BBd{q}_{AB}$, $\EE{o}_{AB}$,
and $\BB{o}_{AB}$. The numerical coefficients are
determined in two steps. First, we construct an improved metric by
appending the terms of order $\R^{-3}$, and we substitute it into the  
vacuum field equations. (In this step the $v$-dependence of the
quadrupole tidal potentials must be taken into account, again keeping
in mind that the $v$-derivatives are suppressed by a factor of order
$\R^{-1}$ relative to the spatial derivatives. It is still possible
momentarily to switch off the $\phi$-dependence of the metric by
adopting $\EE{q}_0$, $\BB{q}_0$, $\EE{o}_0$, and $\BB{o}_0$ as the
only nonvanishing components of the tidal moments.) The first step
leaves two coefficients undetermined, a common multiplicative factor 
in front of $\E_{abc}$, and another common factor in front of
$\B_{abc}$. These are determined in the second step, in which we
compute the frame components $C_{a0b0|c}$ and $C_{abc0|d}$ of the
covariant derivatives of the Weyl tensor in the limit $r\to 0$, and
demand that the results agree with Eqs.~(\ref{tidalmoment_3}). The
metric is now known through order $\R^{-3}$. 

We complete the computation of the metric with the determination of
$A_4$, $A_{4A}$, and $A_{4AB}$. Here the potentials must scale as
$\R^{-4}$, and Eq.~(\ref{sloppyscale}) implies that they must be
constructed from $\ddot{\E}_{ab}$, $\ddot{\B}_{ab}$, $\dot{\E}_{abc}$,
$\dot{\B}_{abc}$, $\E_{abcd}$, and $\B_{abcd}$. That is not all,
however, because at order $\R^{-4}$ we must also include potentials
that are generated by quadratic combinations of $\E_{ab}$ and
$\B_{ab}$; the list of potentials is long, and this makes the
computations much more involved than in the previous cases. The 
possible building blocks are listed in Tables~\ref{tab:E_ang},
\ref{tab:B_ang}, \ref{tab:EE_ang}, \ref{tab:BB_ang},
\ref{tab:EBeven_ang}, and \ref{tab:EBodd_ang}. We express $A_4$ as
a linear combination of $\EEdd{q}$, $\EEd{o}$, $\EE{h}$, $\PP{m}$,
$\PP{q}$, $\PP{h}$, $\QQ{m}$, $\QQ{q}$, $\QQ{h}$, $\GG{d}$, and
$\GG{o}$. We express $A_{4A}$ as a linear combination of 
$\EEdd{q}_A$, $\EEd{o}_A$, $\EE{h}_A$, $\BBdd{q}_A$, $\BBd{o}_A$,
$\BB{h}_A$, $\PP{q}_A$, $\PP{h}_A$, $\QQ{q}_A$, $\QQ{h}_A$,
$\GG{d}_A$, $\GG{o}_A$, $\HH{q}_A$, and $\HH{h}_A$. And we express
$A_{4AB}$ as a linear combination of $\EEdd{q}_{AB}$, $\EEd{o}_{AB}$,
$\EE{h}_{AB}$, $\BBdd{q}_{AB}$,  $\BBd{o}_{AB}$, $\BB{h}_{AB}$,
$\PP{q}_{AB}$, $\PP{h}_{AB}$, $\QQ{q}_{AB}$, $\QQ{h}_{AB}$,
$\GG{d}_{AB}$, $\GG{o}_{AB}$, $\HH{q}_{AB}$, and $\HH{h}_{AB}$. In
addition to all this the angular components of the metric contain a
bilinear term proportional to $A_{2AB} A^{AB}_2 = \frac{1}{9} 
(\EE{q}_{AB} - \BB{q}_{AB})(\EEup{q}{AB} - \BBup{q}{AB})$, which can
be simplified with the help of Eqs.~(\ref{square_identities}). The many
numerical coefficients that appear in the metric at order $\R^{-4}$
are determined as we did previously. Substitution of the metric in the
vacuum field equations determines all but two coefficients, the common 
multiplicative factors in front of $\E_{abcd}$ and $\B_{abcd}$. These
are determined by computing the frame components $C_{a0b0|cd}$ and
$C_{abc0|de}$ of the second covariant derivatives of the Weyl tensor
in the limit $r\to 0$, and demanding that the results agree with
Eqs.~(\ref{tidalmoment_4}). The metric is now known through order
$\R^{-4}$, and the end result is displayed in
Eqs.~(\ref{background_metric_ang}). The quasi-Cartesian representation
of Eqs.~(\ref{background_metric_cart}) can be obtained directly from 
this.   

This completes the derivation of the background metric. An alternative
derivation is presented in Appendix~\ref{app:alternative}. 

\section{Black-hole metric: Derivation} 
\label{sec:blackhole_derivation} 

We next turn to a derivation of the black-hole metric of
Eqs.~(\ref{blackhole_metric_ang}), a vacuum perturbation of
the Schwarzschild solution 
\begin{equation} 
ds^2 = -f\, dv^2 + 2\, dvdr + r^2\, d\Omega^2, 
\label{Schwarzschild} 
\end{equation} 
where $f = 1-2M/r$ and $d\Omega^2 = \Omega_{AB} d\theta^A d\theta^B = 
d\theta^2 + \sin^2\theta\, d\phi^2$. To construct the perturbation we
rely heavily on the formalism of Martel and Poisson
\cite{martel-poisson:05}, and we implement the light-cone gauge of
Preston and Poisson \cite{preston-poisson:06b}. The computations that 
lead to the black-hole metric are extremely lengthy, and are presented
in four highly technical subsections. In Secs.~\ref{subsec:lin_prep}
and \ref{subsec:lin_field} we construct the linear piece of the metric 
perturbation, and in Secs.~\ref{subsec:nonlin_prep} and
\ref{subsec:nonlin_field} we turn to the bilinear piece. The results
are collected in Sec.~\ref{subsec:conclusion}.   

\subsection{Linear perturbation: Preparation} 
\label{subsec:lin_prep} 

\subsubsection{Form of the metric perturbation, 
and considerations of gauge} 

The metric of Eqs.~(\ref{background_metric_ang}) specifies the
asymptotic conditions (when $r \gg 2M$) for the metric of a tidally
deformed black hole. Focusing our attention on the linearly perturbed
piece of the metric, we write it as 
\begin{widetext} 
\begin{subequations} 
\label{metric_linear} 
\begin{align} 
g_{vv} &= -f - r^2 \ee{q}{1} \EE{q} 
+ \frac{1}{3} r^3 \ee{q}{2} \EEd{q}
- \frac{2}{21} r^4 \ee{q}{3} \EEdd{q} 
- \frac{1}{3} r^3 \ee{o}{1} \EE{o} 
+ \frac{1}{6} r^4 \ee{o}{2} \EEd{o} 
- \frac{1}{12} r^4 \ee{h}{1} \EE{h}
+ \mbox{bilinear}, \\ 
g_{vr} &= 1, \\ 
g_{vA} &=
-\frac{2}{3} r^3 \bigl( \ee{q}{4} \EE{q}_A 
  - \bb{q}{4} \BB{q}_A \bigr)  
+ \frac{1}{3} r^4 \bigl( \ee{q}{5} \EEd{q}_A 
  - \bb{q}{5} \BBd{q}_A \bigr)  
- \frac{8}{63} r^5 \bigl( \ee{q}{6} \EEdd{q}_A 
  - \bb{q}{6} \BBdd{q}_A \bigr) 
- \frac{1}{4} r^4 \bigl( \ee{o}{4} \EE{o}_A 
  - \bb{o}{4} \BB{o}_A \bigr) 
\nonumber \\ & \quad \mbox{} 
+ \frac{1}{6} r^5 \bigl( \ee{o}{5} \EEd{o}_A 
  - \bb{o}{5} \BBd{o}_A \bigr) 
- \frac{1}{15} r^5 \bigl( \ee{h}{4} \EE{h}_A 
  - \bb{h}{4} \BB{h}_A \bigr)
+ \mbox{bilinear}, \\ 
g_{AB} &= r^2 \Omega_{AB} 
- \frac{1}{3} r^4 \bigl( \ee{q}{7} \EE{q}_{AB} 
  - \bb{q}{7} \BB{q}_{AB} \bigr) 
+ \frac{5}{18} r^5 \bigl( \ee{q}{8} \EEd{q}_{AB} 
  - \bb{q}{8} \BBd{q}_{AB} \bigr) 
- \frac{1}{7} r^6 \bigl( \ee{q}{9} \EEdd{q}_{AB} 
  - \bb{q}{9} \BBdd{q}_{AB} \bigr) 
\nonumber \\ & \quad \mbox{} 
- \frac{1}{6} r^5 \bigl( \ee{o}{7} \EE{o}_{AB} 
  - \bb{o}{7} \BB{o}_{AB} \bigr) 
+ \frac{3}{20} r^6 \bigl( \ee{o}{8} \EEd{o}_{AB} 
  - \bb{o}{8} \BBd{o}_{AB} \bigr) 
- \frac{1}{20} r^6 \bigl( \ee{h}{7} \EE{h}_{AB} 
  - \bb{h}{7} \BB{h}_{AB} \bigr)
+ \mbox{bilinear}, 
\end{align} 
\end{subequations} 
\end{widetext} 
where the undetermined radial functions $\ee{q}{n}(r)$,
$\ee{o}{n}(r)$, $\ee{h}{n}(r)$, $\bb{q}{n}(r)$, $\bb{o}{n}(r)$, and
$\bb{h}{n}(r)$ are all required to approach unity when $2M/r \to
0$. This ansatz for the metric is motivated by the facts that (i) it 
reduces to the Schwarzschild metric when the tidal fields are turned
off; (ii) it reduces to the (linear piece of the) background metric
when $M \to 0$ (so that the radial functions all become equal to
unity); and (iii) its expansion in terms of tidal potentials
constitutes a decomposition of the metric perturbation into a complete
basis of spherical-harmonic modes. An important aspect of the metric
is that its $v$-dependence is assumed to be slow (with a time scale of
the order of $\R$) and contained in the tidal moments; we rule out
time-dependent processes that take place over time scales comparable
to $2M$.     

We write the black-hole metric as $g_{\alpha\beta} =
\hat{g}_{\alpha\beta} + p_{\alpha\beta}$, in which
$\hat{g}_{\alpha\beta}$ is the Schwarzschild metric of
Eq.~(\ref{Schwarzschild}) and $p_{\alpha\beta}$ is the tidal
perturbation. In Eq.~(\ref{metric_linear}) the perturbation is
presented in the Preston-Poisson light-cone gauge, which enforces the
conditions $p_{vr} = p_{rr} = p_{rA} = 0$. The even-parity sector of
the perturbation is  
\begin{subequations} 
\label{pert_even} 
\begin{align} 
p_{vv} &= \sum_{lm} h^{lm}_{vv} Y^{lm}, \\ 
p_{vA} &= \sum_{lm} j^{lm}_v Y^{lm}_A, \\ 
p_{AB} &= r^2 \sum_{lm} \bigl( K^{lm} \Omega_{AB} Y^{lm} 
+ G^{lm} Y^{lm}_{AB} \bigr), 
\end{align}
\end{subequations}  
where the reduced perturbations $h^{lm}_{vv}$, $j^{lm}_v$, $K^{lm}$,
and $G^{lm}$ depend on $v$ and $r$ only. Preston and Poisson show that
for vacuum perturbations, the gauge can be refined to also enforce
$K^{lm} = 0$. This leaves $h^{lm}_{vv}$, $j^{lm}_v$, and $G^{lm}$ as
nonvanishing perturbations, and these can be read off from
Eqs.~(\ref{metric_linear}). The odd-parity sector of the perturbation
is  
\begin{subequations}
\label{pert_odd} 
\begin{align}  
p_{vA} &= \sum_{lm} h^{lm}_v X^{lm}_A, \\
p_{AB} &= \sum_{lm} h^{lm}_2 X^{lm}_{AB}, 
\end{align} 
\end{subequations}
in which $h^{lm}_v$ and $h^{lm}_2$ depend on $v$ and $r$ only; these
also can be read off from Eqs.~(\ref{metric_linear}).  

Preston and Poisson show that the residual gauge freedom that remains
within the even-parity sector of the perturbation is a family
characterized by an arbitrary function $a^{lm}(v)$ --- one function
for each $l$ and $m$. Under such a gauge transformation the
perturbations change according to 
\begin{subequations}
\begin{align}  
h_{vv} &\to h_{vv} + l(l+1) \frac{M}{r^2} a 
- 2 \biggl[ {\textstyle \frac{1}{2}} l(l+1) 
  - 1 + \frac{3M}{r} \biggr] \dot{a} 
\nonumber \\ & \qquad \mbox{}
+ 2 r \ddot{a}, \\
j_v &\to j_v - \bigl[ {\textstyle \frac{1}{2}} l(l+1) 
   - f \bigr] a + \frac{2r^2}{l(l+1)} \ddot{a}, \\ 
G &\to G - \frac{2}{r} a + \frac{4}{l(l+1)} \dot{a}, 
\end{align} 
\end{subequations} 
where the $lm$ label was omitted for ease of notation. In addition,
Preston and Poisson show that the residual gauge freedom that remains
within the odd-parity sector of the perturbation is a family
characterized by an arbitrary function $\alpha^{lm}(v)$ --- one 
function for each $l$ and $m$. Under such a gauge transformation the
perturbations change according to   
\begin{subequations}
\begin{align}  
h_v &\to h_v - r^2 \dot{\alpha}, \\ 
h_2 &\to h_2 - 2r^2 \alpha. 
\end{align}
\end{subequations} 

We shall use the residual gauge freedom to specialize the light-cone
gauge to a {\it horizon-locking gauge} defined by the requirements
$p_{vv} = p_{vr} = p_{vA} = 0$ at $r=2M$. Since the light-cone gauge
already enforces $p_{vr} = 0$ everywhere, the horizon-locking
gauge requires  
\begin{equation} 
h^{lm}_{vv} = j^{lm}_{v} = 0 = h^{lm}_v \qquad 
\mbox{at $r=2M$}. 
\label{horizon_lock_gauge1} 
\end{equation} 
Because the residual gauge freedom is limited to two functions
$a^{lm}(v)$ and $\alpha^{lm}(v)$, it may seem doubtful that the three
conditions of Eq.~(\ref{horizon_lock_gauge1}) can be
imposed. Nevertheless, we shall see that the Einstein field equations
do allow the specialization of the light-cone gauge to the
horizon-locking gauge. We saw in Sec.~\ref{subsec:horizon_position}
that the conditions of Eq.~(\ref{horizon_lock_gauge1}) ensure that the 
horizon keeps its coordinate description $r=2M$ in the perturbed 
spacetime. 

The horizon-locking gauge, together with the vacuum field equations, 
imply the existence of a useful constraint on the values of $h^{lm}_v$
and $h^{lm}_2$ at $r=2M$. We examine the perturbation equation 
$P_{lm}^r = 0$ (in the notation of Martel and Poisson) and evaluate it 
at $r=2M$. Making use of the statement $h^{lm}_v = 0$ returns 
\begin{equation} 
0 = -\frac{(l-1)(l+2)}{8M^2} \frac{\partial h_2}{\partial v} 
- \frac{\partial}{\partial v} \frac{\partial h_v}{\partial r}. 
\end{equation} 
Integration with respect to $v$, keeping $r$ anchored at $r=2M$, 
yields
\begin{equation} 
h^{lm}_2(v,r=2M) = -\frac{8M^2}{(l-1)(l+2)} 
\frac{\partial h^{lm}_v}{\partial r} \biggr|_{r=2M}. 
\label{horizon_lock_gauge2} 
\end{equation} 
The constant of integration was set equal to zero to respect the
ansatz of Eqs.~(\ref{metric_linear}), which implies that there is no
$v$-independent term in the perturbation. We shall refer to
Eq.~(\ref{horizon_lock_gauge2}) as the {\it horizon-locking
constraint}; it refers to the odd-parity sector of the perturbation
only.  

\subsubsection{Redefinition of tidal moments} 

In addition to making use of the residual gauge freedom, the form of
the metric perturbation can be adjusted by redefining the tidal
moments $\E_{ab}$, $\E_{abc}$, $\B_{ab}$, and $\B_{abc}$ according to 
$\E \to \E + p_1 M \dot{\E} + p_2 M^2 \ddot{\E} + \cdots$
and $\B \to \B + q_1 M \dot{\B} + q_2 M^2 \ddot{\B} + \cdots$. As we
shall see below, these redefinitions have the effect of inducing
changes in the radial functions $\ee{q}{n}$, $\ee{o}{n}$, $\bb{q}{n}$, 
and $\bb{o}{n}$. 

\subsubsection{Even-parity sector; $l=2$} 

According to Eqs.~(\ref{metric_linear}) and Table~\ref{tab:E_ang}, the
even-parity, $l=2$ piece of the perturbation is described by 
\begin{subequations}
\label{lin_even_l2} 
\begin{align}  
h_{vv} &= -r^2 e_1 \E + \frac{1}{3} r^3 e_2 \dot{\E} 
- \frac{2}{21} r^4 e_3 \ddot{\E} + \cdots, \\ 
j_{v} &= -\frac{1}{3} r^3 e_4 \E + \frac{1}{6} r^4 e_5 \dot{\E} 
- \frac{4}{63} r^5 e_6 \ddot{\E} + \cdots, \\ 
G &= -\frac{1}{3} r^2 e_7 \E + \frac{5}{18} r^3 e_8 \dot{\E} 
- \frac{1}{7} r^4 e_9 \ddot{\E} + \cdots, 
\end{align} 
\end{subequations} 
in which $e_n := \ee{q}{n}(r)$ and $\E := \EE{q}_m(v)$. For ease of
notation we have omitted the label $lm = 2m$ on the perturbation
functions; the radial functions do not depend on $m$. Under a residual
gauge transformation generated by the function $a(v) 
= -\frac{1}{6} c_1 M^3 \E - \frac{2}{9} c_2 M^4 \dot{\E} 
- \frac{1}{63} c_3 M^5 \ddot{\E} + \cdots$, the radial functions
change according to    
\begin{subequations}
\label{gauge_even_l2} 
\begin{align}  
e_1 &\to  e_1 + c_1 \frac{M^4}{r^4}, \\
e_2 &\to  e_2 + 2c_1 \frac{M^3}{r^3} + 3c_1 \frac{M^4}{r^4} 
- 4 c_2 \frac{M^5}{r^5}, \\
e_3 &\to  e_3 + \frac{7}{2} c_1 \frac{M^3}{r^3} 
- \frac{28}{3} c_2 \frac{M^4}{r^4} - 14 c_2 \frac{M^5}{r^5} 
+ c_3 \frac{M^6}{r^6}, \\ 
e_4 &\to  e_4 - c_1 \frac{M^3}{r^3} - c_1 \frac{M^4}{r^4}, \\ 
e_5 &\to  e_5 + \frac{8}{3} c_2 \frac{M^4}{r^4} 
+ \frac{8}{3} c_2 \frac{M^5}{r^5}, \\ 
e_6 &\to  e_6 + \frac{7}{8} c_1 \frac{M^3}{r^3} 
- \frac{1}{2} c_3 \frac{M^5}{r^5} - \frac{1}{2} c_3 \frac{M^6}{r^6}, \\ 
e_7 &\to  e_7 - c_1 \frac{M^3}{r^3}, \\ 
e_8 &\to  e_8 - \frac{2}{5} c_1 \frac{M^3}{r^3} 
+ \frac{8}{5} c_2 \frac{M^4}{r^4}, \\ 
e_9 &\to  e_9 + \frac{28}{27} c_2 \frac{M^4}{r^4}
- \frac{2}{9} c_3 \frac{M^5}{r^5}. 
\end{align} 
\end{subequations}
Here the residual gauge freedom was reduced from a functional
family characterized by an arbitrary function $a(v)$ to a
three-parameter family (with parameters $c_1$, $c_2$, and $c_3$). This
loss of generality is a choice that is motivated by the observation
that the even-parity, $l=2$ piece of the perturbed metric should be 
driven by a single function $\E(v)$, so that $a(v)$ should involve
only $\E(v)$ and its derivatives. Below we shall seek choices for
$c_1$, $c_2$, and $c_3$ that enforce the horizon-locking conditions
$e_1 = e_2 = e_3 = e_4 = e_5 = e_6 = 0$ at $r=2M$.   

Implementing the redefinition $\E \to \E - \frac{1}{3} p_1 M \dot{\E}
+ \frac{2}{21} p_2 M^2 \ddot{\E} + \cdots$ in
Eqs.~(\ref{lin_even_l2}) has the effect of changing the identity of
the radial functions $e_1 \cdots e_9$. They become 
\begin{subequations}
\label{redef_even_l2}
\begin{align}  
e_1 &\to  e_1, \\ 
e_2 &\to  e_2 + p_1 \frac{M}{r} e_1, \\ 
e_3 &\to  e_3 + \frac{7}{6} p_1 \frac{M}{r} e_2 
+ p_2 \frac{M^2}{r^2} e_1, \\
e_4 &\to  e_4, \\ 
e_5 &\to  e_5 + \frac{2}{3} p_1 \frac{M}{r} e_4, \\ 
e_6 &\to  e_6 + \frac{7}{8} p_1 \frac{M}{r} e_5 
+ \frac{1}{2} p_2 \frac{M^2}{r^2} e_4, \\
e_7 &\to  e_7, \\ 
e_8 &\to  e_8 + \frac{2}{5} p_1 \frac{M}{r} e_7, \\ 
e_9 &\to  e_9 + \frac{35}{54} p_1 \frac{M}{r} e_8 
+ \frac{2}{9} p_2 \frac{M^2}{r^2} e_7.  
\end{align} 
\end{subequations}
Below we shall seek choices for $p_1$ and $p_2$ that enforce $e_8 
= e_9 = 0$ at $r=2M$; this does not alter the conditions already
imposed by the horizon-locking gauge. With all these choices
implemented, only $e_7$ will be nonvanishing at $r=2M$. 

\subsubsection{Even-parity sector; $l=3$} 

According to Eqs.~(\ref{metric_linear}) and Table~\ref{tab:E_ang}, the
even-parity, $l=3$ piece of the perturbation is described by 
\begin{subequations}
\label{lin_even_l3} 
\begin{align}  
h_{vv} &= -\frac{1}{3} r^3 e_1 \E 
+ \frac{1}{6} r^4 e_2 \dot{\E} + \cdots, \\ 
j_{v} &= -\frac{1}{12} r^4 e_4 \E 
+ \frac{1}{18} r^5 e_5 \dot{\E} + \cdots, \\ 
G &= -\frac{1}{18} r^3 e_7 \E 
+ \frac{1}{20} r^4 e_8 \dot{\E} + \cdots, 
\end{align} 
\end{subequations}
in which $e_n := \ee{o}{n}(r)$ and $\E := \EE{o}_m(v)$. For ease of 
notation we have omitted the label $lm = 3m$ on the perturbation
functions; the radial functions do not depend on $m$. Under a residual
gauge transformation generated by the function $a(v) 
= -\frac{1}{36} c_1 M^4 \E - \frac{1}{72} c_2 M^5 \dot{\E} 
+ \cdots$, the radial functions change according to   
\begin{subequations}
\begin{align}  
e_1 &\to  e_1 + c_1 \frac{M^5}{r^5}, \\
e_2 &\to  e_2 + \frac{5}{3} c_1 \frac{M^4}{r^4} 
+ c_1 \frac{M^5}{r^5} - c_2 \frac{M^6}{r^6}, \\
e_4 &\to  e_4 - \frac{5}{3} c_1 \frac{M^4}{r^4} 
- \frac{2}{3} c_1 \frac{M^5}{r^5},  \\ 
e_5 &\to  e_5 + \frac{5}{4} c_2 \frac{M^5}{r^5} 
+ \frac{1}{2} c_2 \frac{M^6}{r^6}, \\ 
e_7 &\to  e_7 - c_1 \frac{M^4}{r^4}, \\ 
e_8 &\to  e_8 - \frac{5}{27} c_1 \frac{M^4}{r^4} 
+ \frac{5}{9} c_2 \frac{M^5}{r^5}.  
\end{align} 
\end{subequations}
Here (as before) the residual gauge freedom was reduced from a
functional family to a two-parameter family. We shall seek choices for
$c_1$ and $c_2$ that enforce the horizon-locking conditions $e_1
= e_2 = e_4 = e_5 = 0$ at $r=2M$.   

Implementing the redefinition $\E \to \E - \frac{1}{2} p_1 M \dot{\E}
+ \cdots$ in Eqs.~(\ref{lin_even_l3}) has once more the effect of
changing the identity of the radial functions. They become 
\begin{subequations}
\begin{align}  
e_1 &\to  e_1, \\ 
e_2 &\to  e_2 + p_1 \frac{M}{r} e_1, \\ 
e_4 &\to  e_4, \\ 
e_5 &\to  e_5 + \frac{3}{4} p_1 \frac{M}{r} e_4, \\ 
e_7 &\to  e_7, \\ 
e_8 &\to  e_8 + \frac{5}{9} p_1 \frac{M}{r} e_7.  
\end{align} 
\end{subequations}
Below we shall seek a choice for $p_1$ that enforces $e_8 = 0$ at
$r=2M$; this does not alter the conditions already imposed by the
horizon-locking gauge. With all these choices implemented, only $e_7$
will be nonvanishing at $r=2M$.  

\subsubsection{Even-parity sector; $l=4$} 

According to Eqs.~(\ref{metric_linear}) and Table~\ref{tab:E_ang}, the
even-parity, $l=4$ piece of the perturbation is described by 
\begin{subequations}
\label{lin_even_l4} 
\begin{align}  
h_{vv} &= -\frac{1}{12} r^4 e_1 \E + \cdots, \\ 
j_{v} &= -\frac{1}{60} r^5 e_4 \E + \cdots, \\ 
G &= -\frac{1}{120} r^4 e_7 \E + \cdots, 
\end{align} 
\end{subequations}
in which $e_n := \ee{h}{n}(r)$ and $\E := \EE{h}_m(v)$. For ease of
notation we have omitted the label $lm = 4m$ on the perturbation
functions; the radial functions do not depend on $m$. Under a residual
gauge transformation generated by the function $a(v) 
= -\frac{1}{240} c_1 M^5 \E + \cdots$, the radial functions change
according to    
\begin{subequations} 
\begin{align} 
e_1 &\to  e_1 + c_1 \frac{M^6}{r^6}, \\
e_4 &\to  e_4 - \frac{9}{4} c_1 \frac{M^5}{r^5} 
- \frac{1}{2} c_1 \frac{M^6}{r^6}, \\
e_7 &\to  e_7 - c_1 \frac{M^5}{r^5}.  
\end{align} 
\end{subequations} 
We shall seek a choice for $c_1$ that enforces the horizon-locking
conditions $e_1 = e_4 = 0$ at $r=2M$. This leaves $e_7$ as the only
nonvanishing function at $r=2M$. 

\subsubsection{Odd-parity sector; $l=2$} 

According to Eqs.~(\ref{metric_linear}) and Table~\ref{tab:B_ang}, the
odd-parity, $l=2$ piece of the perturbation is described by 
\begin{subequations}
\label{lin_odd_l2} 
\begin{align}  
h_v &= \frac{1}{3} r^3 b_4 \B - \frac{1}{6} r^4 b_5 \dot{\B} 
+ \frac{4}{63} r^5 b_6 \ddot{\B} + \cdots, \\ 
h_2 &= \frac{1}{3} r^4 b_7 \B - \frac{5}{18} r^5 b_8 \dot{\B} 
+ \frac{1}{7} r^6 b_9 \ddot{\B} + \cdots,
\end{align} 
\end{subequations}
in which $b_n := \bb{q}{n}(r)$ and $\B := \BB{q}_m(v)$. For ease of 
notation we have omitted the label $lm = 2m$ on the perturbation
functions; the radial functions do not depend on $m$. Under a residual
gauge transformation generated by the function 
$\alpha(v) = -\frac{1}{6} k_1 M^2 \B + \frac{5}{36} k_2 M^3
\dot{\B} - \frac{1}{14} k_3 M^4 \ddot{\B} + \cdots$, the radial
functions change according to   
\begin{subequations}
\label{gauge_odd_l2} 
\begin{align}  
b_4 &\to  b_4, \\
b_5 &\to  b_5 - k_1 \frac{M^2}{r^2}, \\
b_6 &\to  b_6 - \frac{35}{16} k_2 \frac{M^3}{r^3}, \\
b_7 &\to  b_7 + k_1 \frac{M^2}{r^2}, \\
b_8 &\to  b_8 + k_2 \frac{M^3}{r^3}, \\
b_9 &\to  b_9 + k_3 \frac{M^4}{r^4}. 
\end{align} 
\end{subequations}
Here the residual gauge freedom was reduced from a functional 
family characterized by an arbitrary function $\alpha(v)$ to a
three-parameter family (with parameters $k_1$, $k_2$, and $k_3$). This 
loss of generality is a choice that is motivated by the observation
that the odd-parity, $l=2$ piece of the perturbed metric should be 
driven by a single function $\B(v)$, so that $\alpha(v)$ should
involve only $\B(v)$ and its derivatives.  

Below we shall seek choices for $k_1$, $k_2$, and $k_3$ that enforce
the horizon-locking conditions $b_4 = b_5 = b_6 = 0$ at $r=2M$. Since 
$b_4$ is gauge invariant, this will be achieved if and only if $b_4$
automatically vanishes at $r=2M$. Using Eqs.~(\ref{lin_odd_l2}) we
find that the horizon-locking constraint of
Eq.~(\ref{horizon_lock_gauge2}) implies  
\begin{subequations}
\label{hor_lock_l2}
\begin{align}  
b_7(2M) &= -M b'_4(2M), \\
b_8(2M) &= -\frac{3}{5} M b'_5(2M), \\ 
b_9(2M) &= -\frac{4}{9} M b'_6(2M), 
\end{align} 
\end{subequations} 
in which a prime indicates differentiation with respect to $r$. The
last equation will be used to determine $k_3$.   

Implementing the redefinition $\B \to \B - \frac{1}{2} q_1 M \dot{\B} 
+ \frac{4}{21} q_2 M^2 \ddot{\B} + \cdots$ in Eqs.~(\ref{lin_odd_l2})
has the effect of changing the identity of the radial functions $b_4
\cdots b_9$. They become  
\begin{subequations}
\label{redef_odd_l2} 
\begin{align}  
b_4 &\to b_4, \\ 
b_5 &\to b_5 + q_1 \frac{M}{r} b_4, \\ 
b_6 &\to b_6 + \frac{21}{16} q_1 \frac{M}{r} b_5 
+ q_2 \frac{M^2}{r^2} b_4, \\
b_7 &\to b_7, \\ 
b_8 &\to b_8 + \frac{3}{5} q_1 \frac{M}{r} b_7, \\ 
b_9 &\to b_9 + \frac{35}{36} q_1 \frac{M}{r} b_8 
+ \frac{4}{9} q_2 \frac{M^2}{r^2} b_7.   
\end{align} 
\end{subequations}
Below we shall seek choices for $q_1$ and $q_2$ that enforce $b_8 
= b_9 = 0$ at $r=2M$; this does not alter the conditions already
imposed by the horizon-locking gauge. With all these choices
implemented, only $b_7$ will be nonvanishing at $r=2M$. 

\subsubsection{Odd-parity sector; $l=3$} 

According to Eqs.~(\ref{metric_linear}) and Table~\ref{tab:B_ang}, the
odd-parity, $l=3$ piece of the perturbation is described by 
\begin{subequations}
\label{lin_odd_l3} 
\begin{align}  
h_v &= \frac{1}{12} r^4 b_4 \B 
- \frac{1}{18} r^5 b_5 \dot{\B} + \cdots, \\ 
h_2 &= \frac{1}{18} r^5 b_7 \B 
- \frac{1}{20} r^6 b_8 \dot{\B} + \cdots, 
\end{align} 
\end{subequations}
in which $b_n := \bb{o}{n}(r)$ and $\B := \BB{o}_m(v)$. For ease of 
notation we have omitted the label $lm = 3m$ on the perturbation
functions; the radial functions do not depend on $m$. Under a residual
gauge transformation generated by the function $\alpha(v) 
= -\frac{1}{36} k_1 M^3 \B + \frac{1}{40} k_2 M^4 \dot{\B} 
+ \cdots$, the radial functions change according to   
\begin{subequations}
\begin{align}  
b_4 &\to b_4, \\
b_5 &\to b_5 - \frac{1}{2} k_1 \frac{M^3}{r^3}, \\
b_7 &\to b_7 + k_1 \frac{M^3}{r^3}, \\
b_8 &\to b_8 + k_2 \frac{M^4}{r^4}.   
\end{align} 
\end{subequations}
Here the residual gauge freedom was reduced to a two-parameter
family. 

Below we shall seek choices for $k_1$ and $k_2$ that enforce the
horizon-locking conditions $b_4 = b_5 = 0$ at $r=2M$. Since $b_4$ is
gauge invariant, this will be achieved if and only if $b_4$
automatically vanishes at $r=2M$. Using Eqs.~(\ref{lin_odd_l3})
we find that the horizon-locking constraint of
Eq.~(\ref{horizon_lock_gauge2}) implies   
\begin{subequations} 
\label{hor_lock_l3}
\begin{align}  
b_7(2M) &= -\frac{3}{5} M b'_4(2M), \\ 
b_8(2M) &= -\frac{4}{9} M b'_5(2M), 
\end{align} 
\end{subequations} 
in which a prime indicates differentiation with respect to $r$. The
last equation will be used to determine $k_2$.  

Implementing the redefinition $\B \to \B - \frac{2}{3} q_1 M \dot{\B} 
+ \cdots$ in Eqs.~(\ref{lin_odd_l3}) has the effect of changing
the identity of the radial functions. They become 
\begin{subequations}
\begin{align}  
b_4 &\to b_4, \\ 
b_5 &\to b_5 + q_1 \frac{M}{r} b_4, \\ 
b_7 &\to b_7, \\ 
b_8 &\to b_8 + \frac{20}{27} q_1 \frac{M}{r} b_7. 
\end{align} 
\end{subequations}
Below we shall seek a choice for $q_1$ that enforces $b_8 = 0$ at
$r=2M$; this does not alter the conditions already imposed by the
horizon-locking gauge. With all these choices implemented, only $b_7$
will be nonvanishing at $r=2M$.  

\subsubsection{Odd-parity sector; $l=4$} 

According to Eqs.~(\ref{metric_linear}) and Table~\ref{tab:B_ang}, the
odd-parity, $l=4$ piece of the perturbation is described by 
\begin{subequations}
\label{lin_odd_l4} 
\begin{align}  
h_v &= \frac{1}{60} r^5 b_4 \B + \cdots, \\ 
h_2 &= \frac{1}{120} r^6 b_7 \B + \cdots,
\end{align} 
\end{subequations}
in which $b_n := \bb{h}{n}(r)$ and $\B := \BB{h}_m(v)$. For ease of
notation we have omitted the label $lm = 4m$ on the perturbation 
functions; the radial functions do not depend on $m$. Under a residual
gauge transformation generated by the function $\alpha(v) 
= -\frac{1}{240} k_1 M^4 \B + \cdots$, the radial functions change
according to     
\begin{subequations}
\begin{align}  
b_4 &\to b_4, \\
b_7 &\to b_7 + k_1 \frac{M^4}{r^4}.  
\end{align} 
\end{subequations}
Here the residual gauge freedom was reduced to a one-parameter 
family. The horizon-locking condition is $b_4 = 0$ at $r=2M$, and this
will be achieved if and only if $b_4$ automatically vanishes at
$r=2M$. Using Eqs.~(\ref{lin_odd_l4}) we find that the
horizon-locking constraint of Eq.~(\ref{horizon_lock_gauge2}) implies   
\begin{equation} 
b_7(2M) = -\frac{4}{9} M b'_4(2M), 
\end{equation} 
in which a prime indicates differentiation with respect to $r$. This
equation will be used to determine $k_1$. 

\subsection{Linear perturbation: Field equations} 
\label{subsec:lin_field} 

\subsubsection{Even-parity sector}

We describe in detail the method by which the perturbation equations
are integrated in the quadrupole case $l=2$; the same strategy is
employed for the other multipoles. We rely on the formalism of
black-hole perturbation theory outlined in Martel and Poisson
\cite{martel-poisson:05}. The method unfolds in a number of steps.   

In the first step the forms $h_{vv} = -r^2 e_1(r) {\E}(\epsilon v)  
+ \cdots$, $j_v = -\frac{1}{3} r^3 e_4(r) {\E}(\epsilon v) 
+ \cdots$, and $G = -\frac{1}{3} r^2 e_7(r) {\E}(\epsilon v) 
+ \cdots$ are inserted within the even-parity field equations. Here 
$\epsilon$ is a book-keeping parameter that reminds us that
derivatives with respect to $v$ are considered to be small; in this
first step the field equations are expanded to order $\epsilon^0$, and
all terms in $\dot{\cal E}$ are neglected. In the notation of Martel
and Poisson, the equation $Q^{vv} = 0$ is automatically satisfied, and
the equations $Q^{vr} = 0$, $Q^{rr} = 0$, and $Q^v = 0$ give rise to a 
system of three independent differential equations for the three
unknown radial functions; the other field equations are related to
these by the Bianchi identities. The general solution to the system of 
equations is easily obtained, and it depends on three
integration constants. Imposing regularity at the event horizon
determines one constant, and removes all terms
proportional to $\log(r-2M)$ from the radial functions. One of the two
remaining integration constants is an overall multiplicative factor
that is chosen so that when $r \gg 2M$, the functions $e_1$, $e_4$,
and $e_7$ all approach unity. The remaining constant is equivalent to
the parameter $c_1$ that appears in Eqs.~(\ref{gauge_even_l2}); it 
characterizes the residual gauge freedom that is still contained within
the light-cone class of gauges. This last constant can be selected by
simultaneously enforcing $e_1 = e_4= 0$ at $r=2M$; this is the
horizon-locking condition of Eq.~(\ref{horizon_lock_gauge1}). The
functions $e_1$, $e_4$, and $e_7$ are now completely determined, and
the constant $c_1$ has been chosen.  

In the second step the forms 
\begin{subequations} 
\begin{align} 
h_{vv} &= -r^2 e_1(r) \E(\epsilon v)   
+ \frac{1}{3} r^3 e_2(r) \dot{\E}(\epsilon v) + \cdots, \\ 
j_v &= -\frac{1}{3} r^3 e_4(r) {\E}(\epsilon v) 
+ \frac{1}{6} r^4 e_5(r) \dot{\E}(\epsilon v) + \cdots, \\ 
G &= -\frac{1}{3} r^2 e_7(r) {\E}(\epsilon v) 
+ \frac{5}{18} r^3 e_8(r) \dot{\E}(\epsilon v) + \cdots
\end{align} 
\end{subequations} 
are inserted within the even-parity field equations. The functions
$e_1$, $e_4$, and $e_7$ are already known from the first step, and the
goal of the second step is to determine $e_2$, $e_5$, and $e_8$. The 
overdot still indicates differentiation with respect to $v$; the terms
in $\dot{\E}$ are therefore of order $\epsilon$, and higher
derivatives are of order $\epsilon^2$ and higher. In the second step
the field equations are expanded to order $\epsilon$, and all terms in  
$\ddot{\cal E}$ are neglected. As in the first step, the equations 
$Q^{vr} = 0$, $Q^{rr} = 0$, and $Q^v = 0$ give rise to a system of
three independent differential equations for the three unknown radial
functions. The general solution is easily obtained, and it depends on
three integration constants. As before, imposing regularity at the
event horizon determines one of these, and removes all
terms proportional to $\log(r-2M)$ from the radial functions. One of
the two remaining constants is equivalent to the parameter $c_2$ that
appears in Eqs.~(\ref{gauge_even_l2}); it characterizes the residual
gauge freedom that is still contained within the light-cone class of 
gauges. The other constant is equivalent to the parameter $p_1$ that
appears in Eqs.~(\ref{redef_even_l2}); it  corresponds to a
redefinition of the quadrupole tidal moment ${\E}$. The constant $c_2$
is selected by simultaneously enforcing $e_2 = e_5= 0$ at $r=2M$; this
is once more the horizon-locking condition of
Eq.~(\ref{horizon_lock_gauge1}). The constant $p_1$ is selected by
also enforcing $e_8 = 0$ at $r=2M$. The functions $e_2$, $e_5$, and
$e_8$ are now completely determined, and the constants $c_2$ and $p_1$
have been chosen.  

In the third and last step Eqs.~(\ref{lin_even_l2}) are inserted
within the field equations, which are expanded to order $\epsilon^2$. 
The general solution to the system of differential equations for
$e_3$, $e_6$, and $e_9$ is easily obtained, and as always it depends
on three integration constants. The two that remain after imposing
regularity at the event horizon are equivalent to the parameters $c_3$
and $p_2$ that appear in Eqs.~(\ref{gauge_even_l2}) and
(\ref{redef_even_l2}). The constant $c_3$ is selected by
simultaneously enforcing $e_3 = e_6 = 0$ at $r=2M$. The constant 
$p_2$ is selected by also enforcing $e_9 = 0$ at $r=2M$. The functions
$e_3$, $e_6$, and $e_9$ are now completely determined, and the
constants $c_3$ and $p_2$ have been chosen.  

The task of solving the linearized field equations in the even-parity,
$l=2$ sector is now completed, and we proceed in the same fashion for
the $l=3$ and $l=4$ sectors. The radial functions obtained here are
listed in Table~\ref{tab:linear_functions}.   

\subsubsection{Odd-parity sector}

Essentially the same strategy is adopted for the odd-parity sector of
the metric perturbation. We describe the details for the quadrupole
($l=2$) sector. 

In the first step we insert the expressions $h_v = \frac{1}{3} r^3
b_4(r) {\B}(\epsilon v) + \cdots$ and $h_2 = \frac{1}{3} r^4
b_7(r) {\B}(\epsilon v) + \cdots$ within the odd-parity field
equations, which are expanded to order $\epsilon^0$. In the notation
of Martel and Poisson, the $P^v = 0$ and $P^r = 0$ equations give rise
to a system of two differential equations for the two unknowns $b_4$
and $b_7$. The general solution to this system depends on three
constants of integration. The first is determined by demanding
regularity at $r=2M$. The second is an overall
multiplicative factor that is chosen so that $b_4$ and $b_7$ approach
unity when $r \gg 2M$. The third and final constant of integration is
equivalent to the parameter $k_1$ that appears in
Eq.~(\ref{gauge_odd_l2}); it characterizes the residual gauge freedom
that is still contained within the light-cone class of gauges. We find
that $b_4$ automatically vanishes at $r=2M$, and we set the value of
$k_1$ by imposing the first horizon-locking constraint of
Eqs.~(\ref{hor_lock_l2}). The functions $b_4$ and $b_7$ are now
completely determined, and the constant $k_1$ has been chosen.     

In the second step we insert the expressions 
\begin{subequations} 
\begin{align} 
h_v &= \frac{1}{3} r^3 b_4(r) {\B}(\epsilon v) 
- \frac{1}{6} r^4 b_5(r) \dot{\B}(\epsilon v) + \cdots, \\ 
h_2 &= \frac{1}{3} r^4 b_7(r) {\B}(\epsilon v) 
- \frac{5}{18} r^5 b_8(r) \dot{\B}(\epsilon v) + \cdots
\end{align}
\end{subequations} 
within the odd-parity field equations, which are expanded to order 
$\epsilon$. As in the first step, the general solution to the
system of differential equations for $b_5$ and $b_8$ depends on three
constants of integration. The first is determined by imposing
regularity at $r=2M$. The second constant is equivalent to the gauge
parameter $k_2$ that appears in Eqs.~(\ref{gauge_odd_l2}), and the
third is equivalent to the parameter $q_1$ that appears in
Eqs.~(\ref{redef_odd_l2}); this corresponds to a redefinition of the
quadrupole tidal moment ${\B}$. We find that $b_5$ automatically
vanishes at $r=2M$, and we choose $k_2$ so that the second
horizon-locking constraint of Eqs.~(\ref{hor_lock_l2}) is
satisfied. Finally, we choose $q_1$ so that $b_8$ also vanishes at
$r=2M$. The functions $b_5$ and $b_8$ are now completely determined,
and the constants $k_2$ and $q_1$ have been chosen. 

In the third and final step we insert Eqs.~(\ref{lin_odd_l2}) within
the odd-parity field equations, which are now expanded to order 
$\epsilon^2$. As before the general solution to the system of
differential equations for $b_6$ and $b_9$ depends on three constants
of integration. The first is set by imposing regularity at $r=2M$. The
second is equivalent to the gauge parameter $k_3$ that appears in
Eqs.~(\ref{gauge_odd_l2}), and the third is equivalent to the
parameter $q_2$ that appears in Eqs.~(\ref{redef_odd_l2}). We find
that $b_6$ automatically vanishes at $r=2M$. We choose $k_3$ so that
the third of Eqs.~(\ref{hor_lock_l2}) is satisfied, and $q_2$ so that
$b_9$ also vanishes at $r=2M$. The functions $b_6$ and $b_9$ are now
completely determined, and the constants $k_3$ and $q_2$ have been
chosen.  
 
The task of solving the linearized field equations in the odd-parity,
$l=2$ sector is now completed, and we proceed in the same fashion for
the $l=3$ and $l=4$ sectors. The radial functions obtained here are
listed in Table~\ref{tab:linear_functions}.   

\subsubsection{Light-cone gauge and horizon-locking condition} 

The perturbation obtained in this section is presented in the
light-cone gauge introduced in Sec.~\ref{subsec:lin_prep}. The gauge
is completely fixed: The residual gauge freedom that was initially
left over was removed by imposing the horizon-locking conditions of 
Eqs.~(\ref{horizon_lock_gauge1}) and (\ref{horizon_lock_gauge2}). In
terms of the even-parity radial functions, this means that
$\ee{q}{1}$, $\ee{q}{2}$, $\ee{q}{3}$, $\ee{q}{4}$, $\ee{q}{5}$,
$\ee{q}{6}$, $\ee{o}{1}$, $\ee{o}{2}$, $\ee{o}{4}$, $\ee{o}{5}$,
$\ee{h}{1}$, and $\ee{h}{4}$ all vanish at $r=2M$. In terms of the
odd-parity radial functions, we have that $\bb{q}{1}$, $\bb{q}{2}$,
$\bb{q}{3}$, $\bb{q}{4}$, $\bb{q}{5}$, $\bb{q}{6}$, $\bb{o}{1}$,
$\bb{o}{2}$, $\bb{o}{4}$, $\bb{o}{5}$, $\bb{h}{1}$, and $\bb{h}{4}$
all vanish at $r=2M$. In addition, the freedom described in
Sec.~\ref{subsec:lin_prep} to redefine the tidal moments was exploited
to force $\ee{q}{8}$, $\ee{q}{9}$, $\ee{o}{8}$, $\bb{q}{8}$,
$\bb{q}{9}$, and $\bb{o}{8}$ also to vanish at $r=2M$. With these
conditions, the only radial functions that do not vanish at $r=2M$ are
$\ee{q}{7}$, $\ee{o}{7}$, $\ee{h}{7}$, $\bb{q}{7}$, $\bb{o}{7}$, and
$\bb{h}{7}$; their values at $r=2M$ are listed in the caption of
Table~\ref{tab:linear_functions}.  

\subsection{Bilinear perturbation: Preparation} 
\label{subsec:nonlin_prep} 

\subsubsection{Form of the metric perturbation, and considerations of
  gauge} 

We next move on to the bilinear piece of the perturbed metric, which
is generated by the linear quadrupole terms. The relevant pieces of
the perturbed metric are 
\begin{widetext} 
\begin{subequations}
\label{metric_nonlinear}  
\begin{align} 
g_{vv} &= -f + \mbox{linear} 
+ \frac{1}{15} r^4 \bigl( \pp{m}{1} \PP{m} + \qq{m}{1} \QQ{m} \bigr)  
+ \frac{2}{15} r^4 \g{d}{1} \GG{d}
+ \frac{2}{7} r^4 \bigl( \pp{q}{1} \PP{q} + \qq{q}{1} \QQ{q} \bigr)  
\nonumber \\ & \quad \mbox{} 
+ \frac{2}{3} r^4 \g{o}{1} \GG{o} 
- \frac{1}{3} r^4 \bigl( \pp{h}{1} \PP{h} + \qq{h}{1} \QQ{h} \bigr)
+ \cdots, \\ 
g_{vr} &= 1, \\ 
g_{vA} &= \mbox{linear} 
- \frac{8}{75} r^5 \g{d}{2} \GG{d}_A 
+ \frac{8}{21} r^5 \hh{q}{2} \HH{q}_A
+ \frac{4}{105} r^5 \bigl( \pp{q}{2} \PP{q}_A 
   + 11 \qq{q}{2} \QQ{q}_A \bigr)  
\nonumber \\ & \quad \mbox{} 
+ \frac{2}{5} r^5 \g{o}{2} \GG{o}_A 
+ r^5 \hh{h}{2} \HH{h}_A 
- \frac{2}{15} r^5 \bigl( \pp{h}{2} \PP{h}_A 
   + \qq{h}{2} \QQ{h}_A \bigr) 
+ \cdots, \\ 
g_{AB} &= r^2 \Omega_{AB} + \mbox{linear} 
+ \frac{8}{225} r^6 \Omega_{AB} 
   \bigl( \pp{m}{3} \PP{m} + \qq{m}{3} \QQ{m} \bigr)
+ \frac{32}{225} r^6 \Omega_{AB}\, \g{d}{3} \GG{d} 
\nonumber \\ & \quad \mbox{} 
- \frac{16}{105} r^6 \Omega_{AB} 
   \bigl( \pp{q}{3} \PP{q} + \qq{q}{3} \QQ{q} \bigr) 
- \frac{3}{14} r^6 
   \bigl( \pp{q}{4} \PP{q}_{AB} - \qq{q}{4} \QQ{q}_{AB} \bigr) 
+ \frac{3}{7} r^6 \hh{q}{3} \HH{q}_{AB}
\nonumber \\ & \quad \mbox{} 
- \frac{8}{45} r^6 \Omega_{AB}\, \g{o}{3} \GG{o} 
+ r^6 \g{o}{4} \GG{o}_{AB} 
+ \frac{2}{45} r^6 \Omega_{AB} 
   \bigl( \pp{h}{3} \PP{h} + \qq{h}{3} \QQ{h} \bigr)
+ r^6 \bigl( \pp{h}{4} \PP{h}_{AB} + \qq{h}{4} \QQ{h}_{AB} \bigr) 
+ r^6 \hh{h}{3} \HH{h}_{AB} 
+ \cdots.
\end{align}
\end{subequations} 
\end{widetext}  
The bilinear metric perturbation is decomposed into multipole
moments, and with the help of Tables~\ref{tab:EE_ang},
\ref{tab:BB_ang}, \ref{tab:EBeven_ang}, and \ref{tab:EBodd_ang},  it
can be expressed as in Eqs.~(\ref{pert_even}) and (\ref{pert_odd}), in 
terms of scalar, vector, and tensor harmonics. The metric of
Eq.~(\ref{metric_nonlinear}) includes all the tidal potentials that
can contribute at bilinear order, including $\PP{q}$, $\HH{h}_A$,
$\GG{o}_{AB}$, $\PP{h}_{AB}$, $\QQ{h}_{AB}$, and $\HH{h}_{AB}$ that
were missing from Eqs.~(\ref{background_metric_ang}). All radial
functions are required to approach unity as $r \to \infty$, except for 
$\pp{q}{1}$, $\g{o}{4}$, $\pp{h}{4}$, $\qq{h}{4}$, $\hh{h}{2}$, and
$\hh{h}{3}$, which are expected to approach zero as a power of 
$2M/r$.   

The form of the metric perturbation can be altered by a
residual gauge transformation that preserves the light-cone nature of
the coordinate system. Because the gauge freedom was completely
exhausted in the linear problem (by enforcing the horizon-locking
conditions), the freedom that is left over in the bilinear problem is 
purely bilinear --- the coordinate shifts are necessarily
proportional to $\E \E$, $\B \B$, or $\E \B$. Under these
circumstances the treatment of bilinear gauge transformations is
identical to the linear treatment, and the residual gauge freedom is
described by Preston and Poisson \cite{preston-poisson:06b}. In this 
context the residual gauge freedom is wider than in the linear
problem: We can no longer enforce the tracefree condition $\Omega^{AB}
p_{AB} = 0$, which was maintained in the linear problem.     

According to Preston and Poisson, the residual gauge freedom in the
even-parity sector of the metric perturbation is described by
\begin{subequations}
\label{gauge_nonlinear_even}
\begin{align} 
h_{vv} &\to h_{vv} + 2 r \ddot{a} 
+ 2\Bigl(1 - \frac{3M}{r} \Bigr) \dot{a} - 2 \dot{b} 
+ \frac{2M}{r^2} b, \\ 
j_{v} &\to j_{v} + f a - b - r^2 \dot{c}, \\ 
K &\to K + 2 \dot{a} + \frac{l(l+1)}{r} a - \frac{2}{r} b 
+ l(l+1) c, \\ 
G &\to G - \frac{2}{r} a - 2c,  
\end{align} 
\end{subequations} 
where $a(v)$, $b(v)$, and $c(v)$ are the generators of the gauge
transformation. The transformation is specific to each $lm$ mode, and
the complete gauge transformation is obtained by summing over
the relevant multipoles ($l=\{0,1,2,3,4\}$).  

The residual gauge freedom in the odd-parity sector of the metric
perturbation is described by 
\begin{subequations}
\label{gauge_nonlinear_odd}
\begin{align}  
h_{v} &\to h_{v} - r^2 \dot{\alpha}, \\ 
h_2 &\to h_2 - 2 r^2 \alpha, 
\end{align}
\end{subequations}  
where $\alpha(v)$ is the generator of the gauge transformation. This 
transformation also is specific to each $lm$ mode, and the complete
gauge transformation is obtained by summing over the relevant
multipoles ($l=\{2,4\}$).    

As we did in the linear problem, we shall exploit the residual gauge
freedom to specialize the light-cone gauge to a {\it horizon-locking
gauge} defined by the requirements 
\begin{equation} 
h^{lm}_{vv} = j^{lm}_{v} = 0 = h^{lm}_v \qquad 
\mbox{at $r=2M$}. 
\end{equation} 
These conditions, together with the field equations, give rise to the
same {\it horizon-locking constraint} as in the linear problem: 
\begin{equation} 
h^{lm}_2(v,r=2M) = -\frac{8M^2}{(l-1)(l+2)} 
\frac{\partial h^{lm}_v}{\partial r} \biggr|_{r=2M}. 
\label{hor_lock_nonlin1}
\end{equation} 
This comes about for the following reason: In the notation of Martel
and Poisson \cite{martel-poisson:05}, we find that while $P^r_{lm}$ is
no longer identically zero in the bilinear problem (because the
linear, quadrupole piece of the metric perturbation produces an
effective energy-momentum tensor for the bilinear metric
perturbation), its value turns out to be zero at $r=2M$; the
derivation leading to Eq.~(\ref{hor_lock_nonlin1}) is therefore the
same as the one leading to Eq.~(\ref{horizon_lock_gauge2}) in the
linear problem.    

In addition to this first set of horizon-locking constraints, another
set arises as a consequence of the even-parity perturbation equations,
in particular the $Q^{rr}_{lm}$ equation (again in the notation of
Martel and Poisson). When evaluated at $r=2M$, this equation implies   
\begin{equation} 
\partial_v K^{lm}(v,r=2M) = 4M Q^{rr}_{lm}(v,r=2M)
\label{hor_lock_nonlin2}
\end{equation}
when we also impose the horizon-locking conditions $h^{lm}_{vv} =
j^{lm}_v = 0$ at $r=2M$. The right-hand side of this equation is not
zero, and the bilinear $Q^{rr}_{lm}$ can be computed from the linear
pieces of the metric perturbation. Calculation reveals that the
contributions to $Q^{rr}_{lm}$ that originate from terms involving
$\dot{\E}_{ab}$ and $\dot{\B}_{ab}$ in the linear perturbation vanish 
at $r=2M$, so that $Q^{rr}_{lm}(r=2M)$ comes entirely from
terms that involve $\E_{ab}$ and $\B_{ab}$. The end result is that   
$Q^{rr}_{lm}(r=2M)$ is equal to a quantity quadratic in $\E_{ab}$ and
$\B_{ab}$ that is differentiated with respect to $v$. Equation
(\ref{hor_lock_nonlin2}) can therefore be integrated with respect to
$v$, and this gives rise to conditions on the value of $K^{lm}$ at
$r=2M$. A detailed examination of Eq.~(\ref{hor_lock_nonlin2}) shows
that these are equivalent to the {\it horizon-locking constraints} 
\begin{equation} 
\pp{m}{3} = \qq{m}{3} = \pp{q}{3} = \qq{q}{3} = \pp{h}{3} = \qq{h}{3}
= \frac{5}{16} 
\label{hor_lock_even_nonlin}
\end{equation} 
and 
\begin{equation} 
\g{d}{3} = \g{o}{3} = -\frac{5}{16} 
\label{hor_lock_odd_nonlin}
\end{equation}
on the value of the radial functions at $r=2M$.  

\subsubsection{Redefinitions} 

In addition to making use of the residual gauge freedom, the form of
the metric perturbation can be adjusted by redefining the black-hole
mass parameter $M$ and the tidal moments $\E_{ab}$, $\B_{ab}$,
$\E_{abc}$, and $\B_{abc}$. We consider the changes   
\begin{subequations} 
\label{redef_nonlin} 
\begin{align}  
M &\to M + \frac{1}{15} m_1 M^5 \E_{pq} \E^{pq} 
+ \frac{1}{15} m_2 M^5 \B_{pq} \B^{pq}, \\ 
\E_{ab} &\to \E_{ab} 
- \frac{2}{7} m_3 M^2 \E_{p\langle a} \E^p_{\ b\rangle} 
- \frac{2}{7} m_4 M^2 \B_{p\langle a} \B^p_{\ b\rangle}, \\ 
\B_{ab} &\to \B_{ab} 
+ \frac{4}{7} m_5 M^2 \E_{p\langle a} \B^p_{\ b\rangle}, \\
\E_{abc} &\to \E_{abc} 
- 2m_6 M \epsilon_{pq\langle a} \E^p_{\ b} \B^q_{\ c\rangle}, \\ 
\B_{abc} &\to \B_{abc} 
\end{align} 
\end{subequations}  
where the parameters $m_n$ are dimensionless constants, and where the 
factors of $\frac{1}{15}$, $\frac{2}{7}$, $\frac{4}{7}$, and $2$ were
inserted for convenience. The redefinition of $M$ formally introduces
a time dependence in the black-hole mass parameter; this can be
ignored at the level of accuracy maintained in this work, because
according to the first of Eqs.~(\ref{redef_nonlin}),  $\dot{M} =
O(M^5/\R^5)$. The impact of the redefinitions on the radial
functions will be examined below. We observe that there is no need to
consider a change such as $\E_{abcd} \to \E_{abcd} 
+ m_9 \E_{\langle ab} \E_{cd \rangle} 
+ m_{10} \B_{\langle ab} \B_{cd \rangle}$, because this redefinition 
does not involve $M$; this ambiguity can be resolved at the level of  
the background metric. 

The redefinitions of Eqs.~(\ref{redef_nonlin}) are compatible with 
the parity rules spelled out in Sec.~\ref{subsec:tidal_parity}. The
rules forbid, for example, the presence of a $\E \B$ term in the shift
in $\E_{ab}$, and the presence of $\E^2$ and $\B^2$ terms in the shift
in $\B_{ab}$. Similarly, a shift in $\B_{abc}$ is ruled out, because
the combinations of $\E$ and $\B$ that could be involved would violate
the parity rules. 

\subsubsection{Even-parity sector; $l=0$} 

It is easy to pick out the monopole piece of the bilinear metric 
perturbation from Eqs.~(\ref{metric_nonlinear}); it involves the
tidal moments $\PP{m} := \E_{pq} \E^{pq}$ and 
$\QQ{m} := \B_{pq} \B^{pq}$. Examining the gauge transformation of
Eqs.~(\ref{gauge_nonlinear_even}), we find that the changes
for $l=0$ concern $h_{vv}$ and $K$ only, and that those are generated
by two functions of time, $\dot{a}(v)$ and $b(v)$. Restricting the
gauge freedom as we have done in the linear problem, we write 
$\dot{a} = \frac{1}{15} M^4 (c_1 \PP{m} + d_1 \QQ{m})$ and 
$b = \frac{1}{15} M^5 (c_2 \PP{m} + d_2 \QQ{m})$, with $c_n$, $d_n$
denoting dimensionless constants. Neglecting time derivatives,
substitution within Eqs.~(\ref{gauge_nonlinear_even}) and comparison
with Eqs.~(\ref{metric_nonlinear}) reveals that the gauge
transformation produces the following changes in the radial functions:   
\begin{subequations} 
\label{gauge_nonlin_even_l0}
\begin{align} 
\pp{m}{1} &\to \pp{m}{1} + 2c_1 \frac{M^4}{r^4} - 6c_1 \frac{M^5}{r^5} 
+ 2c_2 \frac{M^6}{r^6}, \\ 
\pp{m}{3} &\to \pp{m}{3} + \frac{15}{4} c_1 \frac{M^4}{r^4} 
- \frac{15}{4} c_2 \frac{M^5}{r^5}, \\ 
\qq{m}{1} &\to \qq{m}{1} + 2d_1 \frac{M^4}{r^4} - 6d_1 \frac{M^5}{r^5} 
+ 2d_2 \frac{M^6}{r^6}, \\ 
\qq{m}{3} &\to \qq{m}{3} + \frac{15}{4} d_1 \frac{M^4}{r^4} 
- \frac{15}{4} d_2 \frac{M^5}{r^5}.
\end{align}
\end{subequations}  
The redefinition of $M$ in Eqs.~(\ref{redef_nonlin}) produces an
additional change in both $\pp{m}{1}$ and $\qq{m}{1}$:  
\begin{subequations}
\label{redef_nonlin_even_l0}
\begin{align}
\pp{m}{1} &\to \pp{m}{1} + 2m_1 \frac{M^5}{r^5}, \\
\qq{m}{1} &\to \qq{m}{1} + 2m_2 \frac{M^5}{r^5}. 
\end{align}
\end{subequations} 
Below we shall use the horizon-locking condition $\pp{m}{1} = 0$ at
$r=2M$ to determine $c_1$, leaving $\pp{m}{1}$ and $\pp{m}{3}$
dependent on $c_2$ and $m_1$. We shall next impose the horizon-locking 
constraint $\pp{m}{3} = 5/16$ at $r=2M$ to determine $m_1$. These
conditions leave $c_2$ undetermined, and this is chosen so as to
simplify the form of the radial functions. Similarly, we shall use the
horizon-locking condition $\qq{m}{1} = 0$ at $r=2M$ to determine
$d_1$, leaving $\qq{m}{1}$ and $\qq{m}{3}$ dependent on $d_2$ and
$m_2$. We shall next impose the horizon-locking constraint $\qq{m}{3}
= 5/16$ at $r=2M$ to determine $m_2$. These conditions leave $d_2$
undetermined, and this is chosen so as to simplify the form of the
radial functions.  

\subsubsection{Even-parity sector; $l=1$} 

The dipole piece of the bilinear perturbation involves the 
tidal potentials $\GG{d}$ and $\GG{d}_A$ defined in
Table~\ref{tab:EBeven_ang}. The gauge transformation for
$l=1$ concern $h_{vv}$, $j_v$, and $K$ only. It is generated by three
functions of time, which we write as $a = \frac{2}{15} c_1 M^5 \GG{d}$, 
$b = \frac{2}{15} c_2 M^5 \GG{d}$, and $c = \frac{2}{15} c_3
M^4 \GG{d}$, with $c_n$ denoting dimensionless constants. (There is an 
abuse of notation here. The functions $a$, $b$, and $c$ are specific
to each mode $l=1$, $m = \{0,1c,1s\}$; the notation $\GG{d}$ therefore
refers to each harmonic component of the bilinear tidal moment, as
listed in Table~\ref{tab:EBeven_ang}.) Neglecting time derivatives,
substitution within Eqs.~(\ref{gauge_nonlinear_even}) and comparison with
Eqs.~(\ref{metric_nonlinear}) reveals that the gauge transformation
produces the following changes in the radial functions:  
\begin{subequations} 
\label{gauge_nonlin_even_l1}
\begin{align} 
\g{d}{1} &\to \g{d}{1} + 2c_2 \frac{M^6}{r^6}, \\ 
\g{d}{2} &\to \g{d}{2} - \frac{5}{4} (c_1-c_2) \frac{M^5}{r^5} 
+ \frac{5}{2} c_1 \frac{M^6}{r^6}, \\ 
\g{d}{3} &\to \g{d}{3} + \frac{15}{8} c_3 \frac{M^4}{r^4} 
+ \frac{15}{8} (c_1-c_2) \frac{M^5}{r^5}. 
\end{align}
\end{subequations} 
Below we shall use the horizon-locking conditions $\g{d}{1} =
\g{d}{2} = 0$ at $r=2M$ to determine $c_2$, leaving $\g{d}{n}$
dependent on $c_1$ and $c_3$. We shall next impose the horizon-locking 
constraint $\g{d}{3} = -5/16$ at $r=2M$ to determine $c_1$. These
conditions leave $c_3$ undetermined, and this is chosen so as to
simplify the form of the radial functions. 

\subsubsection{Even-parity sector; $l=2$} 

The quadrupole piece of the bilinear perturbation involves the
tidal potentials $\PP{q}$ and $\QQ{q}$, as well as their
vectorial and tensorial counterparts; these are defined in
Tables~\ref{tab:EE_ang} and \ref{tab:BB_ang}. The
residual gauge transformation is generated by 
$a = \frac{2}{7} M^5 (c_1 \PP{q} + d_1 \QQ{q})$, 
$b = \frac{2}{7} M^5 (c_2 \PP{q} + d_2 \QQ{q})$, and 
$c = \frac{2}{7} M^4 (c_3 \PP{q} + d_3 \QQ{q})$, in which $\PP{q}$ and
$\QQ{q}$ stand for the harmonic components of the tidal moments listed
in Tables~\ref{tab:EE_ang} and \ref{tab:BB_ang}. Neglecting
time derivatives, substitution within Eqs.~(\ref{gauge_nonlinear_even}) and
comparison with Eqs.~(\ref{metric_nonlinear}) reveals that the gauge
transformation produces the following changes in the radial functions:  
\begin{subequations} 
\label{gauge_nonlin_even_l2}
\begin{align} 
\pp{q}{1} &\to \pp{m}{1} + 2c_2 \frac{M^6}{r^6}, \\ 
\pp{q}{2} &\to \pp{q}{2} + 15 (c_1-c_2) \frac{M^5}{r^5} 
- 30 c_1 \frac{M^6}{r^6}, \\ 
\pp{q}{3} &\to \pp{q}{3} - \frac{45}{4} c_3 \frac{M^4}{r^4} 
- \frac{15}{4} (3c_1-c_2) \frac{M^5}{r^5}, \\
\pp{q}{4} &\to \pp{q}{4} + \frac{8}{3} c_3 \frac{M^4}{r^4} 
+ \frac{8}{3} c_1 \frac{M^5}{r^5}, \\ 
\qq{q}{1} &\to \qq{m}{1} + 2d_2 \frac{M^6}{r^6}, \\ 
\qq{q}{2} &\to \qq{q}{2} + \frac{15}{11} (d_1-d_2) \frac{M^5}{r^5}  
- \frac{30}{11} d_1 \frac{M^6}{r^6}, \\ 
\qq{q}{3} &\to \qq{q}{3} - \frac{45}{4} d_3 \frac{M^4}{r^4} 
- \frac{15}{4} (3d_1-d_2) \frac{M^5}{r^5}, \\
\qq{q}{4} &\to \qq{q}{4} - \frac{8}{3} d_3 \frac{M^4}{r^4} 
- \frac{8}{3} d_1 \frac{M^5}{r^5}.
\end{align} 
\end{subequations} 
The redefinition of Eq.~(\ref{redef_nonlin}) produces the additional
changes    
\begin{subequations}
\label{redef_nonlin_even_l2}
\begin{align}  
\pp{q}{1} &\to \pp{q}{1} + m_3 \frac{M^2}{r^2} f^2, \\
\pp{q}{2} &\to \pp{q}{2} + 5 m_3 \frac{M^2}{r^2} f, \\ 
\pp{q}{4} &\to \pp{q}{4} 
- \frac{4}{9} m_3 \frac{M^2}{r^2} \biggl(1 - \frac{2M^2}{r^2} \biggr)
\\ 
\qq{q}{1} &\to \qq{q}{1} + m_4 \frac{M^2}{r^2} f^2, \\ 
\qq{q}{2} &\to \qq{q}{2} + \frac{5}{11} m_4 \frac{M^2}{r^2} f, \\ 
\qq{q}{4} &\to \qq{q}{4} 
+ \frac{4}{9} m_4 \frac{M^2}{r^2} \biggl(1 - \frac{2M^2}{r^2} \biggr). 
\end{align} 
\end{subequations} 
Below we shall use the horizon-locking conditions $\pp{q}{1} =
\pp{q}{2} = 0$ at $r=2M$ to determine $c_2$, leaving $\pp{q}{n}$
dependent on $c_1$, $c_3$, and $m_3$. We shall next impose the
horizon-locking constraint $\pp{q}{3} = 5/16$ at $r=2M$ to determine
$c_1$ and the additional condition $\pp{q}{4} = 0$ at $r=2M$ to
determine $m_3$. These conditions leave $c_3$ undetermined, and this
is chosen so as to simplify the form of the radial
functions. Similarly, we shall use the horizon-locking conditions
$\qq{q}{1} = \qq{q}{2} = 0$ at $r=2M$ to determine $d_2$, leaving
$\qq{q}{n}$ dependent on $d_1$, $d_3$, and $m_4$. We shall next impose
the horizon-locking constraint $\qq{q}{3} = 5/16$ at $r=2M$ to
determine $d_1$ and the additional condition $\qq{q}{4} = 0$ at $r=2M$
to determine $m_4$. These conditions leave $d_3$ undetermined, and
this is chosen so as to simplify the form of the radial
functions.  

\subsubsection{Even-parity sector; $l=3$} 

The octupole piece of the bilinear perturbation involves the
tidal potentials $\GG{o}$, $\GG{o}_A$, and $\GG{o}_{AB}$,
which are defined in Table~\ref{tab:EBeven_ang}. The residual gauge 
transformation is generated by $a = \frac{2}{3} c_1 M^5 \GG{o}$, 
$b = \frac{2}{3} c_2 M^5 \GG{o}$, and 
$c = \frac{2}{3} c_3 M^4 \GG{o}$, in which $\GG{o}$ stands for the
harmonic components of the tidal moments listed in
Table~\ref{tab:EBeven_ang}. Neglecting time derivatives, substitution
within Eqs.~(\ref{gauge_nonlinear_even}) and comparison with
Eqs.~(\ref{metric_nonlinear}) reveals that the gauge transformation
produces the following changes in the radial functions:   
\begin{subequations}
\label{gauge_nonlin_even_l3}
\begin{align}  
\g{o}{1} &\to \g{o}{1} + 2c_2 \frac{M^6}{r^6}, \\ 
\g{o}{2} &\to \g{o}{2} + 5 (c_1-c_2) \frac{M^5}{r^5} 
- 10 c_1 \frac{M^6}{r^6}, \\ 
\g{o}{3} &\to \g{o}{3} - 45 c_3 \frac{M^4}{r^4} 
- \frac{15}{2} (6c_1-c_2) \frac{M^5}{r^5},  \\
\g{o}{4} &\to \g{o}{4} - 4 c_3 \frac{M^4}{r^4} 
- 4 c_1 \frac{M^5}{r^5}. 
\end{align}
\end{subequations} 
The redefinition of Eqs.~(\ref{redef_nonlin}) produces the additional
changes   
\begin{subequations}
\label{redef_nonlin_even_l3}
\begin{align}   
\g{o}{1} &\to \g{o}{1} + m_6 \frac{M}{r} 
  \biggl(1 - \frac{M}{r} \biggr) f^2, \\ 
\g{o}{2} &\to \g{o}{2} + \frac{5}{4} m_6 \frac{M}{r} 
  \biggl(1 - \frac{4M}{3r} \biggr) f, \\ 
\g{o}{4} &\to \g{o}{4} + \frac{1}{3} m_6 \frac{M}{r} 
  \biggl(f + \frac{4M^3}{5r^3} \biggr). 
\end{align}
\end{subequations}
Below we shall use the horizon-locking conditions $\g{o}{1} =
\g{o}{2} = 0$ at $r=2M$ to determine $c_2$, leaving $\g{o}{n}$
dependent on $c_1$, $c_3$, and $m_6$. We shall next impose the
horizon-locking constraint $\g{o}{3} -5/16$ at $r=2M$ to determine
$c_1$, and the additional condition $\g{o}{4} = 0$ at $r=2M$ to
determine $m_6$. These conditions leave $c_3$ undetermined, and this
is chosen so as to simplify the form of the radial functions. 

\subsubsection{Even-parity sector; $l=4$} 

The hexadecapole piece of the bilinear perturbation involves the
tidal potentials $\PP{h}$ and $\QQ{h}$, as well as their
vectorial and tensorial counterparts; these are defined in
Tables~\ref{tab:EE_ang} and \ref{tab:BB_ang}. The residual
gauge transformation is generated by 
$a = -\frac{1}{3} M^5 (c_1 \PP{h} + d_1 \QQ{h})$, 
$b = -\frac{1}{3} M^5 (c_2 \PP{h} + d_2 \QQ{h})$, and 
$c = -\frac{1}{3} M^4 (c_3 \PP{q} + d_3 \QQ{q})$, in which $\PP{h}$ and
$\QQ{h}$ denote the harmonic components of the tidal moments listed in
Tables~\ref{tab:EE_ang} and \ref{tab:BB_ang}.  Neglecting time
derivatives, substitution within 
Eqs.~(\ref{gauge_nonlinear_even}) and comparison with
Eqs.~(\ref{metric_nonlinear}) reveals that the gauge transformation
produces the following changes in the radial functions:  
\begin{subequations}
\label{gauge_nonlin_even_l4}
\begin{align}  
\pp{h}{1} &\to \pp{h}{1} + 2c_2 \frac{M^6}{r^6}, \\ 
\pp{h}{2} &\to \pp{h}{2} + 10 (c_1-c_2) \frac{M^5}{r^5} 
- 20 c_1 \frac{M^6}{r^6}, \\ 
\pp{h}{3} &\to \pp{h}{3} - 150 c_3 \frac{M^4}{r^4} 
- 15 (10c_1-c_2) \frac{M^5}{r^5}, \\
\pp{h}{4} &\to \pp{h}{4} + 4 c_3 \frac{M^4}{r^4} 
+ 4 c_1 \frac{M^5}{r^5}, \\ 
\qq{h}{1} &\to \qq{h}{1} + 2d_2 \frac{M^6}{r^6}, \\ 
\qq{h}{2} &\to \qq{h}{2} + 10 (d_1-d_2) \frac{M^5}{r^5} 
- 20 d_1 \frac{M^6}{r^6}, \\ 
\qq{h}{3} &\to \qq{h}{3} - 150 d_3 \frac{M^4}{r^4} 
- 15 (10d_1-d_2) \frac{M^5}{r^5}, \\
\qq{h}{4} &\to \qq{h}{4} + 4 d_3 \frac{M^4}{r^4} 
+ 4 d_1 \frac{M^5}{r^5}. 
\end{align} 
\end{subequations} 
Below we shall use the horizon-locking conditions $\pp{h}{1} =
\pp{h}{2} = 0$ at $r=2M$ to determine $c_2$, leaving $\pp{h}{n}$
dependent on $c_1$ and $c_3$. We shall next impose the horizon-locking 
constraint $\pp{h}{3} = 5/16$ at $r=2M$ to determine $c_1$. These
conditions leave $c_3$ undetermined. We shall find that $\pp{h}{4}
\neq 0$ at $r=2M$, but that its value is independent of $c_3$, which 
is chosen so as to simplify the form of the radial functions.
Similarly, we shall use the horizon-locking conditions $\qq{h}{1} =
\qq{h}{2} = 0$ at $r=2M$ to determine $d_2$, leaving $\qq{h}{n}$
dependent on $d_1$ and $d_3$. We shall next impose the horizon-locking 
constraint $\qq{h}{3} = 5/16$ at $r=2M$ to determine $d_1$. These
conditions leave $d_3$ undetermined. We shall find that $\qq{h}{4}
\neq 0$ at $r=2M$, but that its value is independent of $d_3$, which 
is chosen so as to simplify the form of the radial functions. 

\subsubsection{Odd-parity sector; $l=2$} 

The quadrupole piece of the bilinear perturbation involves the
tidal potentials $\HH{q}_A$ and $\HH{q}_{AB}$ defined in 
Table~\ref{tab:EBodd_ang}. The residual gauge transformation
is generated by $\alpha = \frac{3}{7} k_1 M^4 \HH{q}$, in which
$\HH{q}$ stands for the harmonic components listed in
Table~\ref{tab:EBodd_ang}. Neglecting time derivatives, substitution
within Eqs.~(\ref{gauge_nonlinear_odd}) and comparison with
Eqs.~(\ref{metric_nonlinear}) reveals that the gauge transformation
produces the following changes in the radial functions:  
\begin{subequations} 
\label{gauge_nonlin_odd_l2}
\begin{align}  
\hh{q}{2} &\to \hh{q}{2}, \\ 
\hh{q}{3} &\to \hh{q}{3} - 2k_1\frac{M^4}{r^4}.
\end{align}
\end{subequations} 
The redefinitions of Eq.~(\ref{redef_nonlin}) produces the additional
changes    
\begin{subequations}
\label{redef_nonlin_odd_l2}
\begin{align}  
\hh{q}{2} &\to \hh{q}{2} + m_5 \frac{M^2}{r^2} f, \\ 
\hh{q}{3} &\to \hh{q}{3} + \frac{4}{9} m_5 \frac{M^2}{r^2} 
  \biggl( 1 - \frac{6M^2}{r^2} \biggr). 
\end{align} 
\end{subequations} 
The horizon-locking constraint of Eq.~(\ref{hor_lock_nonlin1}) gives
rise to  
\begin{equation} 
\hh{q}{3}(2M) = -\frac{4}{9} M \frac{d \hh{q}{2}}{dr} \biggr|_{r=2M}. 
\label{horlock_nonlin_l2} 
\end{equation} 
We shall verify that the horizon-locking condition $\hh{q}{2}(2M) = 0$
is automatically satisfied, and use Eq.~(\ref{horlock_nonlin_l2}) to
determine $k_1$. We shall then choose $m_5$ so that $\hh{q}{3}(2M) =
0$. 

\subsubsection{Odd-parity sector; $l=4$} 

The hexadecapole piece of the bilinear perturbation involves the
tidal potentials $\HH{h}_A$ and $\HH{h}_{AB}$ defined in
Table~\ref{tab:EBodd_ang}. The residual gauge transformation
is generated by $\alpha = \frac{1}{6} k_1 M^4 \HH{h}$, in which
$\HH{h}$ denotes the harmonic components listed in
Table~\ref{tab:EBodd_ang}. Neglecting time derivatives, substitution
within Eqs.~(\ref{gauge_nonlinear_odd}) and comparison with
Eqs.~(\ref{metric_nonlinear}) reveals that the gauge transformation 
produces the following changes in the radial functions:    
\begin{subequations}
\label{gauge_nonlin_odd_l4}
\begin{align}  
\hh{h}{2} &\to \hh{h}{2}, \\ 
\hh{h}{3} &\to \hh{h}{3} - 2k_1\frac{M^4}{r^4}.
\end{align} 
\end{subequations} 
The horizon-locking constraint of Eq.~(\ref{hor_lock_nonlin1}) gives
rise to  
\begin{equation} 
\hh{h}{3}(2M) = -\frac{1}{3} M \frac{d \hh{h}{2}}{dr} \biggr|_{r=2M}. 
\label{horlock_nonlin_l4}
\end{equation} 
We shall verify that $\hh{h}{2}(2M) = 0$, and use
Eq.~(\ref{horlock_nonlin_l4}) to determine $k_1$. We shall find that
$\hh{h}{3}$ does not vanish at $r=2M$, but that its value is gauge
invariant.  

\subsection{Bilinear perturbation: Field equations} 
\label{subsec:nonlin_field}

\subsubsection{General strategy} 

The solution to the bilinear problem requires a generalization of the
perturbation formalism employed in the linear problem. The perturbed
metric takes the schematic form $g = \hat{g} + \epsilon p_1 
+ \epsilon^2 p_2 + O(\epsilon^3)$, in which $\hat{g}$ stands for the
Schwarzschild metric, $\epsilon p_1$ is the linear perturbation
calculated previously, and $\epsilon^2 p_2$ the bilinear perturbation
that we now wish to obtain; $\epsilon \sim {\cal R}^{-2}$ is a
perturbative book-keeping parameter. While $\epsilon p_1$ is linear in
the tidal moments $\E_{ab}$ and $\B_{ab}$, $\epsilon^2 p_2$ involves
terms of the schematic form $\E\E$, $\E\B$, and $\B\B$; those are
decomposed into multipole moments and expressed in terms of the
potentials $\PP{m}$, $\PP{q}$, $\PP{h}$, $\QQ{m}$, $\QQ{q}$, $\QQ{h}$,
$\GG{d}$, $\GG{o}$, $\HH{q}$, and $\HH{o}$. Notice that the bilinear
problem involves the quadrupole tidal moments $\E_{ab}$ and $\B_{ab}$
only; at order $\epsilon^2 \sim {\cal R}^{-4}$ there is no need to
involve the octupole and hexadecapole moments that also appear in the
linear perturbation. 

The Einstein tensor for the perturbed metric takes the schematic form
$G[\hat{g}] + \epsilon G_1[p_1] + \epsilon^2 G_1[p_2] 
+ \epsilon^2 G_2[p_1] + O(\epsilon^3)$, where $G[\hat{g}] = 0$ is the
Einstein tensor of the Schwarzschild metric, and $\epsilon G_1[p_1] =
0$ is the first-order perturbation created by the linear perturbation
$\epsilon p_1$. The remaining terms appear at second order. The first 
contribution, $\epsilon^2 G_1[p_2]$, is the Einstein tensor generated
entirely from the second-order perturbation $\epsilon^2 p_2$; $G_1$ is
the same linear differential operator that was encountered in the
linear problem. The second contribution, $\epsilon^2 G_2[p_1]$, is
generated by the first-order perturbation, and it originates in the
nonlinearities of the vacuum field equations. 

The bilinear perturbation problem consists of finding solutions to
the field equations 
\begin{equation} 
G_1[p_2] = -G_2[p_1].
\end{equation} 
These have the same formal structure as the linear field equations
$G_1[p_1] = 0$, except for the fact that there is a source term on the
right-hand side. The problem is tractable because the differential
operator on the left-hand side is the same as in the linear
problem, and because the source term on the right-hand side can be
computed directly from the known solution $\epsilon p_1$. Notice than
time derivatives of $\E_{ab}$ and $\B_{ab}$ can be ignored when
computing $G_1[p_2]$ and $G_2[p_1]$; these contribute at order 
${\cal R}^{-5}$ and higher, and they do not affect the field equations
at order $\epsilon^2 \sim {\cal R}^{-4}$. 

The strategy to solve the bilinear field equations is the same as in
the linear problem. First $p_2$ is decomposed in scalar, vector, and
tensor harmonics, a task that was already accomplished in
Eqs.~(\ref{metric_nonlinear}). Second, the linear differential
operator $G_1[p_2]$ is allowed to act on the perturbation, and the
result is again decomposed in spherical harmonics. Third, the
effective source term $-G_2[p_1]$ is computed and decomposed in
spherical harmonics. And fourth, the perturbation equations are
integrated, one mode at a time. In practice we rely on the
Martel-Poisson perturbation formalism \cite{martel-poisson:05} to
implement this strategy.    

\subsubsection{Even-parity sector}

The even-parity sector of the perturbation $p_2$ involves the terms in
$\PP{m}$, $\PP{q}$, $\PP{h}$, $\QQ{m}$, $\QQ{q}$, $\QQ{h}$, $\GG{d}$,
and $\GG{o}$; these are decomposed as in Eqs.~(\ref{pert_even}). The
source terms $-G_2[p_1]$ are computed from $p_1$ (which, we recall,
involves the potentials $\EE{q}$ and $\BB{q}$ only), and these also
are decomposed as in Eqs.~(\ref{pert_even}). In the notation of Martel
and Poisson, the reduced source functions are denoted $Q^{vv}$,
$Q^{vr}$, $Q^{rr}$, $Q^v$, $Q^r$, $Q^\flat$, and $Q^\sharp$, in which
we suppress the use of the $lm$ labels. Examination of $p_2$ and
calculation of $-G_2[p_1]$ reveal that the relevant multipoles are
$l=\{0, 1, 2, 3, 4\}$.  

The perturbation equations for $l=0$ decouple into two sets, the first
proportional to $\PP{m}$ and involving the radial functions
$\pp{m}{1}$ and $\pp{m}{3}$, the second proportional to $\QQ{m}$ and
involving the radial functions $\qq{m}{1}$ and $\qq{m}{3}$. For each
set of equations the general solution depends on three integration
constants; these were denoted $\{ c_1, c_2, m_1 \}$ and 
$\{ d_1, d_2, m_2 \}$ in Eqs.~(\ref{gauge_nonlin_even_l0}) and
(\ref{redef_nonlin_even_l0}). The constants $c_n$ and $d_n$ are  
chosen so as to enforce the horizon-locking condition $\pp{m}{n} =
\qq{m}{n} = 0$ at $r=2M$. The constants $m_1$ and $m_2$
are chosen so as to simplify the form of the radial functions;
we use this freedom to remove terms proportional to $x^{-5}$ in
$\pp{m}{1}/f$ and $\qq{m}{1}/f$, and terms proportional to $x^{-4}$ in
$\pp{m}{3}/f$ and $\qq{m}{3}/f$. The end results are listed in
Table~\ref{tab:nonlinear_functions}.   

The perturbation equations for $l=1$ involve the dipole potentials
$\GG{d}$ and the three radial functions $\g{d}{n}$. The general
solution to the set of coupled differential equations for the radial
functions depends on three integration constants; these were denoted
$\{ c_1, c_2, c_3 \}$ in Eqs.~(\ref{gauge_nonlin_even_l1}). The
constants $c_1$ and $c_2$ can be chosen so as to enforce the
horizon-locking condition $\g{d}{n} = 0$ at $r=2M$. This leaves $c_3$
undetermined, and it can be chosen so as to simplify the form of the
radial functions; we use this freedom to remove a term proportional to
$x^{-5}$ in $\g{d}{2}/f$. The end results are displayed in
Table~\ref{tab:nonlinear_functions}.    

The perturbation equations for $l=2$ decouple into two sets, the first
proportional to $\PP{q}$ and involving the four radial functions
$\pp{q}{n}$, the second proportional to $\QQ{q}$ and involving the
four radial functions $\qq{q}{n}$. For each
set of differential equations the general solution depends on five
integration constants; one can be eliminated by removing all terms
proportional to $\log(r-2M)$ from the radial functions, and the others
correspond to $\{ c_1, c_2, c_3, m_3 \}$ and $\{ d_1, d_2, 
d_3, m_4 \}$ in Eqs.~(\ref{gauge_nonlin_even_l2}) and
(\ref{redef_nonlin_even_l2}). The constants
$c_1$, $c_2$, and $m_3$ are chosen so as to enforce the
horizon-locking condition $\pp{q}{n} = 0$ at $r=2M$, while $d_1$,
$d_2$, and $m_4$ are chosen to enforce $\qq{q}{n} = 0$ at $r=2M$. This
leaves $c_3$ and $d_3$ undetermined, and those are chosen so as to
simplify the form of the radial functions; we use this freedom to
remove terms proportional to $x^{-5}$ in both $\pp{q}{2}/f$ and
$\qq{q}{2}/f$. The end results are listed in
Table~\ref{tab:nonlinear_functions}.    

The perturbation equations for $l=3$ involve the octupole potentials 
$\GG{o}$ and the four radial functions $\g{o}{n}$. The general
solution to the set of coupled differential equations for the radial
functions depends on five integration constants; one can be eliminated
by removing all terms proportional to $\log(r-2M)$ from the radial
functions, and the remaining constants were denoted 
$\{ c_1, c_2, c_3, m_6 \}$ in Eqs.~(\ref{gauge_nonlin_even_l3}) and 
(\ref{redef_nonlin_even_l3}). The constants $c_1$, $c_2$, and $m_6$
can be chosen so as to enforce the horizon-locking condition $\g{o}{n}
= 0$ at $r=2M$. This leaves $c_3$ undetermined, and it can be chosen
so as to simplify the form of the radial functions; we use this
freedom to remove a term proportional to $x^{-5}$ in $\g{o}{2}/f$. The
end results are displayed in Table~\ref{tab:nonlinear_functions}.   

The perturbation equations for $l=4$ decouple into two sets, the first
proportional to $\PP{h}$ and involving the four radial functions
$\pp{h}{n}$, the second proportional to $\QQ{h}$ and involving the
four radial functions $\qq{h}{n}$. For each
set of differential equations the general solution depends on five
integration constants; one can be eliminated by removing all terms
proportional to $\log(r-2M)$ from the radial functions, and another
must be set so that the radial functions (except for $\pp{h}{4}$ and
$\qq{h}{4}$) approach unity as $r \to \infty$. The remaining three are
denoted $\{ c_1, c_2, c_3 \}$ and $\{ d_1, d_2, d_3 \}$ in
Eqs.~(\ref{gauge_nonlin_even_l4}). The constants $c_1$ and $c_2$ are 
chosen so as to enforce the horizon-locking condition $\pp{h}{1}  
= \pp{h}{2} = \pp{h}{3} = 0$ at $r=2M$, and we find that the value of
$\pp{h}{4}$ at $r=2M$ is independent of $c_3$ (and therefore gauge
invariant); we choose $c_3$ so that a term proportional to $x^{-5}$ is 
removed from $\pp{h}{2}/f$. Similarly, $d_1$ and $d_2$ are chosen so
as to enforce the horizon-locking condition $\qq{h}{1} = \qq{h}{2} =
\qq{h}{3} = 0$ at $r=2M$, and we find that the value of $\qq{h}{4}$ at
$r=2M$ is independent of $d_3$ (and therefore gauge invariant); we
choose $d_3$ so that a term proportional to $x^{-5}$ is removed from
$\qq{h}{2}/f$. The end results are listed in
Table~\ref{tab:nonlinear_functions}.    

\subsubsection{Odd-parity sector}

The odd-parity sector of the perturbation $p_2$ involves the terms in 
$\HH{q}$ and $\HH{o}$; these are decomposed as in
Eq.~(\ref{pert_odd}). The source terms $-G_2[p_1]$ are computed from
$p_1$ and also decomposed as in Eq.~(\ref{pert_odd}). In the
notation of Martel and Poisson, the reduced source functions are
denoted $P^v$, $P^r$, and $P$, in which we suppress the use of the
$lm$ labels. Examination of $p_2$ and calculation of $-G_2[p_1]$
reveal that the relevant multipoles are $l=\{2, 4\}$. 

The perturbation equations for $l=2$ involve the potentials $\HH{q}$
and the radial functions $\hh{q}{2}$ and $\hh{q}{3}$. The general
solution to the set of coupled differential equations for the radial
functions depends on three integration constants; one can be
eliminated by removing all terms proportional to $\log(r-2M)$ from the
radial functions, and the others correspond to  
$\{ k_1, m_5 \}$ in Eqs.~(\ref{gauge_nonlin_odd_l2}) and
(\ref{redef_nonlin_odd_l2}). We observe that $\hh{q}{2} = 0$ at
$r=2M$, and we use Eq.~(\ref{horlock_nonlin_l2}) to set the value of
$k_1$. Finally, we choose $m_5$ to enforce the additional condition
that $\hh{q}{3} = 0$ at $r=2M$. The end results are listed in
Table~\ref{tab:nonlinear_functions}.  
  
The perturbation equations for $l=4$ involve the potentials $\HH{h}$ 
and the radial functions $\hh{h}{2}$ and $\hh{h}{3}$. The general
solution to the set of coupled differential equations for the radial
functions depends on three integration constants. One of these can be
eliminated by removing all terms proportional to $\log(r-2M)$ from the
radial functions, and another must be set so that both $\hh{h}{2}$ and
$\hh{h}{3}$ approach zero as $r \to \infty$. The remaining constant is
denoted $k_1$ in Eq.~(\ref{gauge_nonlin_odd_l4}). We observe that
$\hh{h}{2} = 0$ at $r=2M$, and we use Eq.~(\ref{horlock_nonlin_l2}) to
set the value of $k_1$. The end results are listed in
Table~\ref{tab:nonlinear_functions}. 

\subsection{Conclusion} 
\label{subsec:conclusion} 

The linear piece of the metric perturbation was computed in
Secs.~\ref{subsec:lin_prep} and \ref{subsec:lin_field}, and the
bilinear piece was computed in Secs.~\ref{subsec:nonlin_prep} and 
\ref{subsec:nonlin_field}. The calculation is complete, and after
collecting results we obtain the black-hole metric of
Eqs.~(\ref{blackhole_metric_ang}). The quasi-Cartesian representation
of the metric can immediately be constructed from this, and the
expressions are given in Eqs.~(\ref{blackhole_metric_cart}).  

\begin{acknowledgments} 
This work was supported by the Natural Sciences and Engineering
Research Council of Canada. EP is grateful to the staff of the
Canadian Institute for Theoretical Astrophysics for their kind
hospitality during the time of his research leave from the University
of Guelph; a large of portion of this work was completed during this
time.  
\end{acknowledgments}    

\appendix

\section{Decomposition of tidal potentials in spherical harmonics} 
\label{app:spherical_harmonics} 

We wish to express the angular version of the tidal potentials
listed in Tables~\ref{tab:E_cart}, \ref{tab:B_cart},
\ref{tab:EE_cart}, \ref{tab:BB_cart}, \ref{tab:EBeven_cart}, and
\ref{tab:EBodd_cart} as expansions in scalar, vector, and tensor
harmonics. Before we proceed we record the useful identities 
\begin{subequations}
\label{ang_identities2}  
\begin{align} 
& \epsilon_{AB} = \epsilon_{abc} \Omega^a_A \Omega^b_B \Omega^c,  
\\
& \epsilon_A^{\ B} \Omega^b_B = -\Omega^a_A \epsilon_{ap}^{\ \ \ b}
\Omega^p, \\ 
& D_A D_B \Omega^a = D_B D_A \Omega^a = -\Omega^a \Omega_{AB}. 
\end{align} 
\end{subequations} 
These quantities were all introduced in
Sec.~\ref{subsec:tidal_pot_angular}  of the main text. 

The general structure of the even-parity tidal potentials is
\begin{subequations}
\label{even_structure} 
\begin{align} 
\A^{(l)} &= \A_{k_1 k_2 \cdots k_l} \Omega^{k_1} 
\Omega^{k_2} \cdots \Omega^{k_l}, \\ 
\A^{(l)}_a &= \gamma_a^{\ c} \A_{c k_2 \cdots k_l} 
\Omega^{k_2} \cdots \Omega^{k_l}, \\
\A^{(l)}_{ab} &= 2\gamma_a^{\ c} \gamma_b^{\ d} \A_{c d k_3 \cdots
  k_l} \Omega^{k_3} \cdots \Omega^{k_l} + \gamma_{ab} \A^{(l)},
\end{align} 
\end{subequations}
in which $\A_{k_1 k_2 \cdots k_l}$ is a constant STF tensor of rank
$l$. It is not difficult to show that these satisfy the eigenvalue
equations 
\begin{subequations} 
\label{even_eigen_cart} 
\begin{align} 
& r^2 \gamma^{cd} D_c D_d \A^{(l)} + l(l+1) \A^{(l)} = 0, \\ 
& r^2 \gamma^{cd} D_c D_d \A^{(l)}_a 
+ \bigl[ l(l+1)-1\bigr] \A^{(l)}_a = 0, \\ 
& r^2 \gamma^{cd} D_c D_d \A^{(l)}_{ab} 
+ \bigl[ l(l+1)-4 \bigr] \A^{(l)}_{ab} = 0,
\end{align} 
\end{subequations} 
where $D_a$ is a projected differential operator that acts as follows
on arbitrary tensor fields: $D_a T_{b_1 b_2 \cdots} := \gamma_a^{\ p} 
\gamma_{b_1}^{\ q_1} \gamma_{b_2}^{\ q_2} \cdots 
\partial_p T_{q_1 q_2 \cdots}$. 

The transformed potentials are
\begin{subequations} 
\begin{align} 
\A^{(l)} &= \A_{k_1 k_2 \cdots k_l} \Omega^{k_1} 
\Omega^{k_2} \cdots \Omega^{k_l}, \\ 
\A^{(l)}_A &= \Omega^a_A \A_{a k_2 \cdots k_l} 
\Omega^{k_2} \cdots \Omega^{k_l}, \\
\A^{(l)}_{AB} &= 2\Omega^a_A \Omega^b_B \A_{ab k_3 \cdots k_l}  
\Omega^{k_3} \cdots \Omega^{k_l} + \Omega_{AB} \A^{(l)}, 
\end{align} 
\end{subequations}
and we wish to expand them in the even-parity harmonics of
Eqs.~(\ref{Ylm_even}). These satisfy the eigenvalue equations 
\begin{subequations} 
\label{even_eigen_ang} 
\begin{align} 
& \Omega^{CD} D_C D_D Y^{lm} + l(l+1) Y^{lm} = 0,\\ 
& \Omega^{CD} D_C D_D Y^{lm}_A 
+ \bigl[ l(l+1) - 1 \bigr] Y^{lm}_A = 0, \\ 
& \Omega^{CD} D_C D_D Y^{lm}_{AB} 
+ \bigl[ l(l+1) - 4 \bigr] Y^{lm}_{AB} = 0,
\end{align}
\end{subequations} 
which are in a close correspondence with Eqs.~(\ref{even_eigen_cart}).  

We begin with $\A^{(l)}$, which we decompose as  
\begin{equation} 
\A^{(l)} = \sum_m \A^{(l)}_m Y^{lm}, 
\end{equation}
in terms of harmonic components $\A^{(l)}_m$. There are $2l+1$ real 
terms in the sum, and the $2l+1$ independent components of 
$\A_{k_1 k_2 \cdots k_l}$ are in a one-to-one correspondence with the
coefficients $\A^{(l)}_m$. Returning to the original representation of 
Eq.~(\ref{even_structure}), we find after differentiation that 
$D_A \A^{(l)} = l \Omega^a_A A_{a k_2 \cdots k_l}
\Omega^{k_2} \cdots \Omega^{k_l}$, and we conclude that 
\begin{equation} 
\A^{(l)}_A = \frac{1}{l} D_A \A^{(l)} 
= \frac{1}{l} \sum_m \A^{(l)}_m Y_A^{lm}. 
\end{equation} 
This is the decomposition of $\A^{(l)}_A$ in vectorial,
even-parity harmonics; this equation is valid for $l \neq 0$. 
An additional differentiation using the last of
Eqs.~(\ref{ang_identities2}) reveals that $D_A D_B \A^{(l)} = -l
\Omega_{AB} \A^{(l)} + l(l-1) \Omega^a_A \Omega^b_B 
\A_{abk_3\cdots k_l} \Omega^{k_3} \cdots \Omega^{k_l}$. From this we
conclude that  
\begin{align} 
\A^{(l)}_{AB} &= \frac{2}{l(l-1)} \Bigl[ D_A D_B 
+ \frac{1}{2} l(l+1) \Omega_{AB} \Bigr] \A^{(l)} 
\nonumber \\
&= \frac{2}{l(l-1)} \sum_m \A^{(l)}_m Y_{AB}^{lm}. 
\end{align} 
This is the decomposition of $\A^{(l)}_{AB}$ in tensorial, 
even-parity harmonics; this equation is valid for $l \neq \{0,1\}$. 

We next examine the odd-parity potentials. Their general structure is 
\begin{subequations} 
\label{odd_structure} 
\begin{align} 
\B^{(l)}_a &= \epsilon_{apq} \Omega^p \B^q_{\ k_2\cdots k_l} 
\Omega^{k_2} \cdots \Omega^{k_l}, \\ 
\B^{(l)}_{ab} &= \bigl(\epsilon_{apq} \Omega^p 
\B^q_{\ d k_3 \cdots k_l} \gamma^d_{\ b} 
\nonumber \\ & \quad \mbox{} 
+ \epsilon_{bpq} \Omega^p \B^q_{\ c k_3 \cdots k_l} 
\gamma^c_{\ a} \bigr) \Omega^{k_3} \cdots \Omega^{k_l},
\end{align} 
\end{subequations} 
and they satisfy the eigenvalue equations 
\begin{subequations} 
\label{odd_eigen_cart} 
\begin{align} 
& r^2 \gamma^{cd} D_c D_d \B^{(l)}_a 
+ \bigl[ l(l+1)-1\bigr] \B^{(l)}_a = 0, \\ 
& r^2 \gamma^{cd} D_c D_d \B^{(l)}_{ab} 
+ \bigl[ l(l+1)-4 \bigr] \B^{(l)}_{ab} = 0, 
\end{align} 
\end{subequations} 
the same as in the even-parity case. The potentials become  
\begin{subequations}
\begin{align}  
\B^{(l)}_A &= \Omega^a_A \epsilon_{apq} \Omega^p 
\B^q_{\ k_2\cdots k_l} \Omega^{k_2} \cdots \Omega^{k_l}, \\ 
\B^{(l)}_{AB} &= \bigl( \Omega^a_A \epsilon_{apq} \Omega^p 
\B^q_{\ b k_3 \cdots k_l} \Omega^b_B 
\nonumber \\ & \quad \mbox{} 
+ \Omega^b_B \epsilon_{bpq} \Omega^p \B^q_{\ a k_3 \cdots k_l} 
\Omega^a_A \bigr) \Omega^{k_3} \cdots \Omega^{k_l}
\end{align} 
\end{subequations} 
after transformation to angular coordinates. We wish to decompose
these in the odd-parity harmonics of Eqs.~(\ref{Ylm_odd}), which
satisfy the eigenvalue equations  
\begin{subequations} 
\label{odd_eigen_ang} 
\begin{align} 
& \Omega^{CD} D_C D_D X^{lm}_A 
+ \bigl[ l(l+1) - 1 \bigr] X^{lm}_A = 0, \\ 
& \Omega^{CD} D_C D_D X^{lm}_{AB} 
+ \bigl[ l(l+1) - 4 \bigr] X^{lm}_{AB} = 0,
\end{align}
\end{subequations} 
the same as in the even-parity case. 

We once more begin with $\B^{(l)} 
:= \B_{k_1 \cdots k_l} \Omega^{k_1} \cdots \Omega^{k_l}$ and its
decomposition 
\begin{equation} 
\B^{(l)} = \sum_m \B^{(l)}_m Y^{lm}. 
\end{equation} 
Differentiating the first expression, multiplying this by the
Levi-Civita tensor, and involving the second of
Eqs.~(\ref{ang_identities2}) returns 
$\epsilon_A^{\ B} D_B \B^{(l)} = -l \Omega^a_A \epsilon_{apq}
\Omega^p \B^q_{\ k_2 \cdots k_l} \Omega^{k_2} \cdots
\Omega^{k_l}$. From this we conclude that 
\begin{equation} 
\B^{(l)}_A = \frac{1}{l} \bigl( -\epsilon_A^{\ B} D_B \bigr) \B^{(l)}  
= \frac{1}{l} \sum_m \B^{(l)}_m X_A^{lm}. 
\end{equation} 
This is the decomposition of $\B^{(l)}_A$ in vectorial, 
odd-parity harmonics; this equation is valid for $l \neq 0$. After a
second differentiation we get  
$-\epsilon_A^{\ C} D_B D_C \B^{(l)} = l \epsilon_{AB} \B^{(l)} 
+ l(l-1) \Omega^a_A \epsilon_{apq} \Omega^p \B^q_{\ b k_3 \cdots k_l}
\Omega^b_B \Omega^{k_3} \cdots \Omega^{k_l}$, and after symmetrization
of the indices we obtain 
\begin{align} 
\B^{(l)}_{AB} &= -\frac{1}{l(l-1)} \bigl( \epsilon_A^{\ C} D_B 
+ \epsilon_B^{\ C} D_A \bigr) D_C \B^{(l)}  \nonumber \\ 
&= \frac{2}{l(l-1)} \sum_m \B^{(l)}_m X_{AB}^{lm}. 
\end{align} 
This is the decomposition of $\B^{(l)}_{AB}$ in tensorial, 
odd-parity harmonics; this equation is valid for $l \neq \{0,1\}$.  

\section{Determinant of the horizon metric} 
\label{app:determinant} 

The horizon metric of Eq.~(\ref{horizon_metric}) can be expressed as
$\gamma_{AB} = 4M^2\Omega_{AB} + p_{AB}$. The metric determinant is
given by $\sqrt{\gamma} = 4 M^2\sin\theta(1 +
\frac{1}{2} \varepsilon + \frac{1}{8} \varepsilon^2 - \frac{1}{4}
\varepsilon^A_{\ B} \varepsilon^B_{\ A} + \cdots)$, where
$\varepsilon^A_{\ B} := \frac{1}{4} M^{-2} \Omega^{AC} p_{CB}$ and
$\varepsilon = \varepsilon^A_{\ A}$. Evaluating this gives 
\begin{widetext} 
\begin{align} 
\sqrt{\gamma} &= 4 M^2\sin\theta \biggl[ 1 
+ \frac{8}{45} M^4 \bigl( \PP{m} + \QQ{m} \bigr)
- \frac{32}{45} M^4 \GG{d} 
- \frac{16}{21} M^4 \bigl( \PP{q} + \QQ{q} \bigr) 
+ \frac{8}{9} M^4 \GG{o} 
+ \frac{2}{9} M^4 \bigl( \PP{h} + \QQ{h} \bigr)
\nonumber \\ & \quad \mbox{} 
- \frac{1}{9} M^4 \bigl( \EE{q}_{AB} + \BB{q}_{AB} \bigr) 
  \bigl( \EEup{q}{AB} + \BBup{q}{AB} \bigr) + O(5) \biggr], 
\end{align}
\end{widetext} 
where the indices on $\EE{q}_{AB}$ and $\BB{q}_{AB}$ are raised with
$\Omega^{AB}$. This convention will be used consistently below: all
indices on tidal potentials will be lowered with $\Omega_{AB}$ and
raised with $\Omega^{AB}$. 

The identities 
\begin{subequations} 
\label{square_identities} 
\begin{align}
& \EE{q}_{ab} \EEup{q}{ab}= 
 \EE{q}_{AB} \EEup{q}{AB} = 
\frac{8}{5} \PP{m} - \frac{48}{7} \PP{q} + 2 \PP{h}, \\
& \EE{q}_{ab} \BBup{q}{ab} =
\EE{q}_{AB} \BBup{q}{AB} = 
-\frac{16}{5} \GG{d} + 4 \GG{o}, \\ 
& \BB{q}_{ab} \BBup{q}{ab} = 
\BB{q}_{AB} \BBup{q}{AB} = 
\frac{8}{5} \QQ{m} - \frac{48}{7} \QQ{q} + 2\QQ{h}
\end{align} 
\end{subequations}
can be established by direct computation, by making use of the  
definitions of the tidal potentials provided in
Tables~\ref{tab:E_cart}, \ref{tab:B_cart}, \ref{tab:EE_cart},
\ref{tab:BB_cart}, \ref{tab:EBeven_cart}, and
\ref{tab:EBodd_cart}. Inserting them into our expression for
$\sqrt{\gamma}$ returns the simple expression displayed in
Eq.~(\ref{determinant}). 
 
\section{Calculation of tidal heating} 
\label{app:horizon_details} 

In this appendix we provide calculational details regarding the
heating of the black hole by the tidal interaction. The results
derived here are used in various places in Sec.~\ref{subsec:heating}.  

To a degree of accuracy that is sufficient for our purposes, the
inverse to the horizon metric of Eq.~(\ref{horizon_metric}) is 
\begin{equation} 
\gamma^{AB} = \frac{1}{4} M^{-2} \Omega^{AB} 
+ \frac{1}{6} \bigl( \EEup{q}{AB} 
+ \BBup{q}{AB} \bigr) + M^{-2} O(3).  
\end{equation} 
It is understood that on the right-hand side of
this equation, upper-case Latin indices are raised with $\Omega^{AB}$. 

We use $\gamma^{AB}$ to raise indices on the shear tensor of 
Eq.~(\ref{shear}). This yields 
\begin{widetext} 
\begin{align} 
\sigma^{AB} &= 
- \frac{1}{12} \bigl( \EEdup{q}{AB} + \BBdup{q}{AB} \bigr) 
- \frac{1}{60} M \bigl( \EEdup{o}{AB} + \BBdup{o}{AB} \bigr) 
- \frac{1}{420} M^2 \bigl( \EEdup{h}{AB} + \BBdup{h}{AB} \bigr)
\nonumber \\ & \quad \mbox{}
- \frac{1}{45} M^2 \Omega^{AB} 
   \bigl( \PPd{m} + \QQd{m} \bigr)
+ \frac{4}{45} M^2 \Omega^{AB}\, \GGd{d} 
+ \frac{2}{21} M^2 \Omega^{AB} 
   \bigl( \PPd{q} + \QQd{q} \bigr) 
- \frac{1}{9} M^2 \Omega^{AB}\, \GGd{o} 
\nonumber \\ & \quad \mbox{}
- \frac{1}{36} M^2 \Omega^{AB} 
   \bigl( \PPd{h} + \QQd{h} \bigr)
+ \frac{1}{42} M^2 \PPdup{h}{AB} 
- \frac{5}{63} M^2 \QQdup{h}{AB}  
+ \frac{13}{126} M^2 \HHdup{h}{AB}
+ M^{-3} O(6). 
\label{shear_inverse}  
\end{align} 
\end{widetext} 
To arrive at this result we made use of the identity  
$\A_{AC} \A^C_{\ B}  = \frac{1}{2} \Omega_{AB} \A_{CD} \A^{CD}$
satisfied by any symmetric-tracefree tensor $\A_{AB}$ on the unit
two-sphere, as well as Eqs.~(\ref{square_identities}) from
Appendix~\ref{app:determinant}. We emphasize that while indices on
$\sigma_{AB}$ are raised with the physical horizon metric
$\gamma^{AB}$, indices on all tidal potentials are raised with
$\Omega^{AB}$.  

To evaluate the right-hand side of Eq.~(\ref{Ray_integrated}) we use
Eqs.~(\ref{shear}) and (\ref{shear_inverse}) to construct $\sigma_{AB}
\sigma^{AB}$, which we integrate over the horizon with the help of
Eq.~(\ref{determinant}). The shear tensor is expressed as a multipole
expansion, and its square consists of products of multipole
moments. Some of these products involve moments of the same
order. For example, $\sigma_{AB} \sigma^{AB}$ contains
the term $\frac{1}{9} M^4 (\EEd{q}_{AB} + \BBd{q}_{AB})
(\EEdup{q}{AB} + \BBdup{q}{AB})$, which is a product of quadrupole
moments; such terms survive an angular integration and contribute to
the right-hand side of Eq.~(\ref{Ray_integrated}). Other products
involve moments of different orders, and those integrate to zero. 

To evaluate an angular integral such as $\int \EEd{q}_{AB}
\EEdup{q}{AB}\, d\Omega$, where $d\Omega := \sin\theta\, d\theta
d\phi$, we  recall the definition $\EEd{q}_{AB} 
= \EEd{q}_{ab}\Omega^a_A \Omega^b_B$ and deduce the identity
$\EEd{q}_{AB} \EEdup{q}{AB} = \EEd{q}_{ab} \EEdup{q}{ab}$. We next
substitute the expression for $\EEd{q}_{ab}$ found in
Table~\ref{tab:E_cart} and carry out the integration. These steps are 
repeated for all other relevant products of multipole moments, and we   
obtain 
\begin{subequations}
\begin{align}  
\frac{1}{4\pi} \int \EEd{q}_{AB} \EEdup{q}{AB}\, d\Omega 
&= \frac{8}{5} \dot{\E}_{ab} \dot{\E}^{ab}, \\ 
\frac{1}{4\pi} \int \BBd{q}_{AB} \BBdup{q}{AB}\, d\Omega 
&= \frac{8}{5} \dot{\B}_{ab} \dot{\B}^{ab}, \\ 
\frac{1}{4\pi} \int \EEd{o}_{AB} \EEdup{o}{AB}\, d\Omega 
&= \frac{8}{21} \dot{\E}_{abc} \dot{\E}^{abc},  \\
\frac{1}{4\pi} \int \BBd{o}_{AB} \BBdup{o}{AB}\, d\Omega 
&= \frac{128}{189} \dot{\B}_{abc} \dot{\B}^{abc}.
\end{align}
\end{subequations}  
All other integrations vanish, or contribute to the right-hand-side of
Eq.~(\ref{Ray_integrated}) at order $(M/\R)^9$ and beyond.  

The general solution to Eq.~(\ref{Adeq}) is 
\begin{equation} 
\dot{\cal A}(v) = e^{\kappa_0 v} \dot{\cal A}(0) 
- 8\pi \int_0^v {\cal F}(v') e^{\kappa_0(v- v')}\, dv'. 
\end{equation} 
After three integration by parts this becomes 
\begin{align} 
\frac{\kappa_0}{8\pi}  \dot{\cal A}(v) &= e^{\kappa_0 v} \biggl[   
\frac{\kappa_0}{8\pi}  \dot{\cal A}(0) - {\cal F}(0) 
- \frac{1}{\kappa_0} \dot{\cal F}(0)
- \frac{1}{\kappa_0^2} \ddot{\cal F}(0)\biggr] 
\nonumber \\ & \quad \mbox{} 
+ {\cal F}(v) + \frac{1}{\kappa_0} \dot{\cal F}(v)
+ \frac{1}{\kappa_0^2} \ddot{\cal F}(v)
\nonumber \\ & \quad \mbox{} 
+ \frac{1}{\kappa_0^2} 
  \int_0^v \frac{d^3 {\cal F}}{dv'^3} e^{\kappa_0(v- v')}\, dv'.    
\label{Adot_general} 
\end{align} 
The last term can be neglected, because 
$\kappa_0^{-3} d^3 {\cal F}/dv^3$ is of order $(M/\R)^9$ and 
beyond the accuracy maintained in the computation of the flux
function. The first collection of terms, those that depend on the
initial conditions at $v=0$, grow exponentially over a short time
scale of order $\kappa_0^{-1} = 4M$. We do not expect the black-hole
area to behave in this way; we expect instead that the tidal
interaction will produce a slow growth over a much longer time
scale. To eliminate the run-away solution we demand that the solution
satisfy the initial condition 
\begin{equation} 
\frac{\kappa_0}{8\pi}  \dot{\cal A}(0) = {\cal F}(0) 
+ \frac{1}{\kappa_0} \dot{\cal F}(0)
+ \frac{1}{\kappa_0^2} \ddot{\cal F}(0) + O(9). 
\label{Adot_init} 
\end{equation} 
Under this restriction, Eq.~(\ref{Adot_general}) reduces to 
\begin{equation} 
\frac{\kappa_0}{8\pi}  \dot{\cal A}(v) = {\cal F}(v) 
+ \frac{1}{\kappa_0} \dot{\cal F}(v)
+ \frac{1}{\kappa_0^2} \ddot{\cal F}(v)
+ O(9), 
\label{Adot_v} 
\end{equation} 
and this is just Eq.~(\ref{Adot_sol}). Notice that Eq.~(\ref{Adot_v})
is compatible with the initial conditions of Eq.~(\ref{Adot_init}).  

It is unusual, when dealing with horizons, to impose {\it initial}
conditions on solutions to differential equations, as we have done in
Eq.~(\ref{Adot_init}). The reason, of course, is that event horizons
are always identified by {\it final} conditions. We chose to proceed
in this way because our horizon is not necessarily an event horizon,
as we explained in Sec.~\ref{subsec:horizon_position}. We also
explained that the horizon {\it becomes} an event horizon when the
tidal interaction switches off in the remote future. Under this
condition we may impose the {\it final} condition that 
$\dot{\cal A} = 0$ when  $v = \infty$. The exact solution to
Eq.~(\ref{Adeq}) is then  
\begin{equation} 
\dot{A}(v) = 8\pi \int_v^\infty {\cal F}(v') e^{-\kappa_0(v'-v)}\,
dv',   
\end{equation} 
and this leads once more to Eq.~(\ref{Adot_v}) after three 
integrations by parts. 

\section{Alternative derivation of the background metric} 
\label{app:alternative} 

In this appendix we give a brief sketch of an alternative derivation
of the background metric of Eqs.~(\ref{background_metric_cart}). The
derivation described in Sec.~\ref{sec:background_derivation} relied on
Zhang's observation \cite{zhang:86} that the metric of a vacuum region
of spacetime around a timelike geodesic $\gamma$ is a functional of
two sets of tidal moments $\E_{a_1 a_2 \cdots a_l}$ and 
$\B_{a_1 a_2 \cdots a_l}$; these are STF tensors that depend on proper
time on the world line, and they are related to components of the Weyl
tensor (and its derivatives) evaluated on the world line. Here we
provide a precise definition of the light-cone coordinates and
construct the metric systematically through order $(r/\R)^4$; we
recover the metric of Eqs.~(\ref{background_metric_cart}) and
therefore confirm the validity of Zhang's observation. Our development
follows the general methods developed in Ref.~\cite{poisson:04a}. 

\subsection{Formal definition of light-cone coordinates}  

As in Sec.~\ref{subsec:tidal_def} we consider a timelike geodesic
$\gamma$ described by the parametric relations $z^\alpha(\tau)$, with
$\tau$ denoting proper time. On $\gamma$ we install an orthonormal  
tetrad $(u^{\alpha'}, e^{\alpha'}_a)$ of parallel-transported
vectors. (We use primed indices to refer to points on the world line;
unprimed indices will refer to points off the world line.) For the
time being we do not assume that $\gamma$ is situated in a vacuum
region of spacetime; this assumption will be incorporated at a later
stage. To assign light-cone coordinates $(v,x^a)$ to a point $x$ off
the world line we locate the unique future-directed null geodesic
segment $\beta$ that begins at $x$ and ends at a point $x'$ on the
world line. (The construction requires that $x$ be in the normal
convex neighborhood of $x'$; the coordinates are defined in this
neighborhood only.) The advanced-time coordinate $v$ is the value of
the proper-time parameter $\tau$ at this point:  $x' = z(\tau =
v)$. And the quasi-Cartesian coordinates $x^a$ are defined by 
\begin{equation} 
x^a := -e^a_{\alpha'} \sigma^{\alpha'}(x,x'), 
\label{coord_def} 
\end{equation} 
where $e^a_{\alpha'} := \delta^{ab} g_{\alpha'\beta'} e^{\beta'}_b$
and $\sigma^{\alpha'} := \nabla^{\alpha'} \sigma$ is Synge's world
function $\sigma(x,x')$ \cite{synge:60, dewitt-brehme:60}
differentiated with respect to its second argument. The points $x$ and
$x'$ are related by the condition $\sigma(x,x') = 0$, which indicates
that the points are linked by a null geodesic segment.  

We define 
\begin{equation} 
r := -\sigma_{\alpha'} u^{\alpha'} 
\end{equation} 
and state without proof that $r$ is an affine parameter on $\beta$; it
decreases as the null geodesic approaches the world line. (The proof
of this statement is contained in Ref.~\cite{poisson:04a}.) The
completeness relation $g^{\alpha'\beta'} = -u^{\alpha'} u^{\beta'} 
+ \delta^{ab} e^{\alpha'}_a e^{\beta'}_b$ and the identity
$\sigma^{\alpha'} \sigma_{\alpha'} = 2\sigma = 0$ imply that 
$r^2 = \delta_{ab} x^a x^b$. It is useful to introduce 
\begin{equation} 
\Omega^a := x^a/r, 
\label{Omega_def} 
\end{equation} 
and we use the completeness relation to write 
\begin{equation} 
\sigma^{\alpha'} = r \bigl( u^{\alpha'} - \Omega^a e^{\alpha'}_a
\bigr). 
\label{sigma} 
\end{equation} 
This is a decomposition of the displacement vector
$\sigma^{\alpha'}(x,x')$ in the tetrad $(u^{\alpha'},
e^{\alpha'}_a)$. The vector 
\begin{equation} 
\ell_\alpha := \sigma_\alpha/r 
\label{ell} 
\end{equation} 
is future-directed and tangent to null geodesic segment
$\beta$. Here $\sigma_\alpha := \nabla_\alpha \sigma$ is the world
function $\sigma(x,x')$ differentiated with respect to its first
argument.  

Suppose now that the point $x$ is moved to a neighboring point 
$x + \delta x$. The coordinates of the new point will be $v 
+ \delta v$ and $x^a + \delta x^a$, and to calculate the coordinate
displacements we must locate the new point $x' + \delta x'$ on the
world line, which is liked to $x + \delta x$ by a new geodesic segment
$\beta + \delta \beta$. Using $\delta x^{\alpha'} = u^{\alpha'}
\delta v$ and expanding $\sigma(x+\delta x, x'+\delta x') =
0$ to first order in the displacements, it is easy to show that 
$\delta v =  -\ell_\alpha\, \delta x^\alpha$, so that 
\begin{equation} 
\partial_\alpha v = -\ell_\alpha. 
\label{coord_disp_time} 
\end{equation} 
The definition of the coordinates $x^a$ in terms of the world
function then implies that $\delta x^a = -(e^a_{\alpha'}
\sigma^{\alpha'}_{\ \beta'} u^{\beta'})\, \delta v - e^a_{\alpha'}
\sigma^{\alpha'}_{\ \beta}\,\delta x^\beta$, so that 
\begin{equation} 
\partial_\alpha x^a = \bigl( e^a_{\alpha'}
\sigma^{\alpha'}_{\ \beta'} u^{\beta'} \bigr)\, \ell_\alpha 
- e^a_{\beta'}\sigma^{\beta'}_{\ \alpha}. 
\label{coord_disp_spatial}
\end{equation} 
Here $\sigma^{\alpha'}_{\ \beta'} := \nabla^{\alpha'} \nabla_{\beta'}
\sigma$ is the second covariant derivative of the world function with
respect to $x'$, while $\sigma^{\beta'}_{\ \alpha} := \nabla^{\beta'}
\nabla_{\alpha'} \sigma$ denotes a mixed derivative with respect to  
each argument.    

\subsection{Metric} 

We begin with a computation of the inverse metric, with components 
\begin{subequations} 
\begin{align} 
g^{vv} &= g^{\alpha\beta} \partial_\alpha v \partial_\beta v, \\ 
g^{va} &= g^{\alpha\beta} \partial_\alpha x^a \partial_\beta v, \\ 
g^{ab} &= g^{\alpha\beta} \partial_\alpha x^a \partial_\beta x^b
\end{align} 
\end{subequations} 
Using Eq.~(\ref{coord_disp_time}) and the fact that $\ell_\alpha$ is a
null vector we immediately find that $g^{vv} = 0$. With
Eq.~(\ref{coord_disp_spatial}) we get $g^{va} = e^a_{\alpha'}
\sigma^{\alpha'}_{\ \beta} \ell^\beta$, and we simplify this with the
help of Eq.~(\ref{ell}) and the identity $\sigma^{\alpha'}_{\ \beta}
\sigma^{\beta} = \frac{1}{2} \nabla^{\alpha'} (\sigma_{\beta}
\sigma^{\beta}) = \nabla^{\alpha'} \sigma$. With
Eqs.~(\ref{coord_def}) and (\ref{Omega_def}), this is 
$g^{va} = \Omega^a$. To obtain the components 
\begin{align} 
g^{ab} &= e^a_{\alpha'} e^b_{\beta'} g^{\alpha\beta}
\sigma^{\alpha'}_{\ \alpha} \sigma^{\beta'}_{\ \beta} 
- \bigl( e^a_{\alpha'} \sigma^{\alpha'}_{\ \beta'} u^{\beta'} \bigr)
\Omega^b 
\nonumber \\ & \qquad \mbox{} 
- \bigl( e^b_{\alpha'} \sigma^{\alpha'}_{\ \beta'} u^{\beta'}
\bigr) \Omega^a 
\label{spatial_metric} 
\end{align}
requires a much longer computation, and the result will be expressed
as an expansion in powers of $r$. 

\begin{widetext} 
We rely on the known expansions (see, for example,
Ref.~\cite{anderson-flanagan-ottewill:05})
\begin{align} 
\sigma_{\alpha'\beta'} &= g_{\alpha'\beta'}
- \frac{1}{3} R_{\alpha'\gamma'\beta'\delta'}
  \sigma^{\gamma'}\sigma^{\delta'} 
+ \frac{1}{12} R_{\alpha'\gamma'\beta'\delta';\epsilon'}
  \sigma^{\gamma'}\sigma^{\delta'}\sigma^{\epsilon'} 
\nonumber \\ & \qquad \mbox{} 
- \biggl( \frac{1}{60} R_{\alpha'\gamma'\beta'\delta';\epsilon'\iota'} 
+ \frac{1}{45} R_{\alpha'\gamma'\delta'\mu'} 
  R^{\mu'}_{\ \epsilon'\iota'\beta'} \biggr)  
  \sigma^{\gamma'}\sigma^{\delta'}\sigma^{\epsilon'}\sigma^{\iota'}
+ \cdots, \\
\sigma_{\alpha'\beta} &= g^{\beta'}_{\ \beta} \biggl[
-g_{\alpha'\beta'} 
- \frac{1}{6} R_{\alpha'\gamma'\beta'\delta'}
  \sigma^{\gamma'}\sigma^{\delta'} 
+ \frac{1}{12} R_{\alpha'\gamma'\beta'\delta';\epsilon'}
  \sigma^{\gamma'}\sigma^{\delta'}\sigma^{\epsilon'} 
\nonumber \\ & \qquad \mbox{} 
- \biggl( \frac{1}{40} R_{\alpha'\gamma'\beta'\delta';\epsilon'\iota'} 
+ \frac{7}{360} R_{\alpha'\gamma'\delta'\mu'} 
  R^{\mu'}_{\ \epsilon'\iota'\beta'} \biggr)  
  \sigma^{\gamma'}\sigma^{\delta'}\sigma^{\epsilon'}\sigma^{\iota'}
+ \cdots \biggr] 
\end{align}
for the derivatives of the world function, in which
$g^{\beta'}_{\ \beta}(x,x')$ is the parallel propagator
\cite{synge:60, dewitt-brehme:60}. We make the substitutions in
Eq.~(\ref{spatial_metric}) and express $\sigma^{\alpha'}$ as in
Eq.~(\ref{sigma}).  

The end result of a lengthy computation is 
\begin{subequations}
\label{inverse_metric} 
\begin{align}  
g^{vv} &= 0, \\ 
g^{va} &= \Omega^a, \\ 
g^{ab} &= \delta^{ab} 
+ \frac{1}{3} r^2 \bigl( P^{ab} + P^a \Omega^b + P^b \Omega^a \bigr) 
- \frac{1}{12} r^3 \bigl( 2\dot{P}^{ab} + \dot{P}^a \Omega^b 
   + \dot{P}^b \Omega^a \bigr) 
- \frac{1}{12} r^3 \bigl( 2Q^{ab} + Q^a \Omega^b + Q^b \Omega^a \bigr) 
\nonumber \\ & \qquad \mbox{}
+ \frac{1}{60} r^4 \bigl( 3\ddot{P}^{ab} + \ddot{P}^a \Omega^b 
   + \ddot{P}^b \Omega^a \bigr) 
+ \frac{1}{30} r^4 \bigl( 3\dot{Q}^{ab} + \dot{Q}^a \Omega^b 
   + \dot{Q}^b \Omega^a \bigr) 
+ \frac{1}{60} r^4 \bigl( 3S^{ab} + S^a \Omega^b + S^b \Omega^a \bigr) 
\nonumber \\ & \qquad \mbox{}
+ \frac{1}{60} r^4 \bigl( 3U^{ab} + U^a \Omega^b + U^b \Omega^a \bigr) 
+ \frac{1}{45} r^4 \bigl( 3V^{ab} + V^a \Omega^b + V^b \Omega^a \bigr) 
+ O(r^5). 
\end{align}
\end{subequations} 
The inverse metric is expressed in terms of the potentials  
\begin{subequations}
\begin{align}  
P_{ab} &= R_{a0b0} - \bigl( R_{acb0} + R_{bca0} \bigr) \Omega^c 
+ R_{acbd} \Omega^c \Omega^d  = P_{ba} \\ 
P_{a} &= R_{a0c0} \Omega^c - R_{acd0} \Omega^c \Omega^d 
= P_{ab} \Omega^b, \\ 
P &= R_{c0d0} \Omega^c \Omega^d = P_a \Omega^a, \\
Q_{ab} &= -R_{a0b0|c}\Omega^c 
+ \bigl( R_{acb0|d} + R_{bca0|d} \bigr) \Omega^c\Omega^d
- R_{acbd|e} \Omega^c \Omega^d \Omega^e = Q_{ba} \\ 
Q_{a} &= -R_{a0c0|d} \Omega^c\Omega^d 
+ R_{acd0|e} \Omega^c \Omega^d \Omega^e = Q_{ab} \Omega^b, \\ 
Q &= -R_{c0d0|e} \Omega^c \Omega^d \Omega^e = Q_{a} \Omega^a, \\
S_{ab} &= R_{a0b0|cd}\Omega^c\Omega^d 
- \bigl( R_{acb0|de} + R_{bca0|de} \bigr) \Omega^c\Omega^d\Omega^e
+ R_{acbd|ef} \Omega^c \Omega^d \Omega^e \Omega^f = S_{ba}, \\ 
S_{a} &= R_{a0c0|de} \Omega^c\Omega^d \Omega^e 
- R_{acd0|ef} \Omega^c \Omega^d \Omega^e \Omega^f 
= S_{ab} \Omega^b, \\ 
S &= R_{c0d0|ef} \Omega^c \Omega^d \Omega^e\Omega^f = S_{a} \Omega^a,  \\
U_{ab} &= \bigl( R_{a0m0}R^m_{\ bc0} + R_{b0m0}R^m_{\ ac0} 
+ R_{amb0} R^m_{\ 0c0} + R_{bma0} R^m_{\ 0c0} \bigr) \Omega^c 
\nonumber \\ & \qquad \mbox{} 
+ \bigl( 2R_{a0c0} R_{b0d0} + R_{amc0} R^m_{\ db0} 
+ R_{bmc0} R^m_{\ da0} - 2 R_{a0b0} R_{0c0d} 
- R_{amb0} R^m_{\ cd0} 
- R_{bma0} R^m_{\ cd0}
\nonumber \\ & \qquad \mbox{} 
- R_{acm0} R^m_{\ bd0} - R_{bcm0} R^m_{\ ad0} 
- R_{acbm} R^m_{\ 0d0} - R_{bcam} R^m_{\ 0d0} \bigr) \Omega^c \Omega^d 
\nonumber \\ & \qquad \mbox{} 
+ \bigl( -R_{a0c0} R_{bde0} - R_{b0c0} R_{ade0} 
+ R_{acmd} R^m_{\ be0} + R_{bcmd} R^m_{\ ae0} 
+ R_{acb0} R_{d0e0} 
\nonumber \\ & \qquad \mbox{} 
+ R_{bca0} R_{d0e0} 
+ R_{acbm} R^m_{\ de0} + R_{bcam} R^m_{\ de0} \bigr) 
\Omega^c \Omega^d \Omega^e = U_{ba}, \\
U_a &= \bigl( R_{a0m0} R^m_{\ cd0} + R_{cma0} R^m_{\ 0d0} \bigr) 
\Omega^c \Omega^d - \bigl( R_{acm0} R^m_{\ de0} 
+ R_{acdm} R^m_{\ 0e0} \bigr) \Omega^c \Omega^d \Omega^e 
= U_{ab} \Omega^b, \\     
V_{ab} &= P_{ac} P^{c}_{\ b} - P_a P_b = V_{ba}, \\
V_{a} &= P_{ac} P^c - P_a P, \\ 
V &= P_c P^c - P^2,  
\end{align}
\end{subequations} 
\end{widetext} 
which are defined in terms of the frame components of the Riemann
tensor (and its derivatives) evaluated on the world line. We adopt the
notation introduced in Sec.~\ref{subsec:tidal_def}, and an overdot
indicates differentiation with respect to proper time. For example,
$\dot{P}_{ab} := \dot{R}_{a0b0} - ( \dot{R}_{acb0} + \dot{R}_{bca0} )
\Omega^c + \dot{R}_{acbd} \Omega^c \Omega^d$, where, for example,  
$\dot{R}_{a0b0} := R_{\alpha'\mu'\beta'\nu';\lambda'} e^{\alpha'}_a
u^{\mu'} e^{\beta'}_b u^{\nu'} u^{\lambda'}$. Notice that the
derivative operator acts on the Riemann tensor only. The fact that the
tetrad is parallel transported on the world line implies that the
right-hand side can also be written as 
$(R_{\alpha'\mu'\beta'\nu'} e^{\alpha'}_a u^{\mu'} e^{\beta'}_b 
u^{\nu'})_{;\lambda'} u^{\lambda'}$, and we find that $\dot{R}_{a0b0} 
= dR_{a0b0}/d\tau$. To avoid ambiguities with second derivatives of
the Riemann tensor, we always differentiate with respect to proper
time in the last step; for example, $\dot{R}_{a0b0|c} 
:= R_{\alpha'\mu'\beta'\nu';\gamma'\lambda'} e^{\alpha'}_a
u^{\mu'} e^{\beta'}_b u^{\nu'} e^{\gamma'}_c u^{\lambda'}$. 

The inverse of Eqs.~(\ref{inverse_metric}) is calculated using the
techniques introduced in Sec.~\ref{subsec:kinematic}. We obtain 
\begin{widetext} 
\begin{subequations}
\label{metric} 
\begin{align} 
g_{vv} &= -1 - r^2 P + \frac{1}{3} r^3 \dot{P} + \frac{1}{3} r^3 Q 
- \frac{1}{12} r^4 \ddot{P} - \frac{1}{6} r^4 \dot{Q} 
- \frac{1}{12} r^4 S + \frac{1}{3} r^4 V + O(r^5), \\ 
g_{va} &= \Omega_a + \gamma_a^{\ c}  
\biggl[ -\frac{2}{3} r^2 P_c + \frac{1}{4} r^3 \dot{P}_c
+ \frac{1}{4} r^3 Q_c - \frac{1}{15} r^4 \ddot{P}_c 
- \frac{2}{15} r^4 \dot{Q}_c 
\nonumber \\ & \qquad \mbox{}
- \frac{1}{15} r^4 S_c 
- \frac{1}{15} r^4 U_c + \frac{2}{15} r^4 V_c + O(r^5) \biggr], \\ 
g_{ab} &= \gamma_{ab}  
+ \gamma_a^{\ c} \gamma_b^{\ d} 
\biggl[ -\frac{1}{3} r^2 P_{cd} + \frac{1}{6} r^3 \dot{P}_{cd}
+ \frac{1}{6} r^3 Q_{cd}  
- \frac{1}{20} r^4 \ddot{P}_{cd}
- \frac{1}{10} r^4 \dot{Q}_{cd} 
\nonumber \\ & \qquad \mbox{}
- \frac{1}{20} r^4 S_{cd} - \frac{1}{20} r^4 U_{cd} 
+ \frac{2}{45} r^4 V_{cd} + O(r^5) \biggr], 
\end{align}
\end{subequations} 
\end{widetext} 
 where $\gamma_{ab} := \delta_{ab} - \Omega_a \Omega_b$. We recognize
the structure of Eqs.~(\ref{metric1_cart}), with a clear decomposition
of the metric into longitudinal and transverse pieces. 

\subsection{Decomposition of the Weyl tensor}

At this stage we demand that the Ricci tensor, its first derivatives,
and its second derivatives, all vanish on the world line
$\gamma$. This implies that the Riemann tensor
$R_{\alpha'\beta'\gamma'\delta'}$ and its derivatives are
equal to the Weyl tensor $C_{\alpha'\beta'\gamma'\delta'}$ and its
derivatives. The symmetries of the Weyl tensor, the Bianchi
identities, and the Ricci identities then imply that the Weyl tensor
(and its derivatives) can be expressed in terms of the tidal moments
$\E_{ab}$, $\E_{abc}$, $\E_{abcd}$, $\B_{ab}$, $\B_{abc}$, and
$\B_{abcd}$; these were defined in Sec.~\ref{subsec:tidal_def}. 

The frame components of the Weyl tensor on the world line are given by 
\begin{subequations} 
\label{Weyl_dec} 
\begin{align} 
C_{a0b0} &= \E_{ab}, \\ 
C_{abc0} &= \epsilon_{abp} \B^p_{\ c}, \\ 
C_{abcd} &= -\epsilon_{abp} \epsilon_{cdq} \E^{pq}.  
\end{align} 
\end{subequations} 
The last equation can also be written as  
\begin{equation} 
C_{abcd} = \delta_{ac} \E_{bd} - \delta_{ad} \E_{bc} 
- \delta_{bc} \E_{ad} + \delta_{bd} \E_{ac}, 
\end{equation}
by making use of the general identity $\epsilon_{abp} \epsilon_{cdq} = 
\delta_{ac}(\delta_{bd}\delta_{pq} - \delta_{bq}\delta_{dp}) 
- \delta_{ad}(\delta_{bc}\delta_{pq} - \delta_{bq}\delta_{cp})  
+ \delta_{aq}(\delta_{bc}\delta_{pd} - \delta_{bd}\delta_{cp})$.

The frame components of the first spatial derivatives of the Weyl
tensor are 
\begin{subequations} 
\label{Weyl_first_dec} 
\begin{align} 
C_{a0b0|c} &= \E_{ab|c}, \\ 
C_{abc0|d} &= \epsilon_{abp} \B^p_{\ c|d}, \\ 
C_{abcd|e} &= -\epsilon_{abp} \epsilon_{cdq} \E^{pq}_{\ \ |e},   
\end{align} 
\end{subequations} 
where $\E_{ab|c} := C_{a0b0|c}$ and $\B_{ab|c} := \frac{1}{2}
\epsilon_{apq} C^{pq}_{\ \ b0|c}$. These are related to the tidal
moments by 
\begin{subequations} 
\begin{align} 
\E_{ab|c} &= \E_{abc} 
+ \frac{1}{3} \bigl( \epsilon_{acp} \dot{\B}^p_{\ b} 
+ \epsilon_{bcp} \dot{\B}^p_{\ a} \bigr), \\ 
\B_{ab|c} &= \frac{4}{3} \B_{abc} 
- \frac{1}{3} \bigl( \epsilon_{acp} \dot{\E}^p_{\ b} 
+ \epsilon_{bcp} \dot{\E}^p_{\ a} \bigr). 
\end{align} 
\end{subequations} 

The frame components of the (symmetrized) second spatial derivatives
of the Weyl tensor are 
\begin{subequations} 
\label{Weyl_second_dec} 
\begin{align} 
C_{a0b0|(cd)} &= \E_{ab|(cd)}, \\ 
C_{abc0|(de)} &= \epsilon_{abp} \B^p_{\ c|(de)}, \\ 
C_{abcd|(ef)} &= -\epsilon_{abp} \epsilon_{cdq} \E^{pq}_{\ \ |(ef)},    
\end{align} 
\end{subequations} 
where $\E_{ab|(cd)} := C_{a0b0|(cd)}$ and $\B_{ab|(cd)} := \frac{1}{2} 
\epsilon_{apq} C^{pq}_{\ \ b0|(cd)}$. These are related to the tidal
moments by 
\begin{widetext} 
\begin{subequations}
\begin{align} 
\E_{ab|(cd)} &= 2 \E_{abcd} + \frac{1}{3} \bigl( 
\epsilon_{acp} \dot{\B}^p_{\ bd} + \epsilon_{adp} \dot{\B}^p_{\ bc}
+ \epsilon_{bcp} \dot{\B}^p_{\ ad} + \epsilon_{bdp} \dot{\B}^p_{\ ac}
\bigr) 
\nonumber \\ & \qquad \mbox{} 
+ \frac{4}{21} \delta_{ab} \ddot{\E}_{cd} - \frac{1}{7} \bigl( 
\delta_{ac} \ddot{\E}_{bd} + \delta_{ad} \ddot{\E}_{bc}
+ \delta_{bc} \ddot{\E}_{ad} + \delta_{bd} \ddot{\E}_{ac} \bigr) 
+ \frac{11}{21} \delta_{cd} \ddot{\E}_{ab} 
\nonumber \\ & \qquad \mbox{} 
- \frac{4}{3} \bigl( \E_{ab} \E_{cd} - \B_{ab} \B_{cd} \bigr) 
+ \frac{2}{3} \bigl( \E_{ac} \E_{bd} - \B_{ac} \B_{bd} \bigr) 
+ \frac{2}{3} \bigl( \E_{ad} \E_{bc} - \B_{ad} \B_{bc} \bigr) 
\nonumber \\ & \qquad \mbox{} 
- \frac{22}{21} \delta_{ab} G_{cd}
+ \frac{19}{42} \bigl( \delta_{ac} G_{bd} + \delta_{ad} G_{bc}
+ \delta_{bc} G_{ad} + \delta_{bd} G_{ac} \bigr)
+ \frac{20}{21} \delta_{cd} G_{ab} 
\nonumber \\ & \qquad \mbox{} 
- \frac{4}{21} \bigl( \delta_{ab} \delta_{cd} 
+ \delta_{ac} \delta_{bd} + \delta_{ad} \delta_{bc} \bigr) G, \\
\B_{ab|(cd)} &= \frac{10}{3} \B_{abcd} - \frac{1}{4} \bigl( 
\epsilon_{acp} \dot{\E}^p_{\ bd} + \epsilon_{adp} \dot{\E}^p_{\ bc}
+ \epsilon_{bcp} \dot{\E}^p_{\ ad} + \epsilon_{bdp} \dot{\E}^p_{\ ac}
\bigr) 
\nonumber \\ & \qquad \mbox{} 
+ \frac{4}{21} \delta_{ab} \ddot{\B}_{cd} - \frac{1}{7} \bigl( 
\delta_{ac} \ddot{\B}_{bd} + \delta_{ad} \ddot{\B}_{bc}
+ \delta_{bc} \ddot{\B}_{ad} + \delta_{bd} \ddot{\B}_{ac} \bigr) 
+ \frac{11}{21} \delta_{cd} \ddot{\B}_{ab} 
\nonumber \\ & \qquad \mbox{} 
- \frac{4}{3} \bigl( \E_{ab} \B_{cd} + \B_{ab} \E_{cd} \bigr) 
+ \frac{2}{3} \bigl( \E_{ac} \B_{bd} + \B_{ac} \E_{bd} \bigr) 
+ \frac{2}{3} \bigl( \E_{ad} \B_{bc} + \B_{ad} \E_{bc} \bigr) 
\nonumber \\ & \qquad \mbox{} 
- \frac{22}{21} \delta_{ab} H_{cd}
+ \frac{19}{42} \bigl( \delta_{ac} H_{bd} + \delta_{ad} H_{bc}
+ \delta_{bc} H_{ad} + \delta_{bd} H_{ac} \bigr)
+ \frac{20}{21} \delta_{cd} H_{ab} 
\nonumber \\ & \qquad \mbox{} 
- \frac{4}{21} \bigl( \delta_{ab} \delta_{cd} 
+ \delta_{ac} \delta_{bd} + \delta_{ad} \delta_{bc} \bigr) H. 
\end{align} 
\end{subequations} 
\end{widetext} 
We introduced the notation 
\begin{subequations} 
\begin{align} 
G_{ab} &:= \E_{ap} \E^p_{\ b} - \B_{ap} \B^p_{\ b}, \\
G &:= \E^{pq} \E_{pq} - \B^{pq} \B_{pq}, \\  
H_{ab} &:= \E_{ap} \B^p_{\ b} + \B_{ap} \E^p_{\ b}, \\ 
H &:= 2\E^{pq} \B_{pq}. 
\end{align} 
\end{subequations}  

\subsection{Metric in irreducible form} 

The decomposition of the Weyl tensor and its derivatives in terms of
the tidal moments is substituted within $P_{ab}$, $P_a$, $P$,
$Q_{ab}$, $Q_a$, $Q$, $S_{ab}$, $S_a$, $S$, $U_{ab}$, $U_a$, $V_{ab}$,
$V_a$, and $V$. All of this is next substituted within our previous
expression for the metric tensor. The manipulations required to
simplify the expressions are extremely lengthy (on the order of 35 
pages of small handwriting). We find that the terms that are linear in
the Weyl tensor and its derivatives group themselves automatically
into the irreducible tidal potentials introduced in
Tables~\ref{tab:E_cart} and \ref{tab:B_cart}. 

The organization of the terms that are quadratic in the Weyl tensor
requires more work. For example, our initial expression for the
quadratic terms in $g_{vv}$ is 
\begin{align} 
g_{vv}^{\rm quadratic} &= 
\frac{1}{21} r^4 \bigr( \E_{pq} \E^{pq} - \B_{pq} \B^{pq} \bigr) 
\nonumber \\ & \quad \mbox{} 
+ \frac{2}{21} r^4 \bigl( 2 \E_{ap} \E^p_{\ b} 
  + 5 \B_{ap} \B^p_{\ b} \bigr) \Omega^a \Omega^b 
\nonumber \\ & \quad \mbox{} 
- \frac{1}{3} r^4 \bigl( \E_{ab} \E_{cd} + \B_{ab} \B_{cd} \bigr) 
  \Omega^a \Omega^b \Omega^c \Omega^d 
\nonumber \\ & \quad \mbox{} 
+ \frac{2}{3} r^4 \epsilon_{cpq} \E^p_{\ a} \B^q_{\ b} 
  \Omega^a \Omega^b \Omega^c.  
\end{align} 
To write this in terms of irreducible tidal potentials we write a
product of unit radial vectors such as $\Omega^a \Omega^b \Omega^c
\Omega^d$ as the STF decomposition
\begin{align} 
\Omega^a \Omega^b \Omega^c \Omega^d
&= \Omega^\stf{abcd} + \frac{1}{7} \bigl( \delta^{ab} \Omega^\stf{cd} 
+ \delta^{ac} \Omega^\stf{bd} + \delta^{ad} \Omega^\stf{bc}
\nonumber \\ & \quad \mbox{} 
+ \delta^{bc} \Omega^\stf{ad} + \delta^{bd} \Omega^\stf{ac}
+ \delta^{cd} \Omega^\stf{ab} \bigr) 
\nonumber \\ & \quad \mbox{} 
+ \frac{1}{15} \bigl( \delta^{ab} \delta{^cd} 
+ \delta^{ac} \delta^{bd} + \delta^{ad} \delta^{bc} \bigr),
\end{align}  
and we make the substitution within the metric function. Looking at
the term $\E_{ab} \E_{cd} \Omega^a \Omega^b \Omega^c \Omega^d$, for 
example, we obtain $\E_{ab} \E_{cd} \Omega^\stf{abcd} 
+ \frac{4}{7} \E_{pa} \E^p_{\ b} \Omega^\stf{ab} + \frac{2}{15}
\E_{pq} \E^{pq}$, which can be expressed in the equivalent form 
$\E_{\langle ab} \E_{cd \rangle} \Omega^a \Omega^b \Omega^c \Omega^d 
+ \frac{4}{7} \E_{p\langle a} \E^p_{\ b \rangle} \Omega^a \Omega^b 
+ \frac{2}{15} \E_{pq} \E^{pq}$. According to the definitions listed
in Table~\ref{tab:EE_cart}, this is $\PP{h} + \frac{4}{7} \PP{q} 
+ \frac{2}{15} \PP{m}$, a superposition of hexadecapole, quadrupole,
and monopole tidal potentials. 

Proceeding in a similar way with $g_{va}$ and $g_{ab}$, we eventually
arrive at Eqs.~(\ref{background_metric_cart}). 

\bibliography{../bib/master} 
\end{document}